%% file: gltresub11jhep.tex
\documentclass[11pt,a4paper,twoside]{article}
\usepackage{ifpdf}
\pdfoutput=1
\usepackage{amssymb}
\usepackage{mathrsfs}
\usepackage{amsmath}
\usepackage{rotating}
\usepackage{graphicx}
\usepackage{graphics}
\usepackage{dcolumn}
\usepackage{bm}
\usepackage{color}

\renewcommand{\thefootnote}{\fnsymbol{footnote}}

\begin{document}
\begin{titlepage}
\hfill
\vbox{
    \halign{#\hfil        \cr
    } 
  }  
  \vspace*{10mm}
  \begin{center}
    {\Large{\bf{Naturalness of Neutralino Dark Matter}}} \\
\vspace*{9mm}

{\large Philipp Grothaus}\footnote[1]{Present address: Theoretical
  Particle Physics and Cosmology Group, Physics Department, King’s
  College London, London WC2R 2LS, U.K.}\footnote[2]{E-mail:
  philipp.grothaus@kcl.ac.uk}, 
{\large Manfred Lindner}\footnote[3]{E-mail: lindner@mpi-hd.mpg.de}
and 
{\large Yasutaka Takanishi}\footnote[4]{E-mail: yasutaka@mpi-hd.mpg.de}

\vspace*{.6cm}

{\it  Max-Planck-Institut f\"ur Kernphysik, \\[1mm] 
Saupfercheckweg 1, \\[1mm] D-69117 Heidelberg, Germany}

\vspace*{.7cm}

\end{center}
\begin{abstract}
  We investigate the level of fine-tuning of neutralino Dark Matter
  below \hbox{200~GeV} in the low-energy phenomenological minimal
  supersymmetric Standard Model taking into account the newest results
  from XENON100 and the Large Hadron Collider as well as all other experimental
  bounds from collider physics and the cosmological abundance.
  We find that current and future direct Dark Matter searches
  significantly rule out a large area of the untuned parameter space,
  but solutions survive which do not increase the level of
  fine-tuning.
  As expected, the level of tuning tends to increase for lower
  cross-sections, but regions of resonant neutralino annihilation
  still allow for a band at light masses, where the fine-tuning stays
  small even below the current experimental limits for direct
  detection cross-sections.
  For positive values of the supersymmetric Higgs mass parameter $\mu$
  large portions of the allowed parameter space are excluded, but there
  still exist untuned solutions at higher neutralino masses which will
  essentially be ruled out if XENON1t does not observe a signal.  For
  negative $\mu$ untuned solutions are not much constrained by current
  limits of direct searches and, if the neutralino mass was found
  outside the resonance regions, a negative $\mu$-term would be
  favored from a fine-tuning perspective. Light stau annihilation
  plays an important role to fulfill the relic density condition in
  certain neutralino mass regions.
  Finally we discuss, in addition to the amount of tuning for certain
  regions in the neutralino mass--direct detection cross-section
  plane, the parameter mapping distribution if the allowed
  model parameter space is chosen to be scanned homogeneously
  (randomized).

\end{abstract}
\vskip 3cm

\end{titlepage}
\newpage
\renewcommand{\thefootnote}{\arabic{footnote}}
\setcounter{footnote}{0}
\setcounter{page}{1}
\section{Introduction}
\indent Recently, the XENON100 collaboration has released new results
after analyzing 225 live days of data taking. Limits on the
spin-independent elastic Dark Matter-nucleon cross-section,
$\sigma^{\rm SI}$, have been increased by a factor of roughly four
with \hbox{$2.0\times10^{-9}$~pb} as the minimal value of the upper
limit on $\sigma^{\rm SI}$ at a Dark Matter particle mass of
\hbox{55~GeV}~\cite{xenon2012}. This leads to further tests for dark
matter models.

Furthermore, the ATLAS and CMS collaborations have presented
their analysis of more than \hbox{5 fb${}^{-1}$} of \hbox{7~TeV}, also
including \hbox{8~TeV}, data and claimed close to 5 local sigma level
the existence of a Higgs boson with a mass of approximately
\hbox{125~GeV}~\cite{ATLASseminarHiggs,CMSseminarHiggs}. This fact
fits very well to the minimal supersymmetric standard model (MSSM)
because its prediction of the lightest Higgs boson mass is, when the
LEP limit is taken into account, between \hbox{115 - 135~GeV}
depending on the supersymmetric parameters, see~\hbox{\it
  e.g.}~\cite{Ellis:1990nz,Ellis:1991zd,Okada:1990vk,Haber:1990aw,%
  Drees:1991mx,Degrassi:2002fi,Buchmueller:2009fn}.

The existence of Dark Matter is supported by various
cosmological observations such as gravitational effects on visible
matter in the infrared and gravitational lensing of background
radiation. Its total abundance, that has important implications for
the evolution of the Universe, has been precisely measured by the WMAP
collaboration~\cite{Komatsu:2010fb} during the last decade.  This
requires that a different kind of matter beyond the Standard Model
(SM) of particle physics must be postulated. One of the most popular
and most intensive studied candidate is the so-called weakly
interacting massive particle (WIMP) that may constitute most of the
matter in the Universe. Cosmology provides therefore a good motivation for
Supersymmetry (SUSY), since the MSSM possesses a natural WIMP candidate as
the lightest supersymmetric particle (LSP) is stable due to $R$-parity
conservation~\cite{Goldberg:1983nd,Ellis:1983ew} (for reviews see
\hbox{\it e.g.}~\cite{Jungman:1995df,Bergstrom:2000pn}).

SUSY~(for reviews we refer
to~\cite{Nilles:1983ge,Haber:1984rc,Martin:1997ns}) has moreover the
ability to solve the famous hierarchy problem by introducing
superpartners with opposite spin statistics to each SM particle such
that the loop contributions from superpartners cancel exactly and the
weak scale is stabilized. Since SUSY must be broken, however, these
cancellations are not exact and the non-discovery of SUSY particles
pushes the breaking scale further up. This separation of the weak
scale and of the SUSY breaking scale raises the question how easily
this stability can be maintained. We apply therefore in this paper a
measure of naturalness~\cite{Ellis:1986yg,Barbieri:1987fn}, which was
used for electroweak symmetry breaking, to the Dark Matter sector and
study the level of fine-tuning.

There is a series of studies on Dark Matter in the framework of
simplified variants of the MSSM, the so-called constrained MSSM
(CMSSM), which possess universal supersymmetry breaking mass
parameters at the grand unification scale (for
example~\cite{Farina:2011bh,Buchmueller:2011sw,Buchmueller:2011ab}).
Due to the existence of the grand unification condition on the gaugino
masses in these models, there exists a LEP limit on the lightest
neutralino mass: they must be heavier than \hbox{$46$~GeV}.  According
to reference~\cite{Buchmueller:2011ab} the lightest neutralino mass
must be larger than about \hbox{$200$~GeV} at \hbox{$95$~\% C.L.}
after taking into account all relevant experimental constraints.

Instead of this restricted class of models non-universal gaugino
models within the framework of the phenomenological MSSM (pMSSM) (see
for
example~\cite{AbdusSalam:2009qd,Sekmen:2011cz,Arbey:2011un,AlbornozVasquez:2012px,CahillRowley:2012rv})
have gained much attention. In the pMSSM low-energy input parameters
are used with no high-energy relations between them. These models were
used to explain the possible annual modulation signals of
DAMA/LIBRA~\cite{Bernabei:2010mq} and CoGeNT~\cite{Aalseth:2010vx}
(\hbox{\it
  e.g.}~\cite{Hooper:2002nq,Bottino:2002ry,Dreiner:2009ic,Kuflik:2010ah,%
  Feldman:2010ke,Vasquez:2010ru,Fornengo:2010mk,Calibbi:2011ug,Arbey:2012na}),
as well as the excess of nuclear recoil events reported by
CRESST~\cite{Angloher:2011uu}.  This would be interpreted in terms of
Dark Matter with a mass between roughly \hbox{10~GeV} and
\hbox{30~GeV} and spin-independent cross-section of order
$10^{-4}-10^{-7}$ pb. However, it was shown that light neutralino
Dark Matter scenarios consistent with DAMA/LIBRA, CoGeNT and CRESST
within the pMSSM are disfavored by LHC
constraints~\cite{Calibbi:2011un}. In addition there are discussions
about the validity and natural consistency of these
signals~\cite{Kopp:2011yr} and
XENON100~\cite{xenon2012,Aprile:2011hi,Aprile:2010um} (see
also~\cite{Angle:2007uj}) as well as CDMS~\cite{Ahmed:2010wy}, since
these experiments have excluded these ``would be'' Dark Matter signals
anyway.

We assume therefore that Dark Matter has so far not been detected and
ask how natural or fine-tuned the left-over parameter space
is. Specifically we study in detail the not so well investigated
neutralino mass range less than \hbox{200~GeV}, taking into account
all collider, cosmological and flavor constraints including the recent
results of LHC Higgs researches as well as flavor studies. Over this
complete mass region we find valid scenarios that may have escaped
every experiment so far. We will especially show that it is possible
to fulfill the muon anomalous magnetic moment condition for positive
gaugino masses and a negative supersymmetric Higgs mass parameter, the
$\mu$-term.

This article is organized as follows: in the next section, we define
the fine-tuning measures and fix our notation of the neutralino
sector. Then, in section~\ref{sec:analysis} the method of our
numerical analysis and the SUSY parameter space is discussed. Our
results will be presented in section~\ref{sec:results} including
discussions about the annihilation mechanisms for neutralinos, the
mapping of the level of fine-tuning into the direct detection
cross-section plane, the direct detection cross-section and its
dependence on the sign of the $\mu$-term, how the muon anomalous
magnetic moment can be obtained correctly with a negative $\mu$-term,
and lastly about functional fine-tuning and the parameter mapping
distribution.  Finally, we conclude in section~\ref{sec:conclusions}.

\section{Definition of fine-tuning}
\label{sec:FT}
\indent SUSY needs to be consistent with the electroweak sector of the
SM and has to reproduce the correct Higgs and $Z$-boson masses when
electroweak symmetry is spontaneously broken. The complete scalar
potential reads as:
\begin{eqnarray}
  \label{eq:scalarpotential}
   V  &\!=\!&  (|\mu|^2 + m^2_{H_u}) (|H^0_u|^2 + |H^{+}_{u}|^2)  
+ (|\mu|^2 + m^2_{H_d})(|H^0_d|^2 + |H^{-}_d|^2) 
+ [b (H^{+}_u H^{-}_d - H^0_u H^0_d) + {\it c.c.}]  \nonumber\\
&&+ \frac{1}{8}(g^2 + {g'}^2)(|H^0_u|^2 + |H^+_u|^2 -|H^0_d|^2 - |H^{-}_d|^2)^2 
+ \frac{1}{2} g^2 |H^+_u {H_{d}^{0}}^{*} + H^0_u {H_{d}^{-}}^{*}|^2 \hspace{2mm}, 
\end{eqnarray}
where $\mu$ is the SUSY respecting Higgs mass parameter from the
superpotential, $m^2_{H_u}$ and $ m^2_{H_d}$ mass terms of the two
complex Higgs doublets $H_u$ and $H_d$ from the soft SUSY breaking
part of the Lagrangian, $b$ the bilinear Higgs coupling and $g$ and
$g'$ are the $U(1)$ and $SU(2)$ gauge couplings,
respectively. Minimizing this potential gives the well-known relation
for the $Z$-boson mass at tree-level:
\begin{equation}
\label{eq:Z_mass}
m^2_Z = \frac{|m^2_{H_d} - m^2_{H_u}|}{\sqrt{1 - \sin^22\beta}} 
- m^2_{H_u} - m^2_{H_d} - 2 |\mu|^2 \hspace{2mm},
\end{equation}
where $\tan\beta$ is the ratio of the vacuum expectation values (VEVs)
of the two Higgs doublet fields.

{}From equation~(\ref{eq:Z_mass}) one can immediately see that SUSY
mass parameters of the order of the weak scale are preferred to avoid
tuning of the $Z$-mass already at tree-level. To quantify this, we use
the fine-tuning measure defined as the sensitivity of the
$Z$-mass~\cite{Ellis:1986yg,Barbieri:1987fn}:
\begin{equation}
\label{eq:barbieri}
\Delta p_i\equiv \left| \frac{p_i}{M^2_Z} \frac{\partial M_Z^2(p_i)}
 {\partial p_i} \right|=\left| \frac{\partial \ln M_Z^2(p_i)}
 {\partial \ln p_i} \right| \hspace{2mm}.
\end{equation}
The parameters $p_i$ that determine the $Z$-mass on tree-level are
$\mu$, the two soft Higgs mass parameters ($m_{H_u}$ and $m_{H_d}$)
and the bilinear coupling $b$. We take the total measure of
fine-tuning arising from these parameters as a summation in
quadrature:
\begin{equation}
\label{eq:finetuning}
 \Delta_{\rm tot} \equiv\sqrt{ \sum{}_{p_i= \mu^2, b, m^2_{H_u}, m^2_{H_d}} \left\{\Delta p_i \right\}^2 }\hspace{2mm},
\end{equation}
with the individual $\Delta p_i$'s obtained in reference~\cite{Perelstein:2007nx}
\begin{eqnarray}
\label{eq:ftmu}
\Delta\mu^2  &\!=\!& \frac{4 \mu^2}{m^2_Z} \Bigl(1 + \frac{m^2_A + m^2_Z}{m^2_A} \tan^22\beta \Bigr) 
\hspace{2mm}, \\
\label{eq:ftb}
\Delta b &\!=\!& \Bigl( 1 + \frac{m^2_A}{m^2_Z}\Bigr) \tan^22\beta \hspace{2mm},\\
\label{eq:fthu}
\Delta m^2_{H_u} &\!=\!& \Big|\frac{1}{2} \cos2\beta + \frac{m^2_A}{m^2_Z} \cos^2\beta
- \frac{\mu^2}{m^2_Z}\Big| \Bigl(1 - \frac{1}{\cos2\beta} + \frac{m^2_A + m^2_Z}{m^2_A} \tan^2 2\beta \Bigr) 
\hspace{2mm},\\
\label{eq:fthd}
\Delta m^2_{H_d} & \!=\!&\Big|-\frac{1}{2} \cos2\beta + \frac{m^2_A}{m^2_Z} \sin^2\beta- \frac{\mu^2}
{m^2_Z}\Big|  \Bigl(1 + \frac{1}{\cos2\beta} + \frac{m^2_A + m^2_Z}{m^2_A} \tan^22\beta \Bigr)\hspace{2mm}.
\end{eqnarray}
Here $m_A$ denotes the pseudo-scalar Higgs mass.

The quantity $\Delta_{\rm tot}$ serves as an indicator how well a
specific SUSY scenario avoids an unnaturally large separation of the
electroweak and SUSY breaking scales. Small values are favored since
they are less tuned and are therefore viewed to be more likely than
those with high values of $\Delta_{\rm tot}$.  Note that the
contributions given above do not depend on the stop mass and, hence,
the fine-tuning will not show an exponential increase with the light
Higgs mass. This is different to the fine-tuning measure for the pMSSM
used in reference~\cite{CahillRowley:2012rv},
or~\cite{Perelstein:2007nx} who evaluated $\Delta_{\rm tot}$ including
terms which arise at leading log level. Note that we choose our
definition of the fine-tuning, since we want to avoid the dependence
on a randomly chosen cut-off scale as we do not want to make any
assumptions about high-energy completions.  Therefore, we stick to a
very rough, first estimate of the fine-tuning via a tree-level
definition of $\Delta_{\rm tot}$, as \hbox{\it e.g.}
reference~\cite{Arbey:2011un}. It is important to mention that the
one-loop contribution is, however, not vanishing, from which our total
measure of fine-tuning may increase by some amount.

Apart from the sensitivity of the $Z$-mass, a tuning of the light
Higgs mass (defined analogously to equation~(\ref{eq:barbieri})) has
been discussed in reference~\cite{Hall:2011a}. In their set-up it has
been found to be of order 100 already. Further discussions of
fine-tuning in the (C)MSSM may be found
in~\cite{Ellis:2001zk,Kitano:2006gv,Cassel:2010px,Ghilencea:2012gz}.

In addition to this ``parameter'' fine-tuning an ``equation-tuning''
can appear when cancellations between different terms are a
consequence of model specific relations.  This will be the case for
the direct detection cross-section of neutralino Dark Matter,
$\sigma^{\rm SI}$. To quantify this, we evaluate the sensitivity of
$\sigma^{\rm SI}$ analogously to the $Z$-mass tuning and discuss its
implications shortly in section~\ref{sec:ft_sigma}. For completeness:
\begin{equation}
 \label{eq:ft_sigma1}
\Delta{{\rm f}_i} \equiv \left| \frac{\partial \ln \sigma^{\rm SI}(p_i)}
 {\partial \ln p_i} \right| \hspace{2mm},
\end{equation}
with $p_i = \{ \mu, \tan \beta, M_1, M_2, m_A \}$, see
also~\cite{Perelstein:2011tg}. $M_1$ is the bino and $M_2$ the wino
mass parameter. Additionally we evaluate a sensitivity of the relic
abundance in section~\ref{sec:tuning_omega}.

To fix our notation, we give some details about the neutralino $\widetilde{\chi}^0_1$ which
is a mixed state of the neutral gauginos (bino $\widetilde{B}$ and
neutral wino $\widetilde{W}^0$) and the two neutral higgsinos
(down-type $\widetilde{H}^0_d$ and up-type $\widetilde{H}^0_u$):
\begin{equation}
\widetilde{\chi}^0_1 = N_{11} \widetilde{B} + N_{12}\widetilde{W}^0 + N_{13} \widetilde{H}^0_d
+ N_{14} \widetilde{H}^0_u \hspace{2mm}.
\end{equation}
The coefficients $N_{ij}~(i,j=1,2,3,4)$ are the components of the
mixing matrix that diagonalizes the neutralino mass matrix:
\begin{equation}
M_{\widetilde{N}} =
\begin{pmatrix}
  M_1 & 0 & -M_Z \sin\theta_W\cos\beta & M_Z\sin\theta_W\sin\beta \\
  0 & M_2 & M_Z \cos\theta_W\cos\beta &  -M_Z \cos\theta_W\sin\beta \\
  -M_Z \sin\theta_W\cos\beta & M_Z \cos\theta_W\cos\beta & 0 & -\mu \\
  M_Z \sin\theta_W\sin\beta &  -M_Z \cos\theta_W\sin\beta & -\mu & 0
\end{pmatrix} \hspace{2mm}.
\end{equation}

\section{Numerical analysis of the parameter space}
\label{sec:analysis}
\indent As already mentioned, we study in this work the MSSM
defined at the electroweak scale, the so-called pMSSM, with the eleven free
parameters as in reference~\cite{AlbornozVasquez:2011yq}:
\begin{equation}
  \label{eq:parameters}
  \tan\beta\,, \, M_1 \,, \, M_2 \,, \, M_3 \,, \, M_A \,, \, \mu \,, \,
  m_{{\widetilde \ell}_L}\,, \, m_{{\widetilde \ell}_R}\,, \,   m_{{\widetilde q}_{1,2}}\,, \,
  m_{{\widetilde q}_{3}}\,, \, a_0\,,
\end{equation}
where $\tan\beta$ is the ratio of the VEVs of the two Higgs doublet
fields, $M_{i}~(i=1,2,3)$ the three gauginos masses, $m_A$ the CP-odd
Higgs mass and $\mu$ the Higgs-higgsino mass parameter. We chose
different masses for left- ($m_{{\widetilde \ell}_L}$) and
right-handed sleptons ($m_{{\widetilde \ell}_R}$) but no distinction
in generations. Furthermore, we assume the left- and right-handed
squark masses to be degenerate but different in generations
($m_{{\widetilde q}_{1,2}}$ and $m_{{\widetilde q}_{3}}$).

The trilinear terms are parameterized by $a_0$ in the following way:
\begin{equation}
  \label{eq:Aterms}
  A_t = a_0 Y_t  m_{{\widetilde q}_{3}}\,, ~
   A_b = a_0 Y_b  m_{{\widetilde q}_{3}}\,, ~
    A_\tau = a_0 Y_\tau  \sqrt{ m_{{\widetilde \ell}_L} m_{{\widetilde \ell}_R}} \,.
\end{equation}
This implies that we use non-zero trilinear couplings that are
proportional to the third generation squark masses for $A_t$ and
$A_b$, not as in reference~\cite{AlbornozVasquez:2011yq}, and the
geometric mean of the slepton masses for the leptonic trilinear term.
Since the SM Yukawa couplings of the first two generations are known
to be very small, we can safely neglect $A_u$, $A_d$, $A_e$, \hbox{\it i.e.} 
we set them to zero.

In order to calculate the level of tuning, the parameters of
equation~(\ref{eq:parameters}) are randomly varied in the following
ranges:
\begin{center}
\begin{tabular}{ccc}
 ~$M_1 \in [10, 200]$ GeV,~ & ~$M_2 \in [100, 2000]$ GeV, ~& ~ $M_3 \in [100,4000]$ GeV, ~\\[1mm]
$m_A \in [90, 4000]$ GeV, & $\left|\mu\right| \in [90, 2000]$ GeV, &  $a_0 \in [-4.0, 4.0]$\,,  \\[1mm]
~$m_{{\widetilde q}_{1,2}} \in [400, 4000]$ GeV, ~& ~$m_{{\widetilde q}_3}\in [200, 4000]$ GeV, ~
&  ~ $\tan\beta \in [2, 65]$\,,  ~\\[1mm]
~$m_{{\widetilde \ell}_L} \in [100, 4000]$ GeV, ~
& ~$m_{{\widetilde \ell}_R} \in [60, 4000]$ GeV. ~& ~
\end{tabular}
\end{center}

For every simulated scenario we create a different random number
$\lambda$ for each of the eleven input parameters $x$ using Mersenne
Twister~\cite{MT1,MT2}.  Their values are then given by $x = x_{\rm
  min} + \lambda \left(x_{\rm max} - x_{\rm min}\right)$, where
$x_{\rm min/max} $ is the minimal/maximal value of $x$, respectively,
and forwarded to {\tt SuSpect}. In order to save memory and make the
scan more efficient, we immediately remove all the scenarios that do
not respect the experimental ranges listed in
table~\ref{tab:flavour}. In this way we create four sets of simulated
data: two sets for each sign of $\mu$, one excluding, one including
the measurement of the muon anomalous magnetic moment
$a_\mu$. Excluding $a_\mu$ from the applied cuts, we get a total
number of approximately 372800 (246800) scenarios for positive
(negative) $\mu$. The data sets that include that constraint consist
of 178000 (64000) scenarios.

We emphasize that we have studied not only the case $\mu>0$ but also
$\mu<0$ which will turn out to be important. One might wonder if we
can manage to satisfy the limit of the muon anomalous magnetic moment
with a negative $\mu$-term. This issue will be discussed later in
section~\ref{sec:g2}.

For our analysis we take into account different experimental data from
cosmology, flavor and collider physics. The Dark Matter abundance
constraints arise from the WMAP analysis~\cite{Komatsu:2010fb} of the
determination of the relic density. We set the neutralino relic
density, $\Omega h^2$, within the \hbox{$2~\sigma$}
range~\cite{Dumont:2012ee} $\Omega h^2\in [0.089, 0.136]$, where
experimental and theoretical uncertainties are included.
\begin{table}[t!]
\centering
\begin{tabular}{|c|c|c|}
\hline
Quantity & & Reference(s) \\\noalign{\hrule height 2pt}
$\Omega h^2$ &  $[0.089, 0.136]$ &  \cite{Komatsu:2010fb}  \\ \hline
$m_h$ &  $(121.0,129.0)$ GeV & \cite{ATLASseminarHiggs, CMSseminarHiggs,%
ATLAS:2012ae,Chatrchyan:2012tx,ATLAS:2012ad,%
Chatrchyan:2012tw,ATLAS-CONF-2012-019,CMS-PAS-HIG-12-008} \\ \hline
${\rm Br}(B\to s \gamma)$ & $[2.89, 4.21] \times 10^{-4}$ & \cite{Barberio:2008fa} \\
${\rm Br}(B_s \rightarrow \mu^{+} \mu^{-}) $& $< 4.5 \times 10^{-9}$ & \cite{Aaij:2012ac} \\
${\rm Br}(B_u \rightarrow \tau \bar{\nu})$ & $0.52 < R_{B \tau \nu}  < 2.61$  &\cite{Asner:2010qj} \\
${\rm Br}(K \rightarrow \mu \nu$) & $0.985 < R_{l23} < 1.013$ & \cite{Antonelli:2008jg} \\
$a_\mu$ & $[0.34, 4.81]\times10^{-9}$ &  \cite{Bennett:2006fi}\\ \hline
$\Gamma(Z \to \widetilde{\chi}_1 \widetilde{\chi}_1)$& $< 3$~MeV & \cite{ALEPH:2005ema}\\
$\sigma(e e \to \widetilde{\chi}_1 \widetilde{\chi}_{2,3})$& $< 100$~fb & \cite{ALEPH:2005ema}\\
$\Delta \rho$ & $< 0.002$ &  \cite{ALEPH:2005ema}\\\hline
\end{tabular}
\caption{\label{tab:flavour} The experimental constraints.}
\end{table}
The influence of the direct detection search -- XENON100 (2012) -- of
Dark Matter will be explicitly shown in the plots representing our
results. Results of ATLAS and CMS analyses for a standard model-like
Higgs boson mass~\cite{ATLASseminarHiggs, CMSseminarHiggs,
  ATLAS:2012ae,Chatrchyan:2012tx,ATLAS:2012ad,%
  Chatrchyan:2012tw,ATLAS-CONF-2012-019,CMS-PAS-HIG-12-008} provide
the allowed mass limit: \hbox{$121.0~{\rm GeV} < m_h < 129.0~{\rm
    GeV}$}. Since the recent results of both collaborations are still
preliminary, we take the range not too restrictive. A smaller mass
range would not affect our results except that the numerical
simulations would be more time-consuming.  Also the light gluino mass
and light squark masses of the first two generations are excluded by
the analyses of the signature of missing transverse
energy~\cite{Aad:2011ib,Chatrchyan:2011zy}, see
also~\cite{atlas_conf,parker:2012}, so we allow their masses to be
greater than \hbox{800~GeV} and \hbox{1~TeV}, respectively.

We also take into account the constraint coming from pseudo-Higgs
boson searches~\cite{Aad:2011rv,Chatrchyan:2012vp} that have excluded
a significant fraction of the \hbox{$M_A-\tan\beta$} plane at small $M_A$ and
large values of $\tan\beta$. Moreover, LEP constraints are included in
our study: the invisible $Z$-decay width \hbox{$\Gamma(Z\to
\widetilde{\chi}_1\widetilde{\chi}_1)<3$ MeV}~\cite{ALEPH:2005ema}, the
pair production cross-section \hbox{$\sigma(e e \to
\widetilde{\chi}_1\widetilde{\chi}_{2,3})< 100$
fb}~\cite{Abbiendi:2003sc}, \\ 
\hbox{$\Delta \rho < 0.002$}~\cite{Nakamura:2010zzi} and the mass limits of supersymmetric
particles~\cite{ALEPH:2005ema,Nakamura:2010zzi}.

The experimental constraints including flavor and collider physics
applied in our analyses are listed in table~\ref{tab:flavour}, where
$R_{B \tau \nu} $ is the ratio between the SUSY and SM prediction of
the branching ratio \hbox{${\rm Br}(B_u \rightarrow \tau \bar{\nu})$},
$R_{l23}$ the leptonic kaon decay quantity and $a_\mu$ the muon
anomalous magnetic moment. For the latter quantity we use the
\hbox{$3~\sigma$} range~\cite{Bennett:2006fi}, because there are
theoretical uncertainties about hadronic effects.

The supersymmetric spectrum is obtained by {\tt
  SuSpect}~\cite{Djouadi:2002ze} using the default SUSY breaking
scale, the neutralino relic density $\Omega h^2$, the spin-independent
cross-section with protons $\sigma^{\rm SI}$ and the annihilation
channels were calculated by the {\tt micrOMEGAs}
code~\cite{Belanger:2006is, Belanger:2010gh}, while the low-energy
observables (\hbox{${\rm Br}(B\to s \gamma)$}, \hbox{${\rm
    Br}(B_s\rightarrow \mu^{+} \mu^{-}) $}, $R_{B \tau \nu}$,
$R_{l23}$) and $a_\mu$ have been determined by {\tt
  SuperIso}~\cite{Mahmoudi:2008tp}.

In our analysis we use the following values of the quark form-factors in the nucleon
which are the default values in the {\tt micrOMEGAs} package:
\begin{eqnarray}
\label{eq:scalar}
f^p_{d}=0.033\,, \;\; f^p_{u}=0.023\,, \;\; f^p_{s}=0.26\,, \nonumber\\
f^n_{d}=0.042\,, \;\; f^n_{u}=0.018\,, \;\; f^n_{s}=0.26\,. 
\end{eqnarray}
It should be mentioned that the main uncertainty comes from the strange quark
coefficient, and using another set of quark coefficients (the large
corrections to $f^{p/n}_{s}$) can lead to a shift by a factor $2-6$
in the spin independent cross-section~\cite{Belanger:2008sj}.

Note that we do not calculate a $\chi^2$ as a probability measure in
this work but see every scenario that passes the constraints as
equally probable. Only the envelope of our final plots, that mildly
depends on the parameters, has a meaning: The area outside of it can
never be reached with the chosen parameter space. We take the
viewpoint that tuned scenarios are less likely to form valid models of
the pMSSM.

\section{Results}
\label{sec:results}
\subsection{Obtaining the correct relic density}
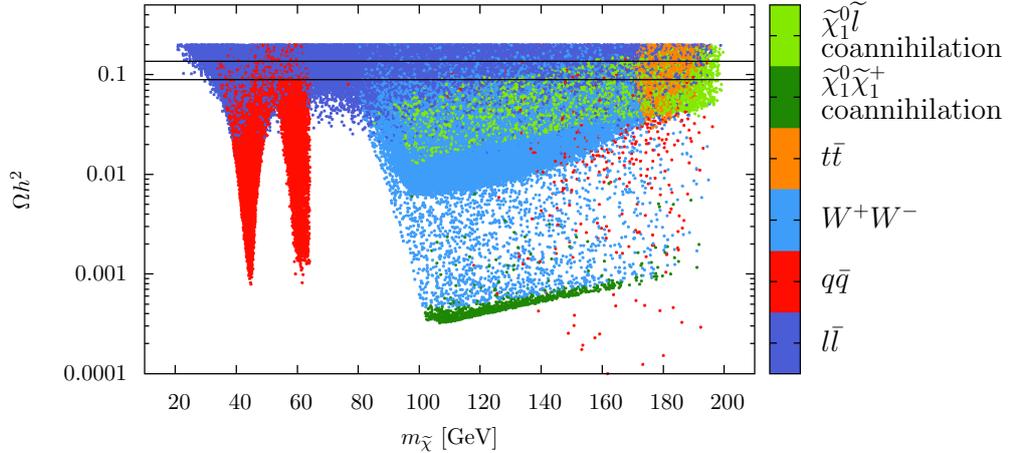
\begin{figure}[!t]
\begin{center}
\resizebox{0.7\columnwidth}{!}{\input{pic/relic_density}}
\caption{The relic density and the different (co)annihilation
  mechanisms presented for negative $\mu$ (see
color coding and table~\ref{tbl:mechanism} for more information about the final
  states). We have applied all collider and flavor constraints.  
The black lines show the upper and lower bounds that will be
  applied on the relic density; $\Omega h^2 \in [0.089,0.136]$. Here $q\bar{q}$
 stands for a light quark and $\ell \bar{\ell}$ for a lepton pair. The plot is similar for positive $\mu$. 
 \label{fig:relic}}
\end{center}
\end{figure}

\indent First we discuss the different mechanisms that bring the relic
density into the cosmological interesting region. The relic density is
basically set by the thermally averaged annihilation cross-section,
$\langle \sigma_{ann} v\rangle$, using a freeze-out mechanism. In
figure~\ref{fig:relic} we plot the relic density on a log scale versus
the neutralino mass as obtained by our simulation with {\tt
  micrOMEGAs} (for simplicity we plot here $\Omega h^2 < 0.2$.), where
the most important annihilation channels are presented by the
indicated color code. In general a very efficient mechanism is needed
to end up with a high enough $\langle \sigma_{ann} v\rangle$ to obtain
the correct relic abundance.

Lepton final states (dark-blue points in figure~\ref{fig:relic}) are
the most dominant annihilation channels and are present in our
complete neutralino mass range. To obtain
these a distinction of left- and right-handed slepton masses is
necessary as we will discuss later. In past studies of the pMSSM the importance of light stau annihilations
at neutralino masses between 60 and 80 GeV has not clearly been pointed out,
compare~\cite{Farina:2011bh,Arbey:2011un,CahillRowley:2012rv} for past scans. 

The two resonant $Z$- and light Higgs-boson annihilations can easily
be seen at around 40 and 60 GeV (red points), respectively. Above a
mass of roughly 80 GeV the neutralino may annihilate into two
$W$-bosons (light-blue). There are further scenarios with dominant
annihilations into light quarks for $m_{\widetilde{\chi}} \gtrsim 80$
GeV hidden behind the lepton final states. The lower branch
(dark-green) corresponds to chargino coannihilations, but for our
chosen range of the lightest neutralino mass this mechanism is too
efficient and never produces enough Dark Matter. These coannihilations
are therefore unimportant for our further discussion, as we are also
taking into account the lower bound from the WMAP measurement. On the
other hand, slepton coannihilations will play an important role at
neutralino masses above 90 GeV (light-green). They are accompanied by
a region of top final states at approximately 180 GeV (orange).
\begin{table}[!t]
\begin{center}
 \begin{tabular}{| c || c| c |}
   \hline
    & initial state &  final states \\\noalign{\hrule height 2pt}
   chargino coannihilation & $\widetilde{\chi}^0_1$ $\widetilde{\chi}^+_1$
   & $s c$, $u d$, $t b$, $e \nu_e$, $\mu \nu_{\mu}$, $Z W$ \\[1.5mm] \hline \hline
   slepton coannihilation & $\widetilde{\chi}^0_1$ $\widetilde{\tau}_1$ & 
$\gamma \tau$, $\tau h$,$ W \nu_{\tau}$, $Z \tau$  \\ \hline
   & $\widetilde{\chi}^0_1$ $\widetilde{e}_R$ & $\gamma e$  \\ \hline
& $\widetilde{\chi}^0_1$ $\widetilde{\nu}_\tau$ & $W \tau$, $Z \nu_\tau$, $\nu_\tau h$  \\ \hline
& $\widetilde{\chi}^0_1$ $\widetilde{\nu}_e$ & $W e$, $Z \nu_e$  \\ \hline
\end{tabular}
\caption{List of the dominant annihilation channels as obtained by our
  simulation. \label{tbl:mechanism}}
\end{center}
\end{table}

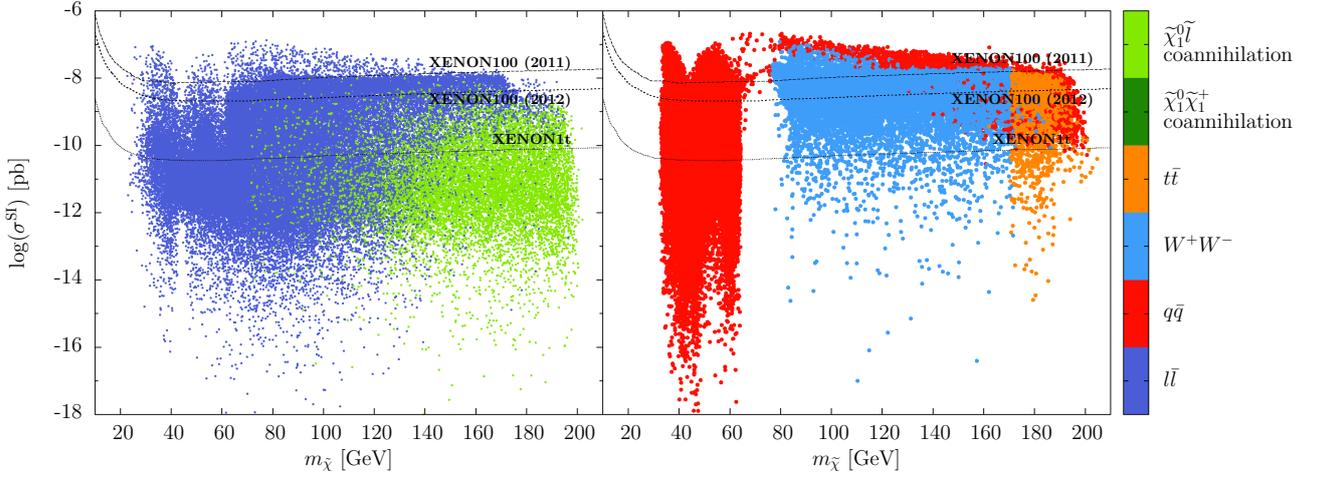
\begin{figure}[!t]
\begin{center}
\resizebox{\columnwidth}{!}{\input{pic/mechanism_xenonplane}}
 \caption{Dominant contribution to the neutralino annihilation in the
   $m_{\widetilde{\chi}} - \sigma^{\rm SI}$ plane for negative $\mu$. The required relic density and all
   collider physics constraints have been applied except the anomalous magnetic moment of the
   muon. Here $q\bar{q}$
 stands for a light quark and $\ell \bar{\ell}$ for a lepton pair. For better visibility we
 display the results in two figures.
 \label{fig:mechanism_xenonplane}}
\end{center}
\end{figure}

When mapping the different models with their mechanisms into the
$m_{\widetilde{\chi}} - \sigma^{\rm SI}$ plane, we find a rather well
ordered picture (all constraints except the muon anomalous magnetic
moment have been applied). We show the case of a negative $\mu$-term
in figure~\ref{fig:mechanism_xenonplane}.

The quark final states at neutralino masses between 40 and 60 GeV are
due to the $s$-channel $Z$-boson and light Higgs boson, $h$,
resonances. The chargino mediated $t$-channel and neutral Higgs
mediated $s$-channel annihilations into $W$-bosons fill a band at
$\sigma^{\rm SI} \simeq 10^{-9}$ pb (light-blue). In this region (red
points above 80 GeV) one also finds quark final states via heavy
neutral Higgs, $H$, and CP-odd Higgs, $A$, exchanges. The most
prominent mechanism, the annihilation into a pair of leptons, is
homogeneously distributed and mainly mediated through $t$-channel
light stau or $s$-channel $Z$, $h$, $A$, $H$ exchange and their
interference terms. Slepton coannihilations are situated in an area
below the $W$-bosons final states. The chargino coannihilations are
not present as argued above\footnote{Some scenarios of chargino
  coannihilations have been present even after cutting the relic
  density, however, these were extremely rare (of order 0.02\%) and we
  removed them from our study.}.

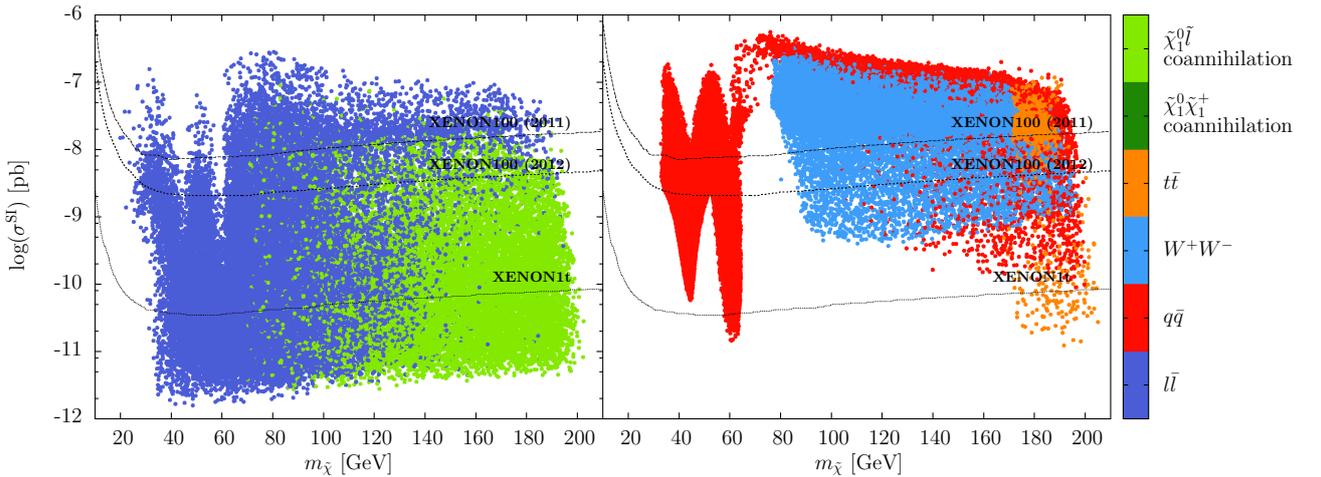
\begin{figure}[!b]
\begin{center}
\resizebox{\columnwidth}{!}{\input{pic/mechanism_xenonplane_pos}}
 \caption{Dominant contribution to the neutralino annihilation in the
   $m_{\widetilde{\chi}} - \sigma^{\rm SI}$ plane for a positive $\mu$ in
 analogy to figure~\ref{fig:mechanism_xenonplane}.
 \label{fig:mechanism_xenonplane_pos}}
\end{center}
\end{figure}

The arrangement of the decay mechanisms holds generally and shows,
aside from a different scale of the direct detection cross-section, no
difference between positive and negative $\mu$ (for more details see
figure~\ref{fig:mechanism_xenonplane_pos}).

Note that we have calculated the annihilation cross-sections of two
Dark Matter into two photons in our models that satisfy all
experimental constraints except the muon anomalous magnetic moment,
and we found that the order of $\langle
\sigma_{\chi\chi\to\gamma\gamma} v\rangle$ is far below the recent
claim of a gamma-ray line in the Fermi-LAT
data~\cite{Weniger:2012tx}. Thus, we conclude that a dark matter
particle mass of order \hbox{130~GeV} and a partial annihilation
cross-section into two photons of approximately \hbox{$1.3 \times
  10^{-27} {\rm cm}^3~{\rm s}^{-1}$} is not compatible with our
models.

\subsection{Fine-tuning and the spin-independent elastic
WIMP nucleon cross-section}
\label{sec:nog2}
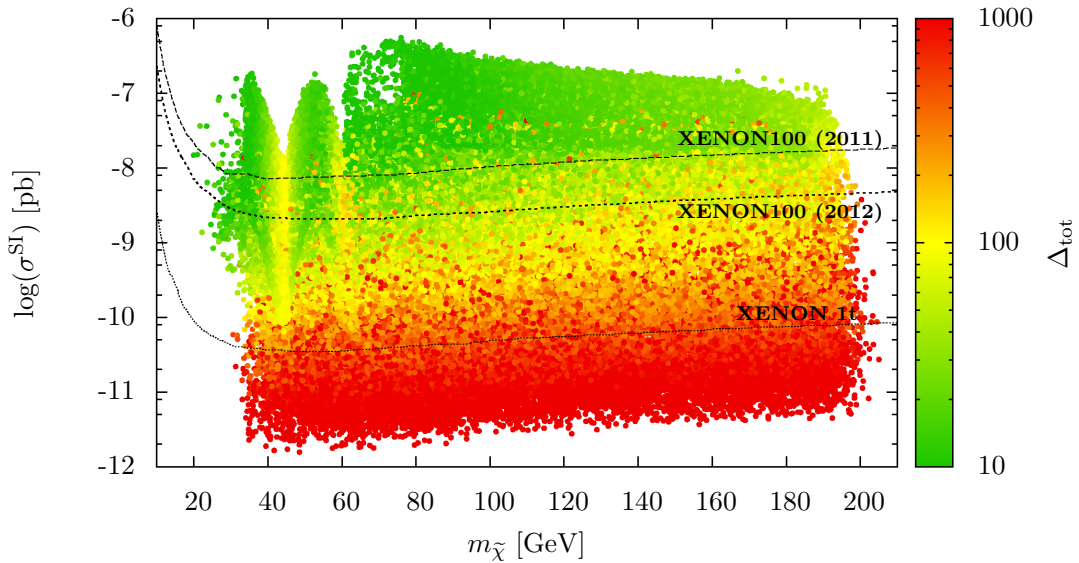
\begin{figure}[t!]
  \begin{center}
  \resizebox{0.85\columnwidth}{!}{\input{pic/pos_nog2_direct}}
  \caption{The level of fine-tuning for our models in the
    $m_{\widetilde\chi}-\sigma^{\rm SI}$ plane. The colored points
    show the fine-tuning $\Delta_{\rm tot}$ for some specific
    parameter combination. Points with $\Delta_{\rm tot} < 10$ are
    given the value 10 and $\Delta_{\rm tot} > 1000$ the value
    1000. The lines represent the exclusion limits of XENON100 (2011)
    and XENON100 (2012), and the prediction for XENON1t,
    respectively. \label{fig:pos_nog2}}
\end{center}
\end{figure}

\indent\ We will discuss the spin-independent elastic WIMP nucleon
cross-section as function of the neutralino mass with respect to the
fine-tuning measure, equation~(\ref{eq:finetuning}). In this section,
we show our results applying all experimental constraints discussed
above except the muon anomalous magnetic moment. To stress its impact
we will consider this quantity separately in section~\ref{sec:g2}.

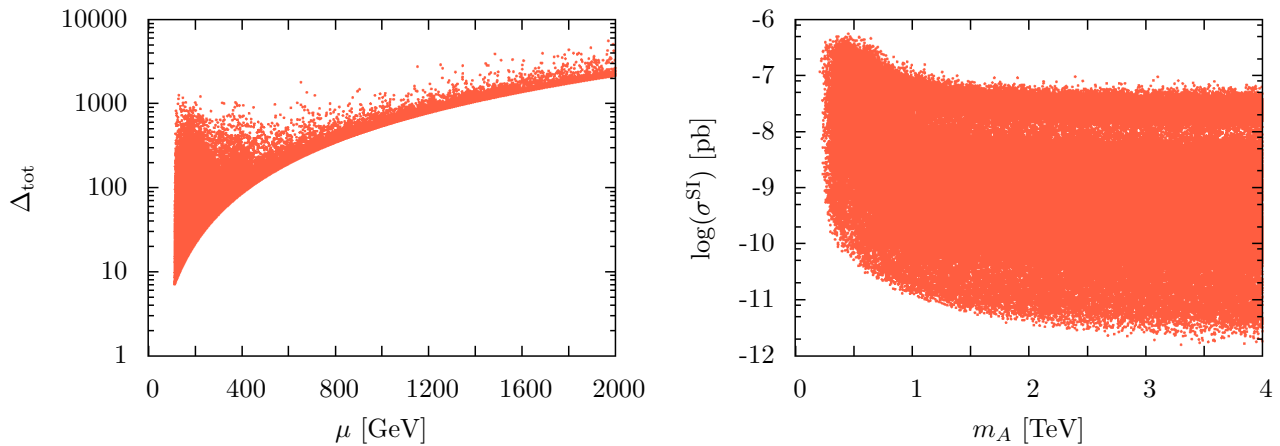
\begin{figure}[t!]
\begin{minipage}[b!]{0.49\textwidth}
    \begin{center}
\resizebox{\columnwidth}{!}{\input{pic/mu_FT}}
   \end{center}
\end{minipage}
\hfill
\begin{minipage}[b!]{0.49\textwidth}
\resizebox{\columnwidth}{!}{\input{pic/ma_sigma}}
\end{minipage}
\caption{Plots motivating our upper bounds on the $\mu$-term (left
  panel) and $m_A$ (right panel). The strong increase of the minimal
  amount of fine-tuning with $\mu$ and the negligible decrease of
  $\sigma^{\rm SI}$ with $m_A$ are visible. They are shown for a
  positive $\mu$-term but are similar for negative values.\label{fig:upper_bounds}}
 \end{figure}

First, positive values of the $\mu$-term will be presented: In
figure~\ref{fig:pos_nog2} low fine-tuned regions can be found near the
$Z$- and $h$-resonances down to $10^{-10}$ pb and for $\sigma^{\rm SI}
\gtrsim 10^{-8}$ pb at masses above \hbox{80~GeV}. A great part of the
latter region is already excluded by the current XENON limit, so that
low fine-tuned Dark Matter preferably appears for positive $\mu$ at
masses between 20 and \hbox{60~GeV}. Moving towards the exact
$Z$-resonance a rise in $\Delta_{\rm tot}$ can be observed, which can
be understood as follows: In order to obtain the correct relic density
the neutralino-neutralino-$Z$-coupling has to be decreased to
compensate for the resonant enhancement of $\langle \sigma_{ann} v
\rangle$. This coupling is determined by the higgsino components of
the neutralino~\cite{Nihei:2002ij}:
\begin{equation}
\langle \sigma_{ann} v \rangle  \propto  |C_A^{{\widetilde{\chi}} {\widetilde{\chi}} Z}|^2 
\propto \left( N_{14}^2 - N_{13}^2\right)^2 \hspace{2mm}.
\end{equation}
Increasing $\mu$ decreases $C_A^{\widetilde{\chi}_1^0 \widetilde{\chi}_1^0 Z}$
(see equations~(\ref{eq:N13}) and~(\ref{eq:N14})) and reproduces the
correct relic abundance close to the exact resonance but at the same
time increases the fine-tuning (compare the left panel of
figure~\ref{fig:upper_bounds}). Note that at the $Z$-resonance
$\mu$-term is fixed due to the relic density condition and the
fine-tuning stays roughly constant when moving towards lower direct
detection cross-sections.

Highly fine-tuned models appearing in figure~\ref{fig:pos_nog2} in the
otherwise low fine-tuned regions show a small $\tan \beta$ and a large
$m_A$. In that case $\Delta b$, equation (\ref{eq:ftb}), becomes very
large.

The models lying out of the $Z$- and $h$-resonances show a clear
tendency towards higher fine-tuning for smaller $\sigma^{\rm SI}$ (see
also~\cite{Perelstein:2011tg}).  To decrease $\sigma^{\rm SI}$, the
higgsino component of the neutralino needs to be reduced which in turn
is achieved through increasing the $\mu$-term. Away from the $Z$- and
$h$-resonances a small neutralino-proton cross-section can therefore
for positive $\mu$ only be obtained at the cost of higher fine-tuning. To decrease
further $\sigma^{\rm SI}$ one could, in principle, increase $\mu$
above 2 TeV and $m_A$ above 4 TeV, but the effect on $\sigma^{\rm SI}$
is small and the fine-tuning becomes unacceptably large (see
figure~\ref{fig:upper_bounds}).

\begin{figure}[t!]
   \begin{center}
\resizebox{0.85\columnwidth}{!}{\input{pic/neg_nog2_direct}}
      \caption{The level of fine-tuning for a negative $\mu$-term in analogy to figure~\ref{fig:pos_nog2}.
        \label{fig:neg_nog2}}
   \end{center}
 \end{figure}
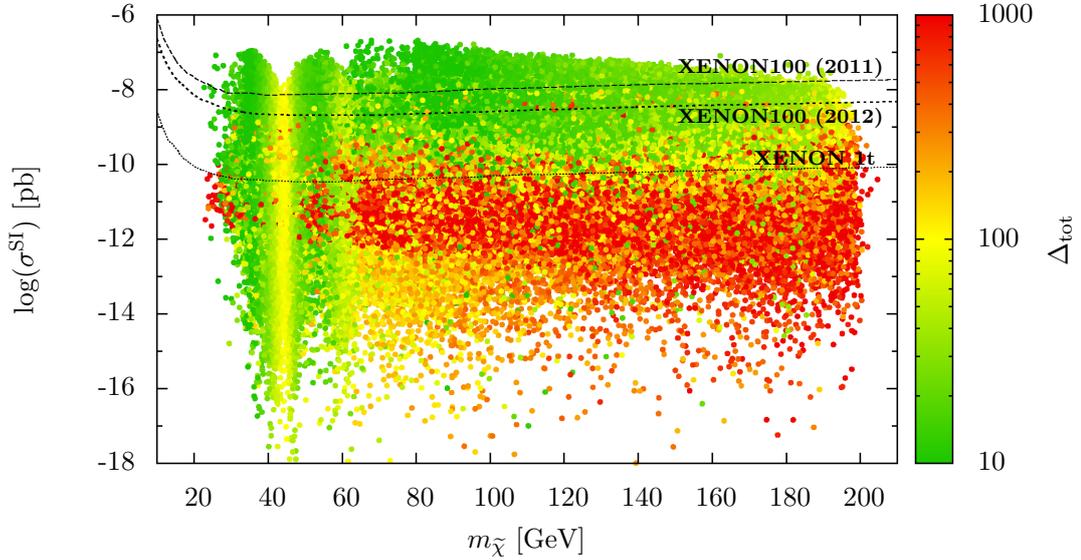

 Now, we will turn to negative values of the $\mu$-term (see
 figure~\ref{fig:neg_nog2}). The big difference is that the parameter
 region has just begun to be probed by direct detection
 experiments. Compared to a positive $\mu$-term the neutralino-proton
 cross-section is shifted to smaller values, so that every mass-region
 still offers models that explain the Dark Matter riddle with low
 electroweak fine-tuning. These shifts due to cancellations between
 light and heavy Higgs contributions to $\sigma^{\rm{SI}}$ are
 possible for specific combinations of the input parameters,
 preferably when the absolute value of $\mu$ is small, and will be
 discussed more detailed in section 4.3. In this way, an interesting
 region occurs for light Dark Matter near the $Z$- and $h$-resonances
 where the fine-tuning stays small even if $\sigma^{\rm SI}$ is
 decreased to tiny values.

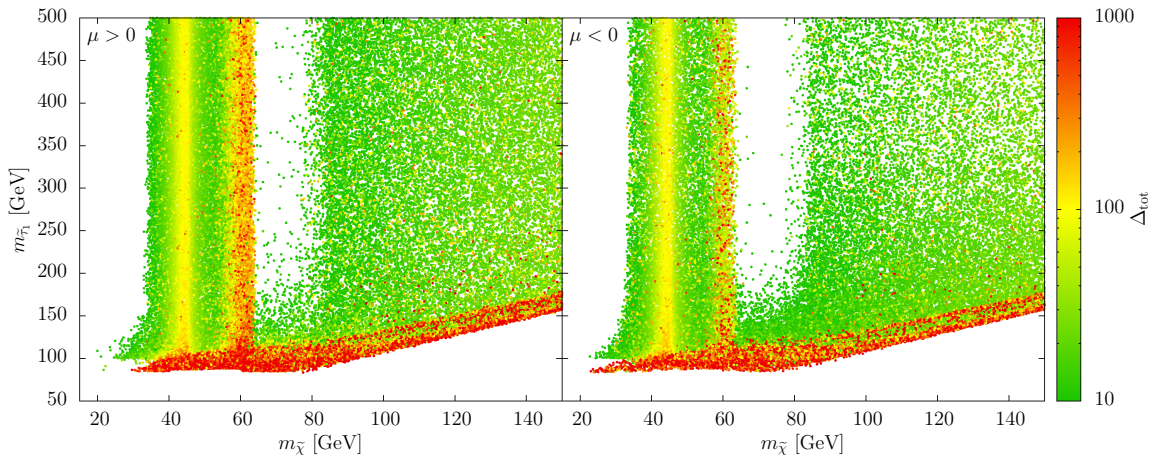
\begin{figure}[t!]
\begin{center}
\resizebox{0.95\columnwidth}{!}{\input{pic/light_staus}}
\caption{Models with very light staus and their level of fine-tuning.
  We show the cases for a positive (left) and negative (right)
  $\mu$-term separately. \label{fig:light_stau}}
\end{center}
 \end{figure}

 Across the complete neutralino mass region, scenarios with large
 fine-tuning show up at $\sigma^{\rm SI} \approx 10^{-11}$ pb below
 the expected XENON1t ``exclusion line''. These come from light stau
 mediated annihilations whose stau mass is strongly influenced by the
 off-diagonal elements of the slepton mass matrix (see
 figures~\ref{fig:mechanism_xenonplane} and~\ref{fig:light_stau}). The
 stau masses are given by:
\begin{eqnarray}
m_{\widetilde{\tau}_{1,2}}^2  &= & \frac{1}{2}  \Big[ m_{\widetilde{l}_L}^2 
+ m_{\widetilde{l}_R}^2 - \frac{1}{2} M_Z^2 \cos 2 \beta + 2 m_\tau^2 \nonumber\\
 &&\pm  \sqrt{\left(m_{\widetilde{l}_L}^2 - m_{\widetilde{l}_R}^2 + (-\frac{1}{2} 
+ 2 \sin^2 \theta_W) M_Z^2 \cos 2 \beta \right)^2 + 4 m_\tau^2 
(A_\tau - \mu \tan \beta)^2} \Big]  \hspace{2mm}.
\label{eq:slepton_mass}
 \end{eqnarray}
 There are two ways to get light stau masses: $(i)$
 $m_{\widetilde{l}_L}\!\sim\!m_{\widetilde{l}_R}$ and $(ii)$
 $m_{\widetilde{l}_L}\gg m_{\widetilde{l}_R}$.  In the first case we
 can approximate $m_{\widetilde{\tau}_{1}}^2$ as:
\begin{equation}
m_{\widetilde{\tau}_{1}}^2  \approx m_{\widetilde{l}_R}^2 + \frac{1}{4}M_Z^2 
- \big|m_\tau (A_\tau - \mu \tan \beta) 
\big| \approx\frac{5}{4} M_Z^2 -  \big|m_\tau (A_\tau - \mu \tan \beta) 
\big|  \hspace{2mm},
\end{equation}
where we used $-\frac{1}{2} + 2 \sin^2 \theta_W \simeq 0$ and assumed
in the last step $m_{\widetilde{l}_R}\!\sim\!  M_Z $. If a large
$\mu$-term is present the $\widetilde{\tau}_1$ mass can be suppressed
(even below \hbox{100~GeV}) when $(A_\tau - \mu \tan \beta) \approx
\frac{M_Z^2}{m_\tau}\approx$ \hbox{10$^4$ GeV}.  The same is true for the
second case. As an approximation for the stau masses we find here:
\begin{equation}
\label{eq:stau_mass_case2}
 m_{\widetilde{\tau}_{1}}^2  \approx  m_{\widetilde{l}_R}^2 
+ \frac{M_Z^2}{4} - \frac{m_\tau^2(A_\tau - \mu \tan \beta)^2}{ 16m_{\widetilde{l}_L}^2} 
\hspace{2mm}.
\end{equation}
One can see that the stau mass is primarily determined by
$m_{\widetilde{l}_R}$ and for $\frac{m_\tau^2(A_\tau - \mu \tan
  \beta)^2}{ 16m_{\widetilde{l}_L}^2} \sim$ \hbox{100$^2$~GeV$^2$} the mass is
suppressed by the off-diagonal elements. This explains why the models
with the lightest possible staus are severely fine-tuned.

 In figure~\ref{fig:light_stau} one can also see that in the
 neutralino mass region $m_{\widetilde{\chi}} \approx [60,80]$~GeV
 exclusively light staus are present. Here, light stau annihilation is
 the only possible mechanism to obtain the correct relic abundance.
 For these scenarios we find that the second case, \hbox{\it i.e.}
 $m_{\widetilde{l}_L}\gg m_{\widetilde{l}_R}$, is dominant to produce
 small $m_{\widetilde{\tau}_{1}}$.

 Below neutralino masses of around 35 GeV we find for both signs of $\mu$ a
 separation of high and low fine-tuned models. Scenarios with a small
 $\mu$-term, \hbox{\it i.e.}  small fine-tuning, posses a significant
 higgsino component so that next to the slepton annihilation an
 annihilation via the $Z$-boson and their interference term is
 dominant for $\langle \sigma_{ann} v\rangle$.  For the points with
 $m_{\widetilde{\tau}_1} < 100$ GeV only stau mediated neutralino
 annihilation is important and we find that
 $m_{\widetilde{l}_L}\!\sim\!  m_{\widetilde{l}_R}$. A detailed study
 on light staus may be found in
 reference~\cite{AlbornozVasquez:2011yq}.

 At the $h$-resonance, $m_{\widetilde{\chi}} \sim 60$ GeV, one can
 find scenarios with high fine-tuning at $\sigma^{\rm SI} \approx$
 \hbox{10$^{-10}$~pb} above the XENON1t ``exclusion limit'' (see
 figure~\ref{fig:neg_nog2}). In contrast to the $Z$-boson resonance,
 the $\mu$-term for $h$-resonant neutralino annihilation is not fixed
 through the relic density condition, because the
 Higgs-fermion-fermion coupling is dependent on a SUSY parameter; in
 the case for annihilation into $b \bar{b}$ the coupling constant is
 proportional to $\cos \beta$. Models that have a small value of $\tan
 \beta$, \hbox{\it i.e.} a rather large $\cos \beta$ and, hence, a
 large Higgs-$b$-$\bar{b}$-coupling, need a high $\mu$-term to
 decrease the neutralino-neutralino-Higgs coupling in order to keep
 the overall neutralino annihilation rate fixed at the value that
 reproduces the correct relic density. Despite their tiny higgsino
 component their direct detection cross-section, $\sigma^{\rm SI}$, is
 not minimal. See section~\ref{sec:cross-section} for further
 discussions on this. (These models can also be found in the vertical
 red stripe at \hbox{60~GeV} in figure~\ref{fig:light_stau}.)

 Note that the LHC phenomenology of the neutralino mass range under
 \hbox{70~GeV} has been studied in reference~\cite{Dreiner:2012}. They
 find that the invisible $h$ decay width may exclude parameter regions
 of small $\mu$ and $M_1$. Taking the most conservative upper bound of
 ${\rm BR}_{\rm inv} < 0.65$ they present how this constraint cuts
 into the low fine tuned regions for a positive $\mu$-term. The
 implications on our study are not significant.  Besides, for a
 negative $\mu$-term the constraints are not competitive at all.

\subsection{The direct detection cross-section and the $\mu$-term}
\label{sec:cross-section}
\begin{figure}[!t]
\begin{center}
\resizebox{0.95\columnwidth}{!}{\input{pic/mu_dependence_sigma}}
\caption{The level of fine-tuning and the direct detection cross-section
for negative (left) and positive (right) values of $\mu$. \label{fig:sigma_mu}}
\end{center}
\end{figure}
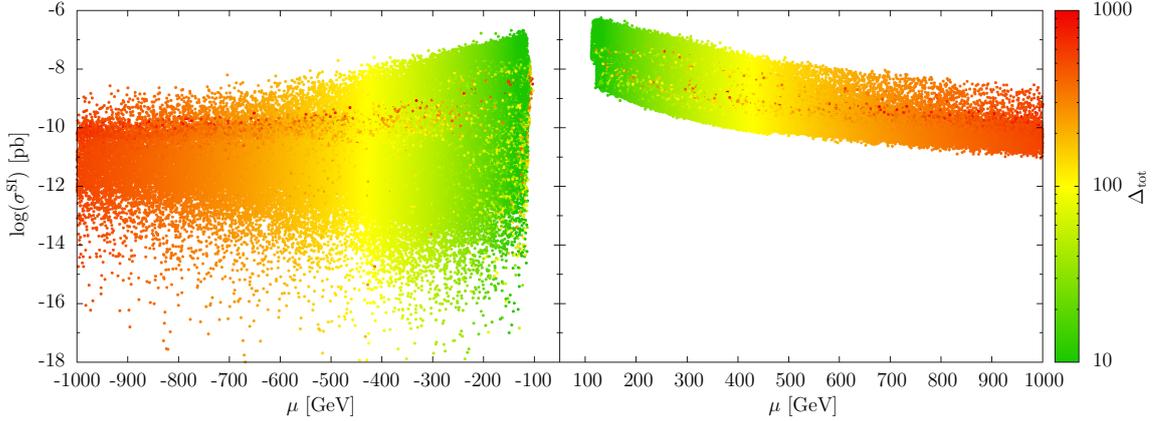
\indent {}From figure \ref{fig:sigma_mu}, one can
immediately notice a clear difference for the spin-independent
neutralino proton cross-section when the sign of $\mu$ is flipped. A
negative value allows for much lower cross-sections for a given
absolute value of $\mu$ than a positive sign.  We want to point out
again, that the red points showing up at small absolute values of
$\mu$ have a large $m_A$ and a small $\tan \beta$.

Let us discuss in detail why the sign of the $\mu$-term plays an
important role for the SI cross-section calculation. The dominating
terms in the SI cross-section come from $t$-channel light and heavy
Higgs boson exchanges. The formula can be found in
references~\cite{Jungman:1995df,Falk:1998xj,Bottino:1998jx}:
\begin{equation}
  \sigma^{\rm SI} \simeq \frac{8 G_{F}^{2}}{\pi} M_{Z}^{2} m_{\text{red}}^{2} 
\left[ \frac{F_{h} I_{h}}{m_{h}^{2}}
    + \frac{F_{H} I_{H}}{m_{H}^{2}} \right]^{2}  \hspace{2mm},
\end{equation}
where $G_{F}$ is Fermi constant and $m_{\text{red}}$ is the neutralino-nucleon reduced mass:
 \begin{equation}
   m_{\text{red}} \equiv \frac{m_{\widetilde{\chi}^{0}_{1}} m_{N}} {m_{\widetilde{\chi}^{0}_{1}} + m_{N}} \hspace{2mm}.
\end{equation}
The functions $F_{h,H}$ and $I_{h,H}$ are defined as follows 
 \begin{eqnarray}
  F_{h} &\equiv&\left( -N_{11} \sin\theta_{W} + N_{12} \cos\theta_{W} \right)  
\left(N_{13} \sin\alpha + N_{14} \cos\alpha \right) \hspace{2mm},\\
 F_{H} &\equiv&\left(-N_{11} \sin\theta_{W} + N_{12} \cos\theta_{W} \right)
       \left( N_{13} \cos\alpha - N_{14} \sin\alpha \right) \hspace{2mm}, \\
 I_{h,H} &\equiv& \sum_{q} k_{q}^{h,H} m_{q} \langle N | \bar{q} q | N \rangle \hspace{2mm}.
\end{eqnarray}
The angle $\alpha$ is the mixing of the mass eigenstates ($h$
and $H$), and the coefficients $k_{q}^{h,H} $ are given by
\begin{center}
\begin{tabular}{cc}
        $k_{u\text{-type}}^{h} = \cos \alpha/ \sin \beta$ \hspace{2mm}, &
       $k_{d\text{-type}}^{h}  = -\sin\alpha/ \cos\beta  $\hspace{2mm}, \\[1mm]
       $k_{u\text{-type}}^{H} = -\sin\alpha/\sin\beta$ \hspace{2mm}, &
       $k_{d\text{-type}}^{H} = -\cos \alpha/\cos \beta$ \hspace{2mm},
\end{tabular}
\end{center}
for the up-type and down-type quarks, respectively. We neglect
threshold corrections for simplicity in our discussion.

To find a good approximation for the elements of the neutralino mixing
matrix, we use the large SUSY scale approximation $(M_i \pm |\mu|)^2
\gg M^2_Z ~(i=1,2)$ from reference~\cite{Choi:2001ww} and find for
the components:
\begin{eqnarray}
  \label{eq:mixings}
N_{12} &\simeq&-M_Z^2 \cos\theta_W \sin\theta_W\frac{M_1 + \mu \sin2 \beta}{(M_1 - M_2)(M_1^2-\mu^2)}\hspace{2mm}, \\
\label{eq:N13}
N_{13} &\simeq& -M_Z \sin\theta_W \frac{M_1 \cos\beta + \mu \sin\beta}{M_1^2 - \mu^2}\hspace{2mm}, \\
\label{eq:N14}
N_{14} &\simeq& M_Z \sin\theta_W \frac{M_1 \sin\beta + \mu \cos\beta}{M_1^2 - \mu^2} \hspace{2mm}.
\end{eqnarray}
The unitary condition on the mixing angles yields 
\begin{equation}
N_{11} =\sqrt{1 - N^2_{12} - N^2_{13} - N^2_{14}} \hspace{2mm}.
\end{equation}

 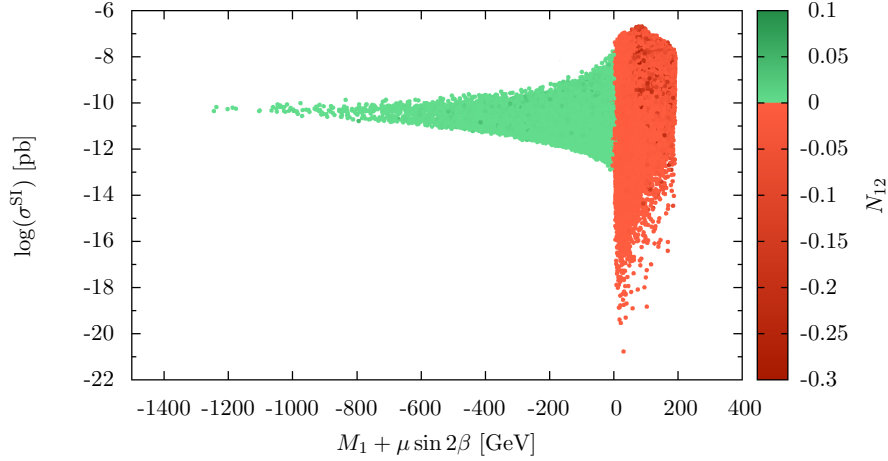
\begin{figure}[t!]
   \begin{center}
\resizebox{0.7\columnwidth}{!}{\input{pic/glt_wino}}
\caption{The dependence of $\sigma^{\rm SI}$ as a function of the quantity
  $M_1 + \mu \sin2 \beta$ and the wino component ($N_{12}$) as obtained by our
  simulations.\label{fig:explaining_sigma}}
   \end{center}
 \end{figure}
 
 These approximations are in good agreement with the simulations. The
 only strong deviation occurs for scenarios with a dominant
 annihilation into $W$-bosons near neutralino masses of around
 \hbox{80~GeV}. For all other annihilation mechanisms they give
 sufficient precision for a reliable qualitative discussion.
 Evaluating the complete expression in the decoupling
 limit\footnote{\hbox{\it I.e.}  we use $\sin^2(\alpha-\beta)
   \simeq1$, such that $\sin \alpha \simeq \cos \beta$, $\cos \alpha
   \simeq - \sin \beta$.}  yields the following formula:
\begin{eqnarray}
  \label{eq:s}
  \sigma^{\rm SI} \simeq\frac{8 G_{F}^{2}}{\pi} m_{\text{red}}^{2} \frac{M_Z^4 \sin^2\theta_W}{(M_1^2-\mu^2)^2} 
\left[ \frac{I_H}{m_H^2} \mu \cos2 \beta + \frac{I_h}{m_h^2}(M_1 + \mu \sin2 \beta) \right]^2
\left(N_{11} \sin\theta_W - N_{12} \cos \theta_W\right)^2 \hspace{1mm}.
\end{eqnarray}
Note, that $\cos2 \beta$ is negative, \hbox{\it i.e.} $\cos 2\beta
\approx -1~(\tan \beta > 2)$, such that both contributions within the
square brackets seem to have a different sign. However, looking at the
$k$-coefficients one can see that $I_h$ and $I_H$ preferably have
opposite signs in the decoupling limit, such that both terms actually 
add up for a positive $\mu$-term.

On the contrary, if $\mu$ is negative, cancellations between both
terms within the square brackets are possible and $\sigma^{\rm SI}$
can be significantly smaller.  However, this is only correct when $M_1
+ \mu \sin 2 \beta$ is positive. Note that in this case the wino
component, equation~(\ref{eq:mixings}), is negative since $M_2 > M_1$
and $|\mu| >M_1$ for the great majority of our models (see
figure~\ref{fig:explaining_sigma}). The boundary at $M_1 + \mu \sin2
\beta \approx$ \hbox{200~GeV} occurs for the maximal values of $M_1$ and
$\tan \beta$ when, at the same time, $\left|\mu\right|$ is small.

For $M_1 + \mu \sin2 \beta < 0$ a cancellation is no longer possible
and the resulting $\sigma^{\rm SI}$ is higher. Most of the
scenarios with a positive wino component in figure
\ref{fig:explaining_sigma} correspond to the earlier mentioned highly
fine-tuned models at the Higgs-resonance. These showed large absolute
values of the $\mu$-term and small $\tan \beta$. Then, $M_1 + \mu
\sin2 \beta$ takes the smallest possible negative values and the 
cross-section is rather high ($\hbox{\it i.e.}~10^{-10}$ pb). Also, those
scenarios with light staus whose mass is suppressed by the
off-diagonal mass matrix elements possess a positive wino component
and hence map accordingly into the direct detection plane.

An expression similar to equation~(\ref{eq:s}) has been found
in~\cite{Kitano:2006ws}, however the wino component has not been
considered there. Even though cancellations that occur between
$N_{11}$ and $N_{12}$ may be neglected, since the bino component is in
general much larger than the wino component, the sign of $N_{12}$ is
crucial for the behavior of $\sigma^{\rm SI}$.

\subsection{The muon anomalous magnetic moment and the $\mu$-term}
\label{sec:g2}
 \begin{figure}[t!]
   \begin{center}
\resizebox{0.85\columnwidth}{!}{\input{pic/pos_g23_direct}}
\caption{The level of fine-tuning
 after inclusion of the muon anomalous magnetic moment for positive
  values of the $\mu$-term. \label{fig:g2_pos}}
   \end{center}
 \end{figure}
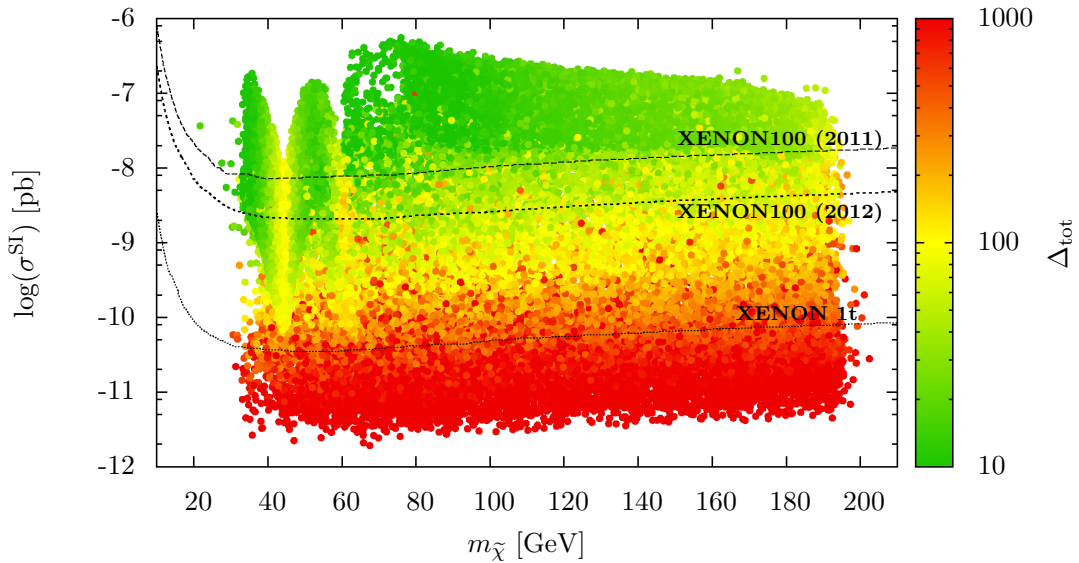
\begin{figure}[t!]
    \begin{center}
\begin{minipage}{0.48\textwidth}
\resizebox{\columnwidth}{!}{\input{pic/assumption_g2}}
\end{minipage}
\hfill
\begin{minipage}{0.49\textwidth}
\resizebox{\columnwidth}{!}{\input{pic/g22_mu}}
\end{minipage}
\caption{The left plot shows the $a_\mu$ condition as a function of
  $M_2/M_1$ and $m_{\widetilde{l}_L}/m_{\widetilde{l}_R}$.  Only the
  green scenarios satisfy 3 $\sigma$ cut of the $a_\mu$
  condition. $m_{\widetilde{l}_L}/m_{\widetilde{l}_R}$ must be much
  greater than one in order to respect $a_\mu$ for a negative
  $\mu$-term. The right panel indicates the behavior of $a_\mu$ when
  moving to smaller values of $\mu$. Green dots satisfy a 2 $\sigma$,
  yellow dots a 3 $\sigma$ cut of $a_\mu$. \label{fig:explaining_g2}}
 \end{center}
 \end{figure}
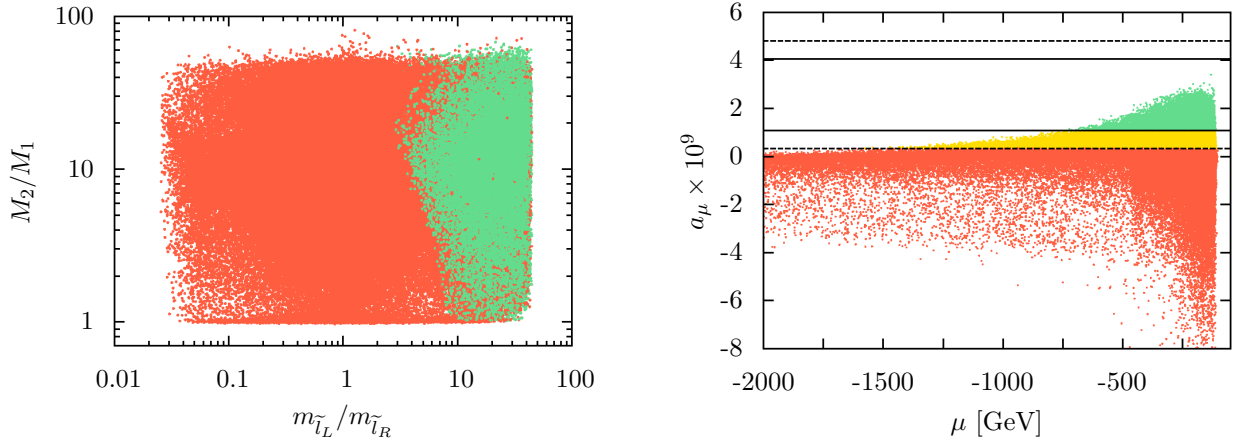

 In figure~\ref{fig:g2_pos} we applied the muon anomalous magnetic
 moment constraint (\hbox{\it i.e.} the deviation from the SM
 expectation) to our models with a positive $\mu$-term. In general it
 is not difficult to fulfill the $a_\mu$ condition.  Compared to
 figure~\ref{fig:pos_nog2} the results essentially do not change. In
 reference~\cite{Cho:2011rk} we find the three most important MSSM
 loop contributions to $a_\mu$:
\begin{eqnarray}
  a_\mu(\widetilde{W}-\widetilde{H},\widetilde{\nu}_\mu)&=&
\frac{g^2}{8 \pi^2}\frac{m_\mu^2M_2\mu\tan\beta}{m^4_{\widetilde{\nu}}}~F_a\left( \frac{M_2^2}{m^2_{\widetilde{\nu}}},
\frac{\mu^2}{m^2_{\widetilde{\nu}}}\right) \label{eq:loop1}\hspace{1mm},\\
  a_\mu(\widetilde{B}, \widetilde{\mu}_L-\widetilde{\mu}_R)&=&\frac{g'^2}{8\pi^2}\frac{m^2_\mu \mu\tan\beta}{M_1^3}~F_b
\left(\frac{m^2_{\widetilde{\mu}_L}}{M_1^2},\frac{m^2_{\widetilde{\mu}_R}}{M_1^2}\right) \label{eq:loop2}\hspace{1mm},\\
  a_\mu(\widetilde{B}-\widetilde{H},\widetilde{\mu}_R)&=&-\frac{g'^2}{8 \pi^2}
\frac{m_\mu^2M_1\mu\tan\beta}{m^4_{\widetilde{\mu}_R}} ~F_b
\left( \frac{M_1^2}{m^2_{\widetilde{\mu}_R}},\frac{\mu^2}{m^2_{\widetilde{\mu}_R}}\right) \label{eq:loop3}\hspace{1mm}.
\end{eqnarray}
Here, the positive defined functions $F_a$ and $F_b$ are given by:
\begin{eqnarray}
 && F_a(x,y) =-\frac{G_3(x)-G_3(y)}{x-y} \hspace{1mm}, \nonumber\\
&& F_b(x,y)=-\frac{G_4(x)-G_4(y)}{x-y}\hspace{1mm}, \nonumber \\
 && G_3(x) =\frac{1}{2(x-1)^3}\left[ (x-1)(x-3) + 2 \ln x \right]\hspace{1mm},  \nonumber\\
&& G_4(x)=\frac{1}{2(x-1)^3}\left[ (x-1)(x+1) - 2 x \ln x \right]\hspace{1mm}. \nonumber
\end{eqnarray}
 It is commonly believed that $a_\mu$ cannot be fulfilled for a
 negative $\mu$-term if $M_1$ and $M_2$ are
 positive~\cite{Kim:2002cy}. However, if equation (\ref{eq:loop3}),
 the bino--higgsino--right-handed smuon loop, dominates over the sum
 of the other contributions, equation~(\ref{eq:loop1}), the
 wino--higgsino--muon sneutrino loop, and equation (\ref{eq:loop2}),
 the bino--left-handed smuon--right-handed smuon loop, the total
 amount of the achieved positive pull of $a_\mu$ can be sufficient to
 correctly deviate from the SM prediction. This situation occurs in
 the limit of $m_{\widetilde{l}_L}/m_{\widetilde{l}_R} \gg 1$, see
 left panel of figure~\ref{fig:explaining_g2}.  In this case the
 $\widetilde{\tau}_1$'s are always light ($\lesssim 400$~GeV) due to
 equation (\ref{eq:stau_mass_case2}).  It is very important to note
 that $a_\mu$ is strongly dependent on the smuon
 parameters. Satisfying $a_\mu$ becomes easier once one abandons the
 slepton mass generation universality since the relic density
 condition, which restricts the stau mass, would then become
 independent of $a_\mu$.

 In our set-up we found that for large negative values of $\mu$ the
 contributions of equations (\ref{eq:loop1}) and (\ref{eq:loop2})
 become important again due to the behavior of the loop functions
 $F_a$ and $F_b$. Because of the negative sign of $\mu$ the pull of
 $a_\mu$ tends to be toward negative values, out of the range that
 respects the experiments. We therefore find a lower limit on $\mu$ of
 about -1500 GeV, see right panel of
 figure~\ref{fig:explaining_g2}. This immediately favors low
 fine-tuned points as applying the $a_\mu$ condition automatically
 cuts out large negative values of $\mu$. A great part of the models
 whose stau mass is strongly influenced by $(A_\tau-\mu\tan\beta)$
 drop out, and so do the highly fine-tuned scenarios from
 Higgs-resonant neutralino annihilation. Comparing to
 figure~\ref{fig:mechanism_xenonplane} one can see that a great part
 of the direct detection plane at low $\sigma^{\rm SI}$ is filled by
 slepton annihilation.
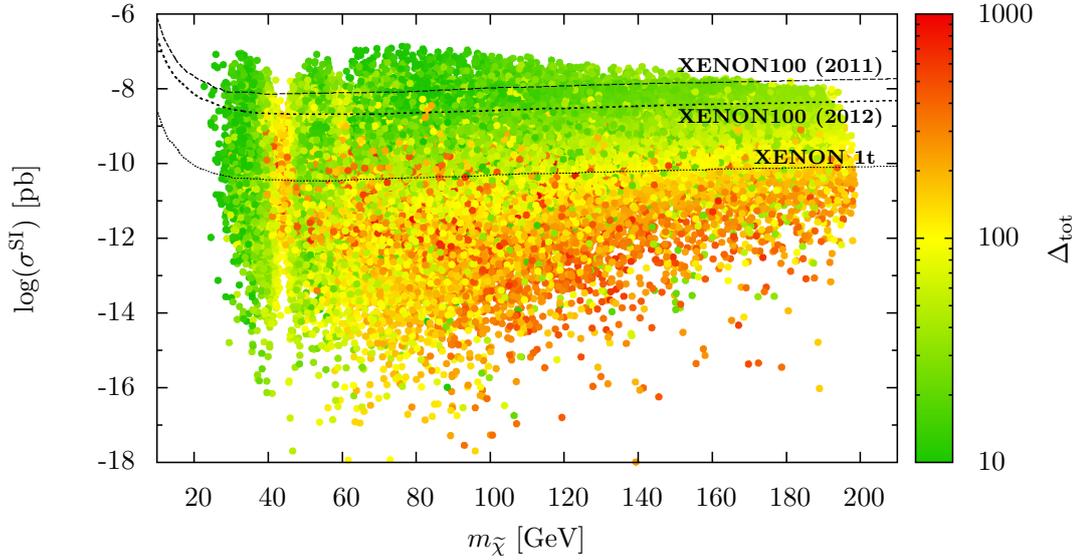
\begin{figure}[t!]
   \begin{center}
\resizebox{0.85\columnwidth}{!}{\input{pic/neg_g23_direct}}
\caption{The level of fine-tuning after inclusion of the muon magnetic
  moment for a negative $\mu$-term in analogy to
  figure~\ref{fig:g2_pos}. \label{fig:g2_neg}}
   \end{center}
 \end{figure}
\begin{figure}[h!]
   \begin{center}
\resizebox{0.85\columnwidth}{!}{\input{pic/neg_g22}}
\caption{The level of fine-tuning after inclusion of the muon magnetic
  moment at a 2 $\sigma$ level for a negative $\mu$-term in analogy to
  figure~\ref{fig:g2_pos}. \label{fig:g2_neg2}}
   \end{center}
 \end{figure}
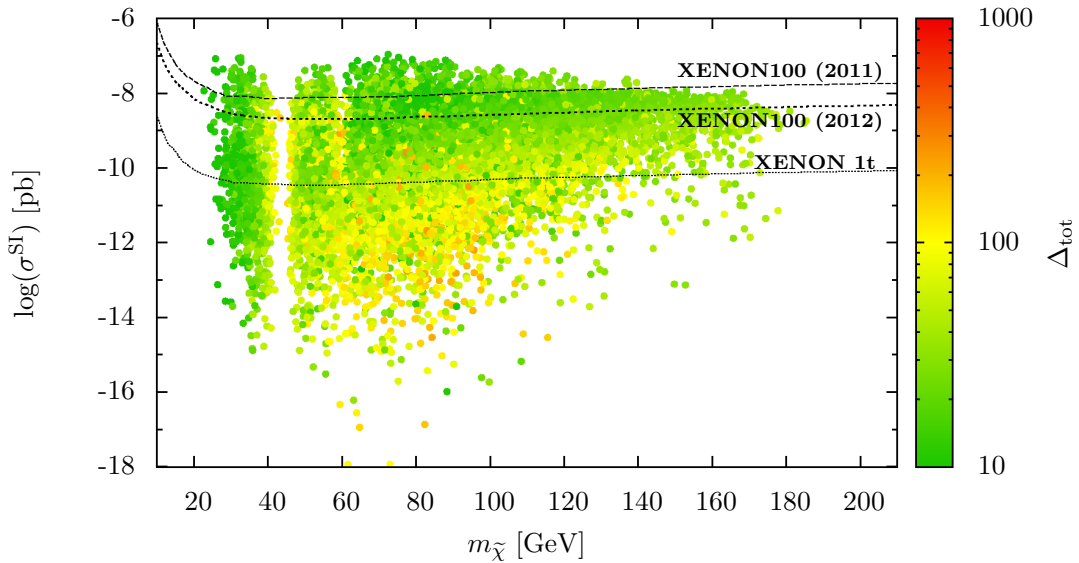

 Taking a more restrictive limit for $a_\mu$ at a 2 $\sigma$ level we
 find an increase of the lower border on $\mu$ and, related to this, a
 blank region in the direct detection plane at the $Z$-resonance
 (compare figure~\ref{fig:g2_neg2}). As argued earlier, around this
 resonance a suppression of the neutralino-neutralino-$Z$-coupling is
 necessary to satisfy the relic density condition.  Because of the
 stronger constraint on $\mu$ from the $a_\mu$ limit, this suppression
 cannot be strong enough and the relic density condition fails to be
 fulfilled. Not only due to uncertainties of hadronic effects, but
 also as our essential results are not changed, we prefer to use the 3
 $\sigma$ limit on $a_\mu$.

The muon anomalous magnetic moment condition has no effect on the low
fine-tuned models and, as argued above, less fine-tuned scenarios are
even preferred (see figure~\ref{fig:g2_neg}).  Note again, that in the
mass range $m_{\widetilde{\chi}} \approx [60,80]$~GeV neutralino
annihilation can proceed via light staus to produce the correct relic
abundance (compare figure~\ref{fig:mechanism_xenonplane} and
references~\cite{Farina:2011bh,Arbey:2011un,CahillRowley:2012rv}.  A
great part of the parameter space with $\Delta_{\rm tot} \lesssim 100$
will be probed by future direct searches, but there are regions left
that will not be tested.

 We emphasize that a negative sign of $\mu$ is by no mean in
 contradiction to the $a_\mu$ condition even though we did not
 distinguish between slepton generations. Therefore, this case should
 be paid more attention to in future studies.

\subsection{The functional tuning of $\sigma^{\rm SI}$}
\label{sec:ft_sigma}
\indent In section~\ref{sec:cross-section} we argued that the direct
detection cross-section may become very small for a negative
$\mu$-term if cancellations between the light and heavy Higgs
contributions are almost exact. This is only possible when the wino
component is negative, or equivalently, for positive $M_1 + \mu \sin 2
\beta$.  In other words: This depends in a very sensitive way
on the functional dependence of $\sigma^{\rm SI}$ on this relation.

To quantify these extremely accurate, tuned cancellations that might
appear in $\sigma^{\rm SI}$, we therefore define a functional
fine-tuning, $ \Delta_{\rm f}$, analogously to
equation~(\ref{eq:finetuning}):
\begin{equation}
 \Delta {{\rm f}_i} \equiv \left| \frac{\partial \ln \sigma^{\rm SI}}{\partial \ln p_i} \right| \hspace{2mm}, 
\hspace{7mm}\Delta_{\rm f} \equiv\sqrt{ \sum{}_{p_i= \mu, \tan \beta, M_1, M_2, m_A} 
\left\{\Delta{{\rm f}_i} \right\}^2 }\hspace{2mm}.
\end{equation}
The overall fine-tuning measure that includes both, the electroweak
and the functional tuning, is then defined as:
\begin{equation}
 \mbox{\Large$\varSigma$}_{\rm FT} \equiv \sqrt{\Delta^2_{\rm tot} + \Delta_{\rm f}^2} \hspace{2mm}.
\end{equation}
In this way we deal with both tunings on equal footing.

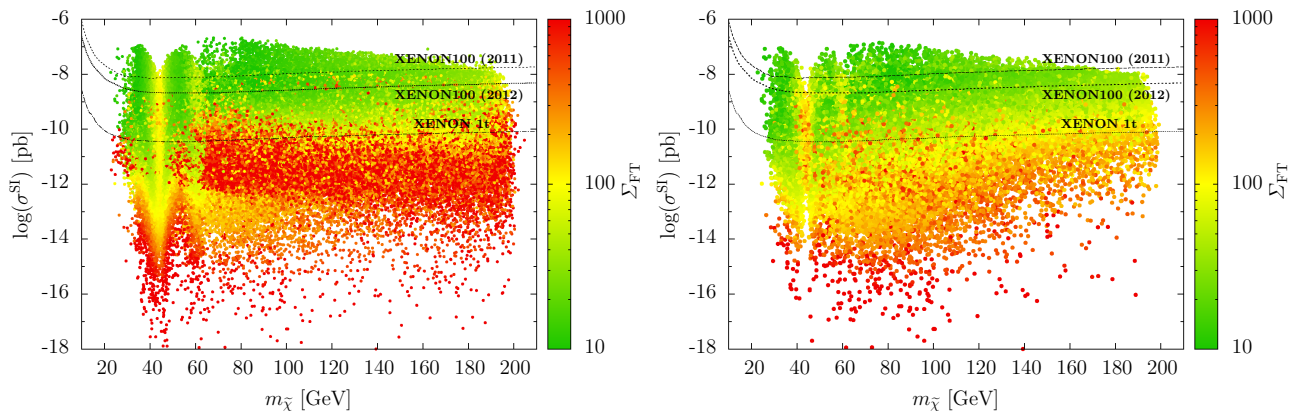
\begin{figure}[!t]
\begin{center}
\begin{minipage}{0.49\textwidth}
\resizebox{\columnwidth}{!}{\input{pic/FT_sigma_neg_nog2_direct}}
\end{minipage}
\hfill
\begin{minipage}{0.49\textwidth}
\resizebox{\columnwidth}{!}{\input{pic/FT_sigma_neg_g23_direct}}
\end{minipage}
\caption{The distribution of the fine-tuning measure
  $\mbox{\Large$\varSigma $}_{\rm FT} $, that includes the functional
  tuning of $\sigma^{\rm SI}$ and the electroweak fine-tuning, in the
  direct detection plane for a negative $\mu$-term. The 3 sigma $a_\mu$
  constraint has been applied in the right panel.\label{fig:ft_sigma}}
\end{center}
\end{figure}

In figure~\ref{fig:ft_sigma} we show how the overall fine-tuning
measure, $\mbox{\Large$\varSigma $}_{\rm FT}$, maps into the direct
detection plane. Of course, this measure only makes sense for a
negative $\mu$-term since no cancellations are possible if $\mu$ is
positive. It is visible comparing to figure~\ref{fig:g2_neg} that for
those scenarios in which the cancellations suppress $\sigma^{\rm SI}$
below approximately $10^{-14}$ pb the functional tuning measure
becomes important and rises the overall tuning into an unacceptable
range. For larger cross-sections the fine-tuning is still dominated by
the sensitivity of the $Z$-mass and no change to the previously
discussed results is observed. Comparing the left- and right-panel of
figure~\ref{fig:ft_sigma} one can again observe the removal of the
highly fine-tuned band containing models with light staus whose mass
is strongly influenced by the off-diagonal stau mass matrix elements,
when the $a_\mu$ constraint is applied.

We see that even when the functional tuning measure is taken into
account scenarios that possess a fine-tuning
$\mbox{\Large$\varSigma$}_{\rm FT}$ lower than 100 are possible in the
pMSSM avoiding all of our applied constraints.

\subsection{The functional tuning of $\Omega h^2$}
\label{sec:tuning_omega}

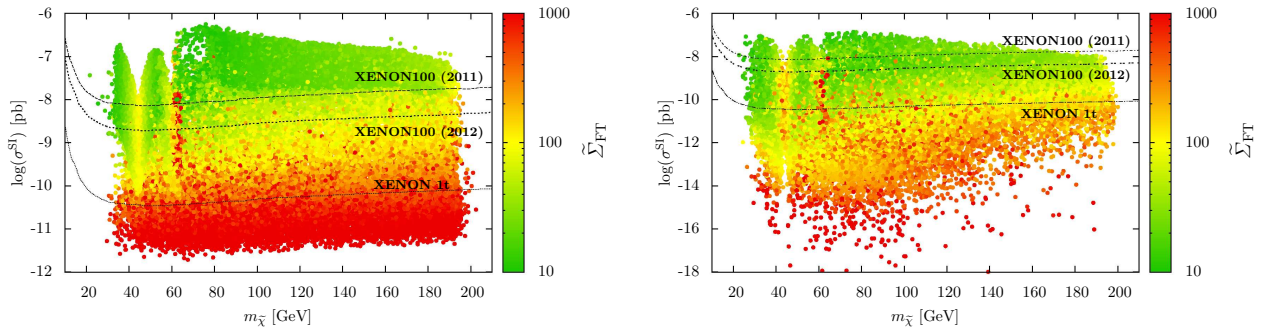
\begin{figure}[!t]
\begin{center}
\begin{minipage}{0.49\textwidth}
\resizebox{\columnwidth}{!}{\input{pic/pos_omega_tuning}}
\end{minipage}
\hfill
\begin{minipage}{0.49\textwidth}
\resizebox{\columnwidth}{!}{\input{pic/neg_omega_tuning}}
\end{minipage}
\caption{The distribution of the fine-tuning measure
  $\mbox{\Large$\widetilde{\varSigma}$}_{\rm FT} $ in the
  direct detection plane for a positive (left panel) and negative (right panel)
  $\mu$-term. The 3 sigma $a_\mu$
  constraint has been applied in both panels.\label{fig:ft_omega}}
\end{center}
\end{figure}

Analogously to the previous section one can also discuss a possible
tuning of the relic abundance, as has been suggested in~\cite{Ellis:2001zk}. We
again define the functional fine-tuning to be:
\begin{equation}
 \Delta {{\rm \tilde{f}}_i} \equiv \left| \frac{\partial \ln \Omega h^2}{\partial \ln p_i} \right| \hspace{2mm}, 
\hspace{7mm}\Delta_{\rm \tilde{f}} \equiv\sqrt{ \sum{}_{p_i} 
\left\{\Delta{{\rm \tilde{f}}_i} \right\}^2 }\hspace{2mm},
\end{equation}
where the sum here runs over all eleven input parameters. The overall tuning
is then given by:
\begin{equation}
 \mbox{\Large$\widetilde{\varSigma}$}_{\rm FT} \equiv \sqrt{\Delta^2_{\rm tot} + \Delta_{\rm f}^2+ \Delta_{\rm \tilde{f}}^2}\hspace{2mm}.
\end{equation}
Remember that for positive $\mu$ we set $\Delta_{\rm f} = 0$ because
there is no ``accidental'' cancellation. From our analysis and
figure~\ref{fig:ft_omega} we observe that the tuning of the relic
density is generally low and plays a subdominant role compared to the
electroweak tuning.  The only tuned solutions occur close to the
$h$-resonance. This is explained by the narrow decay width of the
Higgs boson.

For a scenario to lie inside this narrow resonance and to produce the
correct relic abundance, the parameters need to fulfill certain
relations which leads to a high sensitivity of $\Omega h^2$ at $m_\chi
\approx m_h/2$. Such a tuning does not appear for a larger resonance
as is the case for the $Z$-boson.  Observe also, that, despite of the
narrow Higgs decay width, the resonant region of light Higgs
annihilation starts for Dark Matter masses much below $m_h/2$ since
the kinetic energy of the neutralino contributes to their overall
energy (see section~\ref{sec:nog2}).

Even though not clearly visible for the negative $\mu$ case in
figure~\ref{fig:ft_omega} (simply because the untuned scenarios where
slepton annihilation is the most important annihilation mechanism
outnumber the tuned scenarios) the effect is the same and
independent of the sign of $\mu$.

\subsection{Parameter mapping distribution of our models}
\label{sec:probability}

\begin{figure}[!t]
\begin{center}
\begin{minipage}{0.49\textwidth}
\resizebox{\columnwidth}{!}{\input{pic/probability_distribution_pos_g23}}
\end{minipage}
\hfill
\begin{minipage}{0.49\textwidth}
 \resizebox{\columnwidth}{!}{\input{pic/probability_distribution_neg_g23}}
\end{minipage}
\caption{\label{fig:probability} The level of fine-tuning and the
  density of model points arising from a homogeneously (randomized)
  scan of the parameter space. The case of $\mu>0$ is displayed in the
  left and $\mu<0$ in the right panel.  All experimental constraints
  are taken into account including the muon anomalous magnetic moment. }
\end{center}
\end{figure}
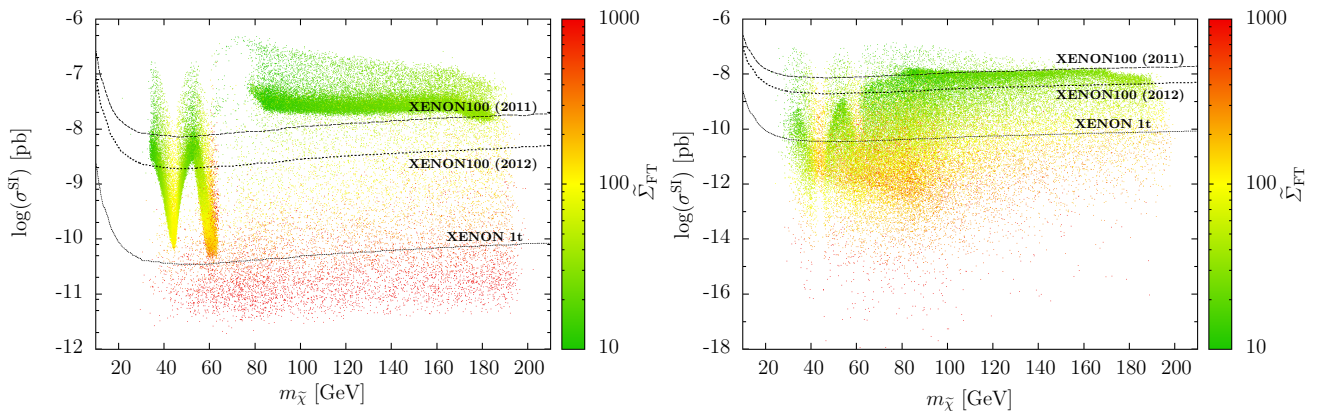

Throughout this paper we have so far studied how much tuning is
required to reach a certain point in the
$m_{\widetilde{\chi}}-\sigma^{\rm SI}$ plane.  In the end we would
also like to briefly look at the mapping between model parameters and
physical quantities, which does not play a role in the previous
results. Within the MSSM it is clear that a statistical interpretation
of the mapping of the model parameters to the
$m_{\widetilde\chi}-\sigma^{\rm SI}$ plane does not make sense, since
there is only one solution in the end. But the mapping is nevertheless
interesting if one thinks beyond, since it shows preferred physical
regions in the $m_{\widetilde\chi}-\sigma^{\rm SI}$ plane under the
assumption that all model parameters have equal probability in an
embedding which explains these parameters.  This is shown in
figure~\ref{fig:probability}, where the color coding shows in addition
the previously discussed fine-tuning measure.  It is immediately
apparent that some regions are more densely populated. For example, we
see that the regions from $Z, h, H$ and $A$ resonant annihilations and
the light chargino mediated annihilation into $W$-bosons have a
stronger weight than light stau annihilations for positive $\mu$ (see
left panel of figure~\ref{fig:probability}). Note, that the light
chargino and heavy and CP-odd Higgs annihilations have now been ruled
out by the XENON100 (2012) update.  For a negative $\mu$-term stau
annihilations are also important, especially in the region
$m_{\widetilde{\chi}} \in [65,80]$~GeV.  Signals from direct searches
should be expected in the untuned, densely populated regions of the
parameter space, which are not yet excluded by data (especially
XENON100). However, we would like to stress again that such a mapping
has no meaning within the MSSM and that the shown plots rest on the
assumption of equal probability in some embedding.

\section{Discussion and conclusions}
\label{sec:conclusions}
\indent In this paper we have analyzed the naturalness of neutralino
Dark Matter in the non-universal gaugino model within the framework of
the minimal supersymmetric extension of the Standard Model. We have
taken into account all cosmological (upper and lower bound on the
relic density), collider and flavor constraints including the results
of XENON100 (2012), LHC data on the Higgs mass, ${\rm
  Br}(B_s\to\mu^+\mu^-)$ and pseudo-Higgs searches. Hereby the soft
supersymmetry breaking terms are parameterized by 11 independent free
parameters, that we have chosen such that the lightest supersymmetric
particle is the lightest neutralino with a mass smaller than 200~GeV.
We studied from the Dark Matter perspective how much fine-tuning is
needed to reach a certain point in the $m_{\widetilde{\chi}} -
\sigma^{\rm SI}$ plane. Therefore we use a parameter fine-tuning
measure (see equation~(\ref{eq:finetuning})) which was used before in
order to study the naturalness of the Higgs -- SUSY breaking scale
separation.

We first presented in figure 1 the contribution to the Dark Matter abundance
$\Omega h^2$ for different annihilation mechanisms as a function of $m_{\widetilde{\chi}}$.
Demanding that neutralinos provide the right amount of Dark Matter, we restrict
the further scans to cases where $\Omega h^2 \in [0.089,0.136]$.

We also investigated the dominant neutralino annihilation mechanisms
(figure~\ref{fig:relic}) and have shown their arrangement in the
$m_{\widetilde{\chi}}-\sigma^{\rm SI}$-plane
(figure~\ref{fig:mechanism_xenonplane}). In case of a signal in a
direct detection experiment the most important annihilation mechanism
can be deduced and we would know which channel at the LHC is promising
for production of neutralino Dark Matter. To avoid the limits of the direct detection
results of XENON100 (2012), we showed that light stau annihilation of
neutralinos in the early Universe play a special role, not only in the
mass range of light neutralinos, $\lesssim30$~GeV, but also between
$ \simeq [60, 80]$~GeV, which has been missed so far in other studies. It is
important to differently parameterize the soft SUSY masses of the
left- ($m_{{\widetilde l}_L}$) and right-handed ($m_{{\widetilde l}_R}$) sleptons.

Note that we scanned the input parameter space in such a way that we
obtain the fine-tuning for every point in the $m_{\tilde\chi} -
\sigma^{\rm SI}$ plane which is accessible.  This implies that the
density of points has no meaning, but that the envelope implies that
certain areas cannot be reached by any input parameter.  With this
method we showed in figure~\ref{fig:pos_nog2} how the electroweak
fine-tuning maps into the direct detection plane and found that a
great part of untuned regions is already excluded by the current
XENON100 (2012) limit when the supersymmetric Higgs mass parameter,
the $\mu$-term, is positive. A general trend for higher fine-tuning
for smaller $\sigma^{\rm SI}$ is then visible. This trend can easily
be understood, since the cross-section departs more and more from its
natural value set by the generic scale. Thus, future direct detection
experiments will push the amount of the electroweak fine-tuning
further up. The only exception occurs for neutralino masses that allow
for resonant annihilations, especially near the $Z$- and $h$-boson
resonances.

This last statement is also valid for a negative $\mu$-term and the
electroweak fine-tuning near the $Z$- and $h$-resonance stays small
independent of the value of $\sigma^{\rm SI}$. Additionally, due to
cancellations between contributions from light and heavy Higgs
exchanges the direct detection cross-section gets shifted to smaller
values, such that the XENON100 (2012) exclusion limit is fulfilled
easily. A negative $\mu$-term is therefore favored from a fine-tuning
perspective (see figure~\ref{fig:neg_nog2}).

In our analytical study (see figure~\ref{fig:explaining_sigma}) we
have discussed the reason why the negative value of the $\mu$-term
allows for these cancellations in the spin-independent cross-section,
and showed that the combination of input parameters $(M_1 + \mu
\sin2\beta)$ and the sign of the wino component are responsible for
decreasing $\sigma^{\rm SI}$ to values that can be as low as
$10^{-20}$ pb (see also figure~\ref{fig:neg_nog2}).  Since these
cancellations might be viewed as an instance of tuning, we reevaluated
the fine-tuning by adding a measure of ``equation-tuning'' and found that
scenarios with $\sigma^{\rm SI} \lesssim 10^{-15}$ pb are always
unbearably tuned (figure~\ref{fig:ft_sigma}). We find the scenario with
the lowest fine-tuning, independent of the sign of $\mu$, at
$m_{\widetilde\chi} \approx$ \hbox{84~GeV} and $\sigma^{\rm SI}
\approx 2.0 \times$ \hbox{10$^{-9}$~pb} just below the new limit.

Even for a negative $\mu$-term we were able to get the correct
positive pull for the muon anomalous magnetic moment, $a_\mu$, to
correctly deviate from the Standard Model
(figure~\ref{fig:explaining_g2}). This has been thought to be very
difficult, but is possible due to the bino--higgsino--right-handed
smuon loop which contributes to $a_\mu$ dominantly when both gaugino
masses are positive ($M_1>0$ and $M_2>0$) and $m_{{\widetilde l}_L}
\gg m_{{\widetilde l}_R}$.  If the latter condition is fulfilled staus
are generally light ($ \lesssim 400$ GeV) and help to respect the
cosmological abundance of Dark Matter by light stau annihilation in
the complete mass region of the neutralino.  It should be stressed
that there is an easy way to satisfy $a_\mu$, namely, if additional
parameters for the smuon masses, \hbox{\it i.e.}  $m_{{\widetilde
    \mu}_{L, R}}$, are introduced. In this case we can avoid the
connection between the relic density and the anomalous muon magnetic
moment and fulfill both conditions without any doubt. Therefore, the
case of a negative sign of the $\mu$-term is equally important and
should be investigated more carefully in future studies.

Note that the density of points is meaningless except in
figure~\ref{fig:probability}, since we do not assign a probability
measure, but determine only the amount of tuning required to reach a
certain point in the $m_{\widetilde{\chi}} - \sigma^{\rm SI}$ plane.
The envelope implies, however, that these points cannot be reached. In
this context it is interesting to note that the cross-section of the
neutralino annihilating into two photons and into a pair of photon and
$Z$-boson is loop suppressed and is therefore much smaller than the
requirement from the claimed \hbox{130~GeV} gamma-ray line in the
Fermi-LAT data. Thus, our models can not explain this
``evidence''. Besides, very light neutralino scenarios consistent with
the DAMA/LIBRA, CoGeNT, CRESST experiments cannot be explained in the
pMSSM especially due to the limits on ${\rm Br}(B_s \rightarrow
\mu^{+} \mu^{-})$ and on the pseudo-Higgs mass-$\tan \beta$-plane.

Finally, in section~\ref{sec:probability} we have discussed in
addition the parameter mapping distribution of our models into the
$m_{\widetilde{\chi}} - \sigma^{\rm SI}$ plane
(figure~\ref{fig:probability}). We found that the $Z$- and $h$-boson
resonant areas become the preferred regions to detect neutralino dark
matter if the $\mu$-term is positive. For negative $\mu$ another
important region is formed by light stau annihilation, that has
avoided direct searches so far.

Note that taking into account the branching ratio of the decay $B_s
\to \mu^+ \mu^-$~\cite{LHCb:2012ct} does not change our discussion and
results because we are in the decoupling regime and $\tan\beta$ is not
too high. There are only few models that do not satisfy the lower
limit of ${\rm Br}(B_s \to \mu^+ \mu^-)$ at 95 \% C.L.  Furthermore,
the strong bounds from LHC for light generation squarks and gluino
masses~\cite{Aad:2012fqa} do not affect the main conclusion of our
discussions, since its contributions to the direct detection and
pair-(co)annihilation cross-sections are typically subdominant.

Note added: After the completion of this work, two papers appeared
which have studied the importance of the $\mu$-term sign for the
direct detection cross-section within the frame work of
MSSM~\cite{Cheung:2012qy} and NMSSM~\cite{Perelstein:2012qg},
respectively. In section~\ref{sec:cross-section} we have discussed the
suppression of $\sigma^{\rm SI}$ in the region (so-called ``blind
spot'') where a particular combination of SUSY parameter $M_1 + \mu
\sin 2\beta$ is small.

\section*{Acknowledgments}
One of us (Y. T.) wishes to thank L.~Calibbi and T.~Ota for
collaboration in early stages of the project and useful
discussions. The work of Y. T.  is supported by the ERC Starting Grant
MANITOP.

\newpage


\end{document}

%% file: pic/relic_density.tex
\begingroup
  \makeatletter
  \providecommand\color[2][]{%
    \GenericError{(gnuplot) \space\space\space\@spaces}{%
      Package color not loaded in conjunction with
      terminal option `colourtext'%
    }{See the gnuplot documentation for explanation.%
    }{Either use 'blacktext' in gnuplot or load the package
      color.sty in LaTeX.}%
    \renewcommand\color[2][]{}%
  }%
  \providecommand\includegraphics[2][]{%
    \GenericError{(gnuplot) \space\space\space\@spaces}{%
      Package graphicx or graphics not loaded%
    }{See the gnuplot documentation for explanation.%
    }{The gnuplot epslatex terminal needs graphicx.sty or graphics.sty.}%
    \renewcommand\includegraphics[2][]{}%
  }%
  \providecommand\rotatebox[2]{#2}%
  \@ifundefined{ifGPcolor}{%
    \newif\ifGPcolor
    \GPcolortrue
  }{}%
  \@ifundefined{ifGPblacktext}{%
    \newif\ifGPblacktext
    \GPblacktextfalse
  }{}%
  \let\gplgaddtomacro\g@addto@macro
  \gdef\gplbacktext{}%
  \gdef\gplfronttext{}%
  \makeatother
  \ifGPblacktext
    \def\colorrgb#1{}%
    \def\colorgray#1{}%
  \else
    \ifGPcolor
      \def\colorrgb#1{\color[rgb]{#1}}%
      \def\colorgray#1{\color[gray]{#1}}%
      \expandafter\def\csname LTw\endcsname{\color{white}}%
      \expandafter\def\csname LTb\endcsname{\color{black}}%
      \expandafter\def\csname LTa\endcsname{\color{black}}%
      \expandafter\def\csname LT0\endcsname{\color[rgb]{1,0,0}}%
      \expandafter\def\csname LT1\endcsname{\color[rgb]{0,1,0}}%
      \expandafter\def\csname LT2\endcsname{\color[rgb]{0,0,1}}%
      \expandafter\def\csname LT3\endcsname{\color[rgb]{1,0,1}}%
      \expandafter\def\csname LT4\endcsname{\color[rgb]{0,1,1}}%
      \expandafter\def\csname LT5\endcsname{\color[rgb]{1,1,0}}%
      \expandafter\def\csname LT6\endcsname{\color[rgb]{0,0,0}}%
      \expandafter\def\csname LT7\endcsname{\color[rgb]{1,0.3,0}}%
      \expandafter\def\csname LT8\endcsname{\color[rgb]{0.5,0.5,0.5}}%
    \else
      \def\colorrgb#1{\color{black}}%
      \def\colorgray#1{\color[gray]{#1}}%
      \expandafter\def\csname LTw\endcsname{\color{white}}%
      \expandafter\def\csname LTb\endcsname{\color{black}}%
      \expandafter\def\csname LTa\endcsname{\color{black}}%
      \expandafter\def\csname LT0\endcsname{\color{black}}%
      \expandafter\def\csname LT1\endcsname{\color{black}}%
      \expandafter\def\csname LT2\endcsname{\color{black}}%
      \expandafter\def\csname LT3\endcsname{\color{black}}%
      \expandafter\def\csname LT4\endcsname{\color{black}}%
      \expandafter\def\csname LT5\endcsname{\color{black}}%
      \expandafter\def\csname LT6\endcsname{\color{black}}%
      \expandafter\def\csname LT7\endcsname{\color{black}}%
      \expandafter\def\csname LT8\endcsname{\color{black}}%
    \fi
  \fi
  \setlength{\unitlength}{0.0500bp}%
  \begin{picture}(8502.00,5668.00)%
    \gplgaddtomacro\gplbacktext{%
    }%
    \gplgaddtomacro\gplfronttext{%
      \csname LTb\endcsname%
      \put(1628,881){\makebox(0,0){\strut{} 20}}%
      \put(2211,881){\makebox(0,0){\strut{} 40}}%
      \put(2794,881){\makebox(0,0){\strut{} 60}}%
      \put(3377,881){\makebox(0,0){\strut{} 80}}%
      \put(3960,881){\makebox(0,0){\strut{} 100}}%
      \put(4542,881){\makebox(0,0){\strut{} 120}}%
      \put(5125,881){\makebox(0,0){\strut{} 140}}%
      \put(5708,881){\makebox(0,0){\strut{} 160}}%
      \put(6291,881){\makebox(0,0){\strut{} 180}}%
      \put(6874,881){\makebox(0,0){\strut{} 200}}%
      \put(4251,551){\makebox(0,0){\strut{}$m_{\widetilde{\chi}}$ [GeV]}}%
      \put(1165,1167){\makebox(0,0)[r]{\strut{} 0.0001}}%
      \put(1165,2128){\makebox(0,0)[r]{\strut{} 0.001}}%
      \put(1165,3088){\makebox(0,0)[r]{\strut{} 0.01}}%
      \put(1165,4049){\makebox(0,0)[r]{\strut{} 0.1}}%
      \put(175,2944){\rotatebox{-270}{\makebox(0,0){\strut{}$\Omega h^2$}}}%
      \put(7734,1462){\makebox(0,0)[l]{\strut{} \Large{$l \bar{l}$}}}%
      \put(7734,2054){\makebox(0,0)[l]{\strut{}  \Large{$q \bar{q}$}}}%
      \put(7734,2647){\makebox(0,0)[l]{\strut{}  \Large{$W^+ W^-$}}}%
      \put(7734,3239){\makebox(0,0)[l]{\strut{}  \Large{$t \bar{t}$}}}%
      \put(7734,3982){\makebox(0,0)[l]{\strut{}  \Large{$\widetilde{\chi}^0_1 \widetilde{\chi}^+_1$}}}%
   \put(7734,3692){\makebox(0,0)[l]{\strut{}  \Large{coannihilation}}}%
      \put(7734,4574){\makebox(0,0)[l]{\strut{} \Large{$\widetilde{\chi}^0_1 \widetilde{\l}$}}}%
 \put(7734,4284){\makebox(0,0)[l]{\strut{} \Large{coannihilation}}}%

    }%
    \gplbacktext
    \put(0,0){\includegraphics{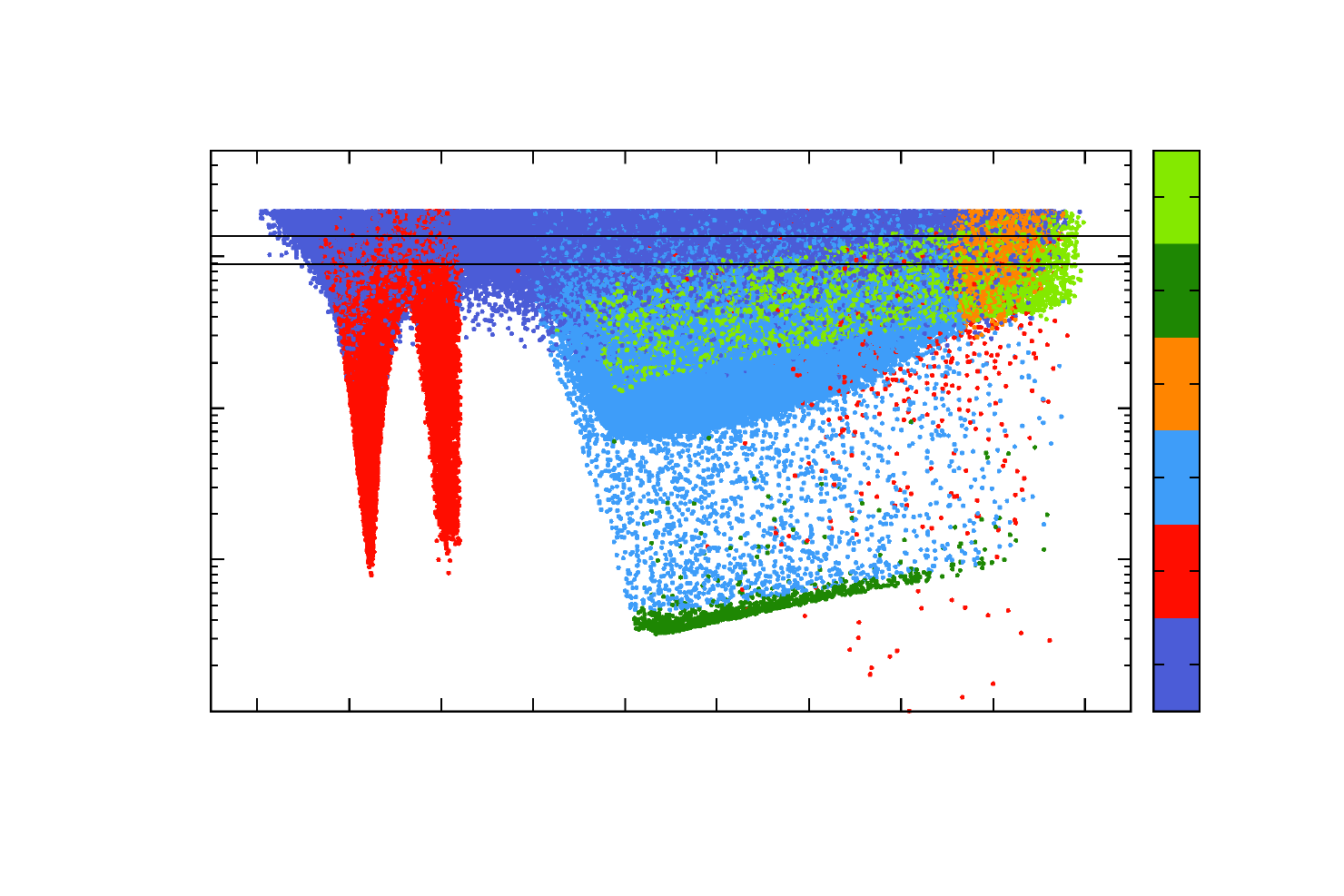}}%
    \gplfronttext
  \end{picture}%
\endgroup

%% file: pic/mechanism_xenonplane.tex
\begingroup
  \makeatletter
  \providecommand\color[2][]{%
    \GenericError{(gnuplot) \space\space\space\@spaces}{%
      Package color not loaded in conjunction with
      terminal option `colourtext'%
    }{See the gnuplot documentation for explanation.%
    }{Either use 'blacktext' in gnuplot or load the package
      color.sty in LaTeX.}%
    \renewcommand\color[2][]{}%
  }%
  \providecommand\includegraphics[2][]{%
    \GenericError{(gnuplot) \space\space\space\@spaces}{%
      Package graphicx or graphics not loaded%
    }{See the gnuplot documentation for explanation.%
    }{The gnuplot epslatex terminal needs graphicx.sty or graphics.sty.}%
    \renewcommand\includegraphics[2][]{}%
  }%
  \providecommand\rotatebox[2]{#2}%
  \@ifundefined{ifGPcolor}{%
    \newif\ifGPcolor
    \GPcolortrue
  }{}%
  \@ifundefined{ifGPblacktext}{%
    \newif\ifGPblacktext
    \GPblacktextfalse
  }{}%
  \let\gplgaddtomacro\g@addto@macro
  \gdef\gplbacktext{}%
  \gdef\gplfronttext{}%
  \makeatother
  \ifGPblacktext
    \def\colorrgb#1{}%
    \def\colorgray#1{}%
  \else
    \ifGPcolor
      \def\colorrgb#1{\color[rgb]{#1}}%
      \def\colorgray#1{\color[gray]{#1}}%
      \expandafter\def\csname LTw\endcsname{\color{white}}%
      \expandafter\def\csname LTb\endcsname{\color{black}}%
      \expandafter\def\csname LTa\endcsname{\color{black}}%
      \expandafter\def\csname LT0\endcsname{\color[rgb]{1,0,0}}%
      \expandafter\def\csname LT1\endcsname{\color[rgb]{0,1,0}}%
      \expandafter\def\csname LT2\endcsname{\color[rgb]{0,0,1}}%
      \expandafter\def\csname LT3\endcsname{\color[rgb]{1,0,1}}%
      \expandafter\def\csname LT4\endcsname{\color[rgb]{0,1,1}}%
      \expandafter\def\csname LT5\endcsname{\color[rgb]{1,1,0}}%
      \expandafter\def\csname LT6\endcsname{\color[rgb]{0,0,0}}%
      \expandafter\def\csname LT7\endcsname{\color[rgb]{1,0.3,0}}%
      \expandafter\def\csname LT8\endcsname{\color[rgb]{0.5,0.5,0.5}}%
    \else
      \def\colorrgb#1{\color{black}}%
      \def\colorgray#1{\color[gray]{#1}}%
      \expandafter\def\csname LTw\endcsname{\color{white}}%
      \expandafter\def\csname LTb\endcsname{\color{black}}%
      \expandafter\def\csname LTa\endcsname{\color{black}}%
      \expandafter\def\csname LT0\endcsname{\color{black}}%
      \expandafter\def\csname LT1\endcsname{\color{black}}%
      \expandafter\def\csname LT2\endcsname{\color{black}}%
      \expandafter\def\csname LT3\endcsname{\color{black}}%
      \expandafter\def\csname LT4\endcsname{\color{black}}%
      \expandafter\def\csname LT5\endcsname{\color{black}}%
      \expandafter\def\csname LT6\endcsname{\color{black}}%
      \expandafter\def\csname LT7\endcsname{\color{black}}%
      \expandafter\def\csname LT8\endcsname{\color{black}}%
    \fi
  \fi
  \setlength{\unitlength}{0.0500bp}%
  \begin{picture}(17006.00,6802.00)%
    \gplgaddtomacro\gplbacktext{%
    }%
    \gplgaddtomacro\gplfronttext{%
      \csname LTb\endcsname%
      \put(2042,395){\makebox(0,0){\strut{} \Large{20}}}%
      \put(2722,395){\makebox(0,0){\strut{} \Large{40}}}%
      \put(3402,395){\makebox(0,0){\strut{} \Large{60}}}%
      \put(4082,395){\makebox(0,0){\strut{} \Large{80}}}%
      \put(4762,395){\makebox(0,0){\strut{} \Large{100}}}%
      \put(5442,395){\makebox(0,0){\strut{} \Large{120}}}%
      \put(6122,395){\makebox(0,0){\strut{} \Large{140}}}%
      \put(6802,395){\makebox(0,0){\strut{} \Large{160}}}%
      \put(7482,395){\makebox(0,0){\strut{} \Large{180}}}%
      \put(8162,395){\makebox(0,0){\strut{} \Large{200}}}%
      \put(5102,65){\makebox(0,0){\strut{}\Large{$m_{\tilde{\chi}}$ [GeV]}}}%
      \put(1530,681){\makebox(0,0)[r]{\strut{} \Large{-18}}}%
      \put(1530,1588){\makebox(0,0)[r]{\strut{} \Large{-16}}}%
      \put(1530,2495){\makebox(0,0)[r]{\strut{} \Large{-14}}}%
      \put(1530,3401){\makebox(0,0)[r]{\strut{} \Large{-12}}}%
      \put(1530,4307){\makebox(0,0)[r]{\strut{} \Large{-10}}}%
      \put(1530,5214){\makebox(0,0)[r]{\strut{} \Large{-8}}}%
      \put(1530,6121){\makebox(0,0)[r]{\strut{} \Large{-6}}}%
      \put(672,3401){\rotatebox{-270}{\makebox(0,0){\strut{}\Large{$\log (\sigma^{\rm SI})$ [pb]}}}}%
    }%

    \gplgaddtomacro\gplbacktext{%
    }%
    \gplgaddtomacro\gplfronttext{%
      \csname LTb\endcsname%
      \put(8844,395){\makebox(0,0){\strut{} \Large{20}}}%
      \put(9524,395){\makebox(0,0){\strut{} \Large{40}}}%
      \put(10204,395){\makebox(0,0){\strut{} \Large{60}}}%
      \put(10884,395){\makebox(0,0){\strut{} \Large{80}}}%
      \put(11564,395){\makebox(0,0){\strut{} \Large{100}}}%
      \put(12244,395){\makebox(0,0){\strut{} \Large{120}}}%
      \put(12924,395){\makebox(0,0){\strut{} \Large{140}}}%
      \put(13604,395){\makebox(0,0){\strut{} \Large{160}}}%
      \put(14284,395){\makebox(0,0){\strut{} \Large{180}}}%
      \put(14964,395){\makebox(0,0){\strut{} \Large{200}}}%
      \put(11904,65){\makebox(0,0){\strut{}\Large{$m_{\widetilde{\chi}}$ [GeV]}}}%
    \put(8084,5421){\makebox(0,0)[r]{\strut{} \rm{ \small \textbf{XENON100 (2011)}}}}%
  \put(8084,4921){\makebox(0,0)[r]{\strut{} \rm{ \small \textbf{XENON100 (2012)}}}}%
 \put(8084,4391){\makebox(0,0)[r]{\strut{} \small{ \rm \textbf{XENON1t}}}}%
    \put(15084,5471){\makebox(0,0)[r]{\strut{} \small{ \rm \textbf{XENON100 (2011)}}}}%
    \put(15084,4921){\makebox(0,0)[r]{\strut{} \small{ \rm \textbf{XENON100 (2012)}}}}%
 \put(14784,4391){\makebox(0,0)[r]{\strut{} \small{ \rm \textbf{XENON1t}}}}%

    }%

    \gplgaddtomacro\gplbacktext{%
    }%
    \gplgaddtomacro\gplfronttext{%
      \csname LTb\endcsname%

      \put(15946,1133){\makebox(0,0)[l]{\strut{} \Large{$l \bar{l}$}}}%
      \put(15946,2040){\makebox(0,0)[l]{\strut{} \Large{$q \bar{q}$}}}%
      \put(15946,2947){\makebox(0,0)[l]{\strut{} \Large{$W^+ W^-$}}}%
      \put(15946,3853){\makebox(0,0)[l]{\strut{} \Large{$t \bar{t}$}}}%
      \put(15946,4910){\makebox(0,0)[l]{\strut{} \Large{$\widetilde{\chi}^0_1 \widetilde{\chi}^+_1$}}}%
      \put(15946,4610){\makebox(0,0)[l]{\strut{} \Large{coannihilation}}}%
      \put(15946,5810){\makebox(0,0)[l]{\strut{} \Large{$\widetilde{\chi}^0_1 \widetilde{\l}$}}}
      \put(15946,5510){\makebox(0,0)[l]{\strut{} \Large{coannihilation}}}%
    }%
    \gplbacktext
    \put(0,0){\includegraphics{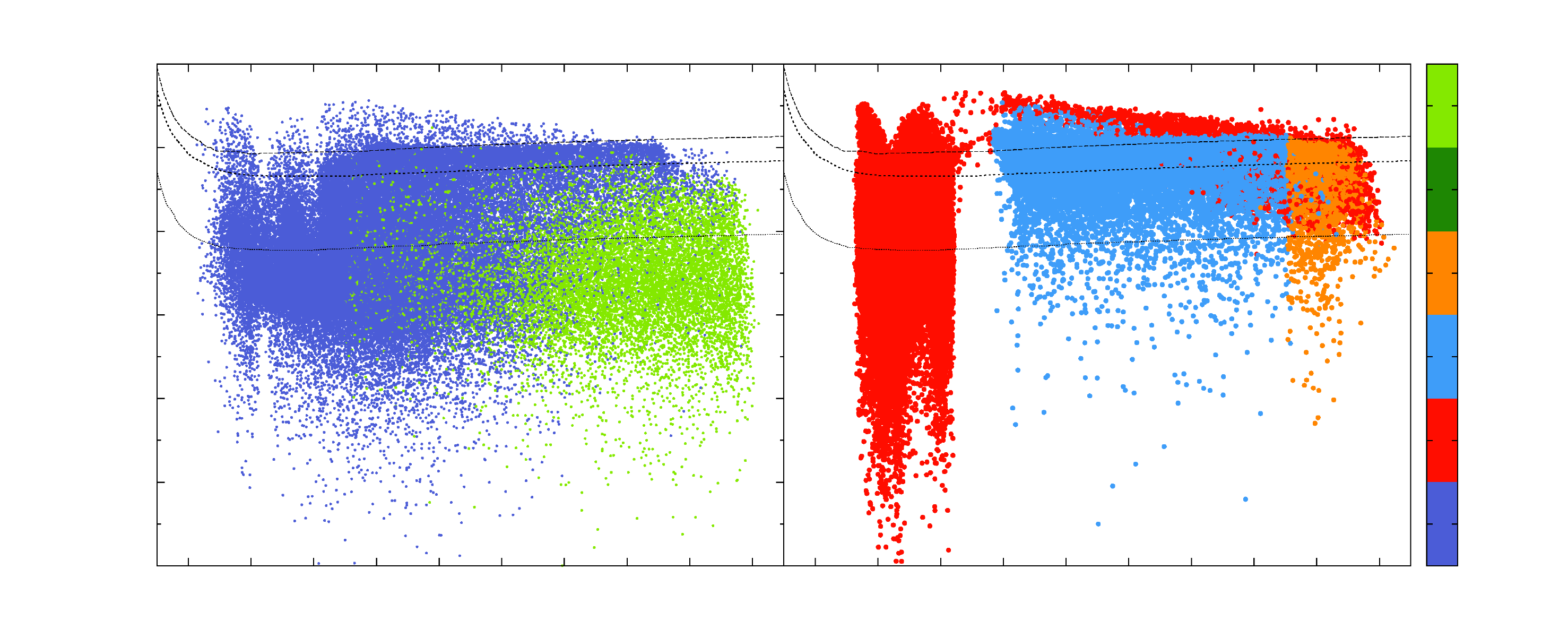}}%
    \gplfronttext
  \end{picture}%
\endgroup

%% file: pic/mechanism_xenonplane_pos.tex
\begingroup
  \makeatletter
  \providecommand\color[2][]{%
    \GenericError{(gnuplot) \space\space\space\@spaces}{%
      Package color not loaded in conjunction with
      terminal option `colourtext'%
    }{See the gnuplot documentation for explanation.%
    }{Either use 'blacktext' in gnuplot or load the package
      color.sty in LaTeX.}%
    \renewcommand\color[2][]{}%
  }%
  \providecommand\includegraphics[2][]{%
    \GenericError{(gnuplot) \space\space\space\@spaces}{%
      Package graphicx or graphics not loaded%
    }{See the gnuplot documentation for explanation.%
    }{The gnuplot epslatex terminal needs graphicx.sty or graphics.sty.}%
    \renewcommand\includegraphics[2][]{}%
  }%
  \providecommand\rotatebox[2]{#2}%
  \@ifundefined{ifGPcolor}{%
    \newif\ifGPcolor
    \GPcolortrue
  }{}%
  \@ifundefined{ifGPblacktext}{%
    \newif\ifGPblacktext
    \GPblacktextfalse
  }{}%
  \let\gplgaddtomacro\g@addto@macro
  \gdef\gplbacktext{}%
  \gdef\gplfronttext{}%
  \makeatother
  \ifGPblacktext
    \def\colorrgb#1{}%
    \def\colorgray#1{}%
  \else
    \ifGPcolor
      \def\colorrgb#1{\color[rgb]{#1}}%
      \def\colorgray#1{\color[gray]{#1}}%
      \expandafter\def\csname LTw\endcsname{\color{white}}%
      \expandafter\def\csname LTb\endcsname{\color{black}}%
      \expandafter\def\csname LTa\endcsname{\color{black}}%
      \expandafter\def\csname LT0\endcsname{\color[rgb]{1,0,0}}%
      \expandafter\def\csname LT1\endcsname{\color[rgb]{0,1,0}}%
      \expandafter\def\csname LT2\endcsname{\color[rgb]{0,0,1}}%
      \expandafter\def\csname LT3\endcsname{\color[rgb]{1,0,1}}%
      \expandafter\def\csname LT4\endcsname{\color[rgb]{0,1,1}}%
      \expandafter\def\csname LT5\endcsname{\color[rgb]{1,1,0}}%
      \expandafter\def\csname LT6\endcsname{\color[rgb]{0,0,0}}%
      \expandafter\def\csname LT7\endcsname{\color[rgb]{1,0.3,0}}%
      \expandafter\def\csname LT8\endcsname{\color[rgb]{0.5,0.5,0.5}}%
    \else
      \def\colorrgb#1{\color{black}}%
      \def\colorgray#1{\color[gray]{#1}}%
      \expandafter\def\csname LTw\endcsname{\color{white}}%
      \expandafter\def\csname LTb\endcsname{\color{black}}%
      \expandafter\def\csname LTa\endcsname{\color{black}}%
      \expandafter\def\csname LT0\endcsname{\color{black}}%
      \expandafter\def\csname LT1\endcsname{\color{black}}%
      \expandafter\def\csname LT2\endcsname{\color{black}}%
      \expandafter\def\csname LT3\endcsname{\color{black}}%
      \expandafter\def\csname LT4\endcsname{\color{black}}%
      \expandafter\def\csname LT5\endcsname{\color{black}}%
      \expandafter\def\csname LT6\endcsname{\color{black}}%
      \expandafter\def\csname LT7\endcsname{\color{black}}%
      \expandafter\def\csname LT8\endcsname{\color{black}}%
    \fi
  \fi
  \setlength{\unitlength}{0.0500bp}%
  \begin{picture}(17006.00,6802.00)%
    \gplgaddtomacro\gplbacktext{%
    }%
    \gplgaddtomacro\gplfronttext{%
      \csname LTb\endcsname%
      \put(2042,395){\makebox(0,0){\strut{} \Large{20}}}%
      \put(2722,395){\makebox(0,0){\strut{} \Large{40}}}%
      \put(3402,395){\makebox(0,0){\strut{} \Large{60}}}%
      \put(4082,395){\makebox(0,0){\strut{} \Large{80}}}%
      \put(4762,395){\makebox(0,0){\strut{} \Large{100}}}%
      \put(5442,395){\makebox(0,0){\strut{} \Large{120}}}%
      \put(6122,395){\makebox(0,0){\strut{} \Large{140}}}%
      \put(6802,395){\makebox(0,0){\strut{} \Large{160}}}%
      \put(7482,395){\makebox(0,0){\strut{} \Large{180}}}%
      \put(8162,395){\makebox(0,0){\strut{} \Large{200}}}%
      \put(5102,65){\makebox(0,0){\strut{}\Large{$m_{\tilde{\chi}}$ [GeV]}}}%
      \put(1530,681){\makebox(0,0)[r]{\strut{} \Large{-12}}}%
      \put(1530,1588){\makebox(0,0)[r]{\strut{}\Large{-11}}}%
      \put(1530,2495){\makebox(0,0)[r]{\strut{} \Large{-10}}}%
      \put(1530,3401){\makebox(0,0)[r]{\strut{} \Large{-9}}}%
      \put(1530,4307){\makebox(0,0)[r]{\strut{} \Large{-8}}}%
      \put(1530,5214){\makebox(0,0)[r]{\strut{} \Large{-7}}}%
      \put(1530,6121){\makebox(0,0)[r]{\strut{} \Large{-6}}}%
      \put(672,3401){\rotatebox{-270}{\makebox(0,0){\strut{} \Large{$\log (\sigma^{\rm SI})$ [pb]}}}}%
    }%

    \gplgaddtomacro\gplfronttext{%
      \csname LTb\endcsname%
      \put(8844,395){\makebox(0,0){\strut{} \Large{20}}}%
      \put(9524,395){\makebox(0,0){\strut{} \Large{40}}}%
      \put(10204,395){\makebox(0,0){\strut{} \Large{60}}}%
      \put(10884,395){\makebox(0,0){\strut{} \Large{80}}}%
      \put(11564,395){\makebox(0,0){\strut{} \Large{100}}}%
      \put(12244,395){\makebox(0,0){\strut{} \Large{120}}}%
      \put(12924,395){\makebox(0,0){\strut{} \Large{140}}}%
      \put(13604,395){\makebox(0,0){\strut{} \Large{160}}}%
      \put(14284,395){\makebox(0,0){\strut{} \Large{180}}}%
      \put(14964,395){\makebox(0,0){\strut{} \Large{200}}}%
      \put(11904,65){\makebox(0,0){\strut{} \Large{$m_{\tilde{\chi}}$ [GeV]}}}%

   \put(15946,1133){\makebox(0,0)[l]{\strut{} \Large{$l \bar{l}$}}}%
      \put(15946,2040){\makebox(0,0)[l]{\strut{} \Large{$q \bar{q}$}}}%
      \put(15946,2947){\makebox(0,0)[l]{\strut{} \Large{$W^+ W^-$}}}%
      \put(15946,3853){\makebox(0,0)[l]{\strut{} \Large{$t \bar{t}$}}}%
      \put(15946,4910){\makebox(0,0)[l]{\strut{} \Large{$\tilde{\chi}^0_1 \tilde{\chi}^+_1$}}}%
      \put(15946,4610){\makebox(0,0)[l]{\strut{} \Large{coannihilation}}}%
      \put(15946,5810){\makebox(0,0)[l]{\strut{} \Large{$\tilde{\chi}^0_1 \tilde{\l}$}}}
      \put(15946,5510){\makebox(0,0)[l]{\strut{} \Large{coannihilation}}}%
    \put(8084,4671){\makebox(0,0)[r]{\strut{} \rm{ \small \textbf{XENON100 (2011)}}}}%
  \put(8084,4101){\makebox(0,0)[r]{\strut{} \rm{ \small \textbf{XENON100 (2012)}}}}%
 \put(8084,2591){\makebox(0,0)[r]{\strut{} \small{ \rm \textbf{XENON1t}}}}%
    \put(15084,4671){\makebox(0,0)[r]{\strut{} \small{ \rm \textbf{XENON100 (2011)}}}}%
    \put(15084,4101){\makebox(0,0)[r]{\strut{} \small{ \rm \textbf{XENON100 (2012)}}}}%
 \put(14784,2591){\makebox(0,0)[r]{\strut{} \small{ \rm \textbf{XENON1t}}}}%
}
    \gplbacktext
    \put(0,0){\includegraphics{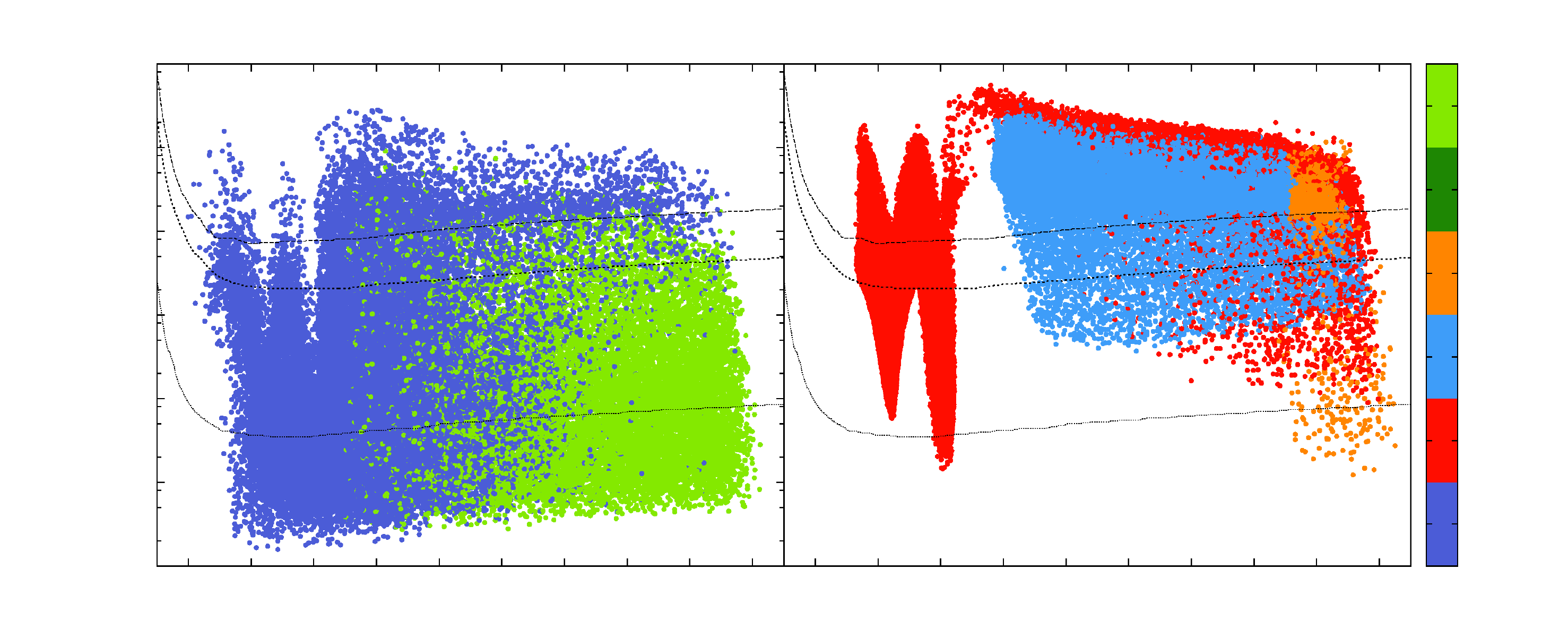}}%
    \gplfronttext
  \end{picture}%
\endgroup

%% file: pic/pos_nog2_direct.tex
\begingroup
  \makeatletter
  \providecommand\color[2][]{%
    \GenericError{(gnuplot) \space\space\space\@spaces}{%
      Package color not loaded in conjunction with
      terminal option `colourtext'%
    }{See the gnuplot documentation for explanation.%
    }{Either use 'blacktext' in gnuplot or load the package
      color.sty in LaTeX.}%
    \renewcommand\color[2][]{}%
  }%
  \providecommand\includegraphics[2][]{%
    \GenericError{(gnuplot) \space\space\space\@spaces}{%
      Package graphicx or graphics not loaded%
    }{See the gnuplot documentation for explanation.%
    }{The gnuplot epslatex terminal needs graphicx.sty or graphics.sty.}%
    \renewcommand\includegraphics[2][]{}%
  }%
  \providecommand\rotatebox[2]{#2}%
  \@ifundefined{ifGPcolor}{%
    \newif\ifGPcolor
    \GPcolortrue
  }{}%
  \@ifundefined{ifGPblacktext}{%
    \newif\ifGPblacktext
    \GPblacktextfalse
  }{}%
  \let\gplgaddtomacro\g@addto@macro
  \gdef\gplbacktext{}%
  \gdef\gplfronttext{}%
  \makeatother
  \ifGPblacktext
    \def\colorrgb#1{}%
    \def\colorgray#1{}%
  \else
    \ifGPcolor
      \def\colorrgb#1{\color[rgb]{#1}}%
      \def\colorgray#1{\color[gray]{#1}}%
      \expandafter\def\csname LTw\endcsname{\color{white}}%
      \expandafter\def\csname LTb\endcsname{\color{black}}%
      \expandafter\def\csname LTa\endcsname{\color{black}}%
      \expandafter\def\csname LT0\endcsname{\color[rgb]{1,0,0}}%
      \expandafter\def\csname LT1\endcsname{\color[rgb]{0,1,0}}%
      \expandafter\def\csname LT2\endcsname{\color[rgb]{0,0,1}}%
      \expandafter\def\csname LT3\endcsname{\color[rgb]{1,0,1}}%
      \expandafter\def\csname LT4\endcsname{\color[rgb]{0,1,1}}%
      \expandafter\def\csname LT5\endcsname{\color[rgb]{1,1,0}}%
      \expandafter\def\csname LT6\endcsname{\color[rgb]{0,0,0}}%
      \expandafter\def\csname LT7\endcsname{\color[rgb]{1,0.3,0}}%
      \expandafter\def\csname LT8\endcsname{\color[rgb]{0.5,0.5,0.5}}%
    \else
      \def\colorrgb#1{\color{black}}%
      \def\colorgray#1{\color[gray]{#1}}%
      \expandafter\def\csname LTw\endcsname{\color{white}}%
      \expandafter\def\csname LTb\endcsname{\color{black}}%
      \expandafter\def\csname LTa\endcsname{\color{black}}%
      \expandafter\def\csname LT0\endcsname{\color{black}}%
      \expandafter\def\csname LT1\endcsname{\color{black}}%
      \expandafter\def\csname LT2\endcsname{\color{black}}%
      \expandafter\def\csname LT3\endcsname{\color{black}}%
      \expandafter\def\csname LT4\endcsname{\color{black}}%
      \expandafter\def\csname LT5\endcsname{\color{black}}%
      \expandafter\def\csname LT6\endcsname{\color{black}}%
      \expandafter\def\csname LT7\endcsname{\color{black}}%
      \expandafter\def\csname LT8\endcsname{\color{black}}%
    \fi
  \fi
  \setlength{\unitlength}{0.0500bp}%
  \begin{picture}(8502.00,5668.00)%
    \gplgaddtomacro\gplbacktext{%
    }%
    \gplgaddtomacro\gplfronttext{%
      \csname LTb\endcsname%
      \put(1628,881){\makebox(0,0){\strut{} 20}}%
      \put(2211,881){\makebox(0,0){\strut{} 40}}%
      \put(2794,881){\makebox(0,0){\strut{} 60}}%
      \put(3377,881){\makebox(0,0){\strut{} 80}}%
      \put(3960,881){\makebox(0,0){\strut{} 100}}%
      \put(4542,881){\makebox(0,0){\strut{} 120}}%
      \put(5125,881){\makebox(0,0){\strut{} 140}}%
      \put(5708,881){\makebox(0,0){\strut{} 160}}%
      \put(6291,881){\makebox(0,0){\strut{} 180}}%
      \put(6874,881){\makebox(0,0){\strut{} 200}}%
      \put(4251,551){\makebox(0,0){\strut{}$m_{\widetilde{\chi}}$ [GeV]}}%
      \put(1165,1167){\makebox(0,0)[r]{\strut{} -12}}%
      \put(1165,1760){\makebox(0,0)[r]{\strut{} -11}}%
      \put(1165,2352){\makebox(0,0)[r]{\strut{} -10}}%
      \put(1165,2944){\makebox(0,0)[r]{\strut{} -9}}%
      \put(1165,3536){\makebox(0,0)[r]{\strut{} -8}}%
      \put(1165,4128){\makebox(0,0)[r]{\strut{} -7}}%
      \put(1165,4721){\makebox(0,0)[r]{\strut{} -6}}%
      \put(307,2944){\rotatebox{-270}{\makebox(0,0){\strut{}$\log (\sigma^{\rm SI})$ [pb]}}}%
      \put(7734,1166){\makebox(0,0)[l]{\strut{} 10}}%
      \put(7734,2943){\makebox(0,0)[l]{\strut{} 100}}%
      \put(7734,4721){\makebox(0,0)[l]{\strut{} 1000}}%
      \put(8460,2943){\rotatebox{-270}{\makebox(0,0){\strut{}$\Delta_{\rm tot}$}}}%
\put(7065,3781){\makebox(0,0)[r]{\strut{} \scriptsize{ \rm \textbf{XENON100 (2011)}}}}%
\put(7065,3200){\makebox(0,0)[r]{\strut{} \scriptsize{ \rm \textbf{XENON100 (2012)}}}}%
 \put(6865,2391){\makebox(0,0)[r]{\strut{} \scriptsize{ \rm \textbf{XENON 1t}}}}%
    }%
    \gplbacktext
    \put(0,0){\includegraphics{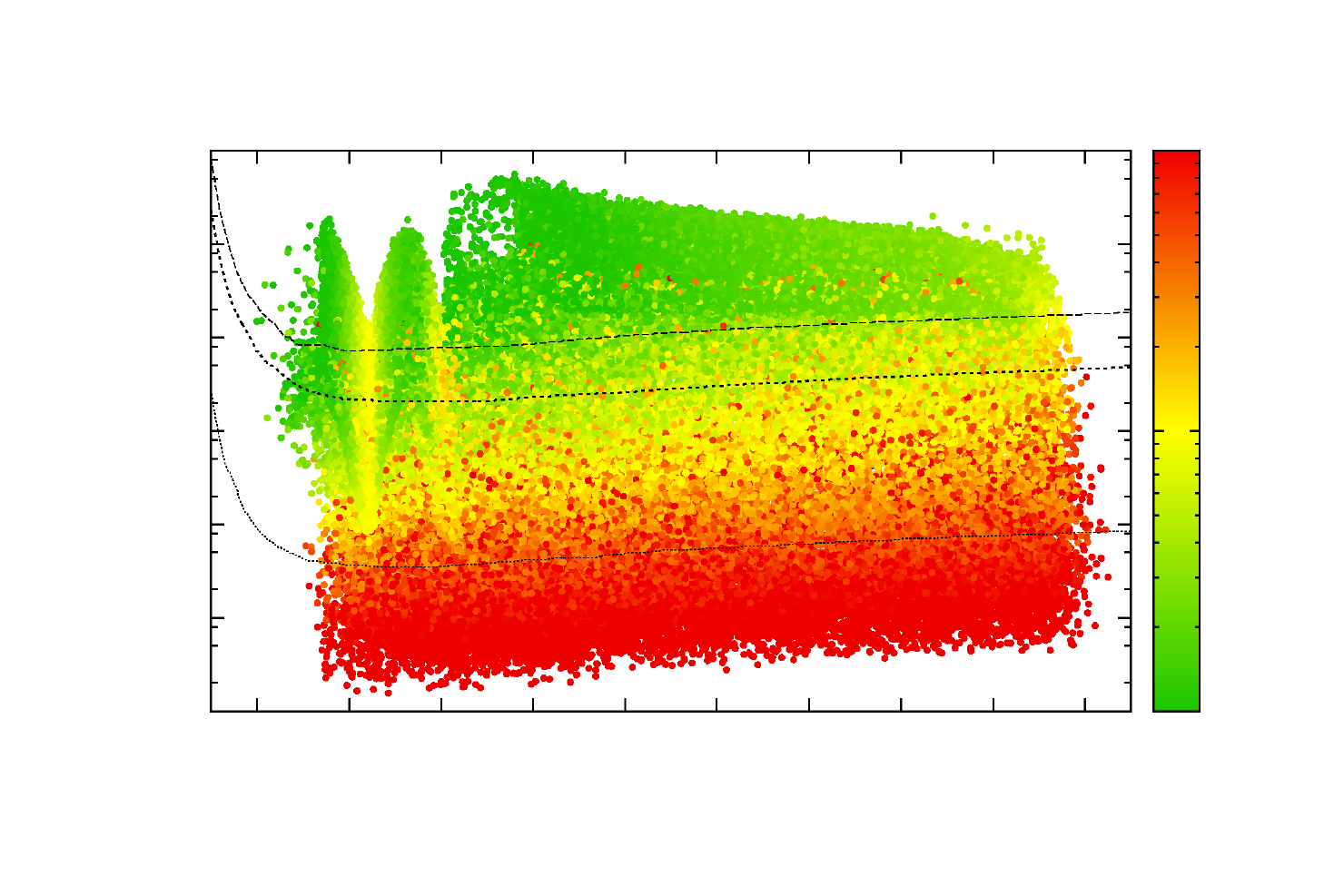}}%
    \gplfronttext
  \end{picture}%
\endgroup

%% file: pic/mu_FT.tex
\begingroup
  \makeatletter
  \providecommand\color[2][]{%
    \GenericError{(gnuplot) \space\space\space\@spaces}{%
      Package color not loaded in conjunction with
      terminal option `colourtext'%
    }{See the gnuplot documentation for explanation.%
    }{Either use 'blacktext' in gnuplot or load the package
      color.sty in LaTeX.}%
    \renewcommand\color[2][]{}%
  }%
  \providecommand\includegraphics[2][]{%
    \GenericError{(gnuplot) \space\space\space\@spaces}{%
      Package graphicx or graphics not loaded%
    }{See the gnuplot documentation for explanation.%
    }{The gnuplot epslatex terminal needs graphicx.sty or graphics.sty.}%
    \renewcommand\includegraphics[2][]{}%
  }%
  \providecommand\rotatebox[2]{#2}%
  \@ifundefined{ifGPcolor}{%
    \newif\ifGPcolor
    \GPcolortrue
  }{}%
  \@ifundefined{ifGPblacktext}{%
    \newif\ifGPblacktext
    \GPblacktextfalse
  }{}%
  \let\gplgaddtomacro\g@addto@macro
  \gdef\gplbacktext{}%
  \gdef\gplfronttext{}%
  \makeatother
  \ifGPblacktext
    \def\colorrgb#1{}%
    \def\colorgray#1{}%
  \else
    \ifGPcolor
      \def\colorrgb#1{\color[rgb]{#1}}%
      \def\colorgray#1{\color[gray]{#1}}%
      \expandafter\def\csname LTw\endcsname{\color{white}}%
      \expandafter\def\csname LTb\endcsname{\color{black}}%
      \expandafter\def\csname LTa\endcsname{\color{black}}%
      \expandafter\def\csname LT0\endcsname{\color[rgb]{1,0,0}}%
      \expandafter\def\csname LT1\endcsname{\color[rgb]{0,1,0}}%
      \expandafter\def\csname LT2\endcsname{\color[rgb]{0,0,1}}%
      \expandafter\def\csname LT3\endcsname{\color[rgb]{1,0,1}}%
      \expandafter\def\csname LT4\endcsname{\color[rgb]{0,1,1}}%
      \expandafter\def\csname LT5\endcsname{\color[rgb]{1,1,0}}%
      \expandafter\def\csname LT6\endcsname{\color[rgb]{0,0,0}}%
      \expandafter\def\csname LT7\endcsname{\color[rgb]{1,0.3,0}}%
      \expandafter\def\csname LT8\endcsname{\color[rgb]{0.5,0.5,0.5}}%
    \else
      \def\colorrgb#1{\color{black}}%
      \def\colorgray#1{\color[gray]{#1}}%
      \expandafter\def\csname LTw\endcsname{\color{white}}%
      \expandafter\def\csname LTb\endcsname{\color{black}}%
      \expandafter\def\csname LTa\endcsname{\color{black}}%
      \expandafter\def\csname LT0\endcsname{\color{black}}%
      \expandafter\def\csname LT1\endcsname{\color{black}}%
      \expandafter\def\csname LT2\endcsname{\color{black}}%
      \expandafter\def\csname LT3\endcsname{\color{black}}%
      \expandafter\def\csname LT4\endcsname{\color{black}}%
      \expandafter\def\csname LT5\endcsname{\color{black}}%
      \expandafter\def\csname LT6\endcsname{\color{black}}%
      \expandafter\def\csname LT7\endcsname{\color{black}}%
      \expandafter\def\csname LT8\endcsname{\color{black}}%
    \fi
  \fi
  \setlength{\unitlength}{0.0500bp}%
  \begin{picture}(5102.00,3968.00)%
    \gplgaddtomacro\gplbacktext{%
    }%
    \gplgaddtomacro\gplfronttext{%
      \csname LTb\endcsname%
      \put(766,508){\makebox(0,0){\strut{} 0}}%
      \put(1531,508){\makebox(0,0){\strut{} 400}}%
      \put(2296,508){\makebox(0,0){\strut{} 800}}%
      \put(3060,508){\makebox(0,0){\strut{} 1200}}%
      \put(3825,508){\makebox(0,0){\strut{} 1600}}%
      \put(4590,508){\makebox(0,0){\strut{} 2000}}%
      \put(2678,178){\makebox(0,0){\strut{}$\mu$ [GeV]}}%
      \put(594,794){\makebox(0,0)[r]{\strut{} 1}}%
      \put(594,1488){\makebox(0,0)[r]{\strut{} 10}}%
      \put(594,2182){\makebox(0,0)[r]{\strut{} 100}}%
      \put(594,2876){\makebox(0,0)[r]{\strut{} 1000}}%
      \put(594,3570){\makebox(0,0)[r]{\strut{} 10000}}%
      \put(-264,2182){\rotatebox{-270}{\makebox(0,0){\strut{}$\Delta_{\rm tot}$}}}%
    }%
    \gplbacktext
    \put(0,0){\includegraphics{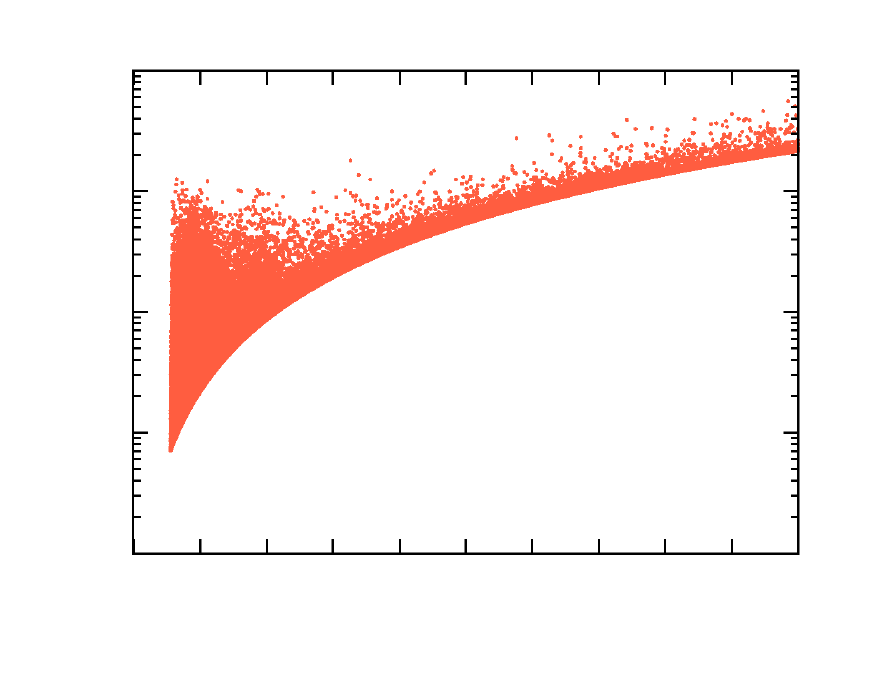}}%
    \gplfronttext
  \end{picture}%
\endgroup

%% file: pic/ma_sigma.tex
\begingroup
  \makeatletter
  \providecommand\color[2][]{%
    \GenericError{(gnuplot) \space\space\space\@spaces}{%
      Package color not loaded in conjunction with
      terminal option `colourtext'%
    }{See the gnuplot documentation for explanation.%
    }{Either use 'blacktext' in gnuplot or load the package
      color.sty in LaTeX.}%
    \renewcommand\color[2][]{}%
  }%
  \providecommand\includegraphics[2][]{%
    \GenericError{(gnuplot) \space\space\space\@spaces}{%
      Package graphicx or graphics not loaded%
    }{See the gnuplot documentation for explanation.%
    }{The gnuplot epslatex terminal needs graphicx.sty or graphics.sty.}%
    \renewcommand\includegraphics[2][]{}%
  }%
  \providecommand\rotatebox[2]{#2}%
  \@ifundefined{ifGPcolor}{%
    \newif\ifGPcolor
    \GPcolortrue
  }{}%
  \@ifundefined{ifGPblacktext}{%
    \newif\ifGPblacktext
    \GPblacktextfalse
  }{}%
  \let\gplgaddtomacro\g@addto@macro
  \gdef\gplbacktext{}%
  \gdef\gplfronttext{}%
  \makeatother
  \ifGPblacktext
    \def\colorrgb#1{}%
    \def\colorgray#1{}%
  \else
    \ifGPcolor
      \def\colorrgb#1{\color[rgb]{#1}}%
      \def\colorgray#1{\color[gray]{#1}}%
      \expandafter\def\csname LTw\endcsname{\color{white}}%
      \expandafter\def\csname LTb\endcsname{\color{black}}%
      \expandafter\def\csname LTa\endcsname{\color{black}}%
      \expandafter\def\csname LT0\endcsname{\color[rgb]{1,0,0}}%
      \expandafter\def\csname LT1\endcsname{\color[rgb]{0,1,0}}%
      \expandafter\def\csname LT2\endcsname{\color[rgb]{0,0,1}}%
      \expandafter\def\csname LT3\endcsname{\color[rgb]{1,0,1}}%
      \expandafter\def\csname LT4\endcsname{\color[rgb]{0,1,1}}%
      \expandafter\def\csname LT5\endcsname{\color[rgb]{1,1,0}}%
      \expandafter\def\csname LT6\endcsname{\color[rgb]{0,0,0}}%
      \expandafter\def\csname LT7\endcsname{\color[rgb]{1,0.3,0}}%
      \expandafter\def\csname LT8\endcsname{\color[rgb]{0.5,0.5,0.5}}%
    \else
      \def\colorrgb#1{\color{black}}%
      \def\colorgray#1{\color[gray]{#1}}%
      \expandafter\def\csname LTw\endcsname{\color{white}}%
      \expandafter\def\csname LTb\endcsname{\color{black}}%
      \expandafter\def\csname LTa\endcsname{\color{black}}%
      \expandafter\def\csname LT0\endcsname{\color{black}}%
      \expandafter\def\csname LT1\endcsname{\color{black}}%
      \expandafter\def\csname LT2\endcsname{\color{black}}%
      \expandafter\def\csname LT3\endcsname{\color{black}}%
      \expandafter\def\csname LT4\endcsname{\color{black}}%
      \expandafter\def\csname LT5\endcsname{\color{black}}%
      \expandafter\def\csname LT6\endcsname{\color{black}}%
      \expandafter\def\csname LT7\endcsname{\color{black}}%
      \expandafter\def\csname LT8\endcsname{\color{black}}%
    \fi
  \fi
  \setlength{\unitlength}{0.0500bp}%
  \begin{picture}(5102.00,3968.00)%
    \gplgaddtomacro\gplbacktext{%
    }%
    \gplgaddtomacro\gplfronttext{%
      \csname LTb\endcsname%
      \put(766,508){\makebox(0,0){\strut{} 0}}%
      \put(1722,508){\makebox(0,0){\strut{} 1}}%
      \put(2678,508){\makebox(0,0){\strut{} 2}}%
      \put(3634,508){\makebox(0,0){\strut{} 3}}%
      \put(4590,508){\makebox(0,0){\strut{} 4}}%
      \put(2678,178){\makebox(0,0){\strut{}$m_A$ [TeV]}}%
      \put(594,794){\makebox(0,0)[r]{\strut{} -12}}%
      \put(594,1257){\makebox(0,0)[r]{\strut{}-11}}%
      \put(594,1720){\makebox(0,0)[r]{\strut{} -10}}%
      \put(594,2182){\makebox(0,0)[r]{\strut{} -9}}%
      \put(594,2644){\makebox(0,0)[r]{\strut{} -8}}%
      \put(594,3107){\makebox(0,0)[r]{\strut{} -7}}%
      \put(594,3570){\makebox(0,0)[r]{\strut{} -6}}%
      \put(4,2182){\rotatebox{-270}{\makebox(0,0){\strut{}$\log (\sigma^{\rm SI})$ [pb]}}}%
    }%
    \gplbacktext
    \put(0,0){\includegraphics{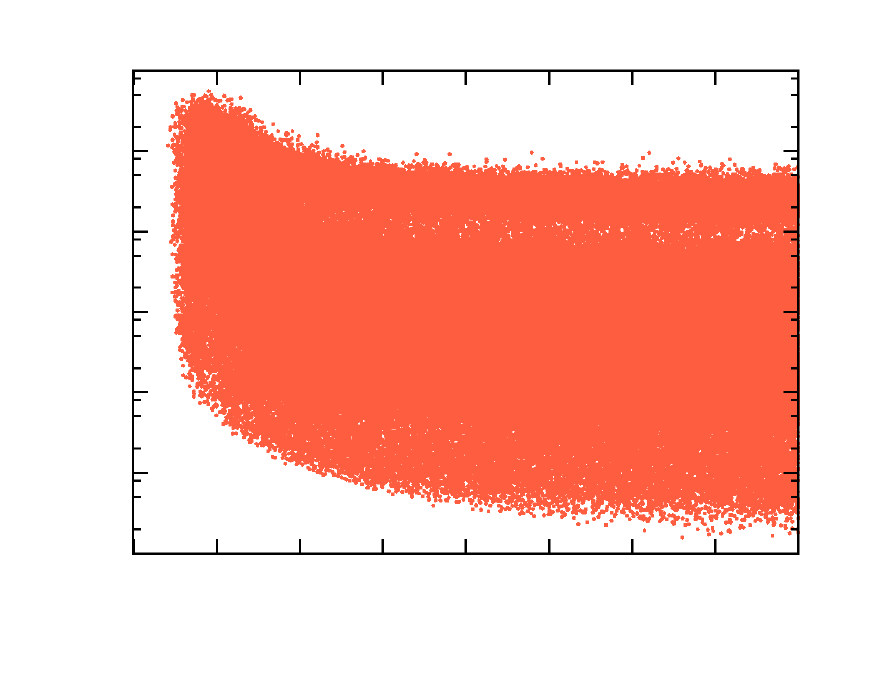}}%
    \gplfronttext
  \end{picture}%
\endgroup

%% file: pic/neg_nog2_direct.tex
\begingroup
  \makeatletter
  \providecommand\color[2][]{%
    \GenericError{(gnuplot) \space\space\space\@spaces}{%
      Package color not loaded in conjunction with
      terminal option `colourtext'%
    }{See the gnuplot documentation for explanation.%
    }{Either use 'blacktext' in gnuplot or load the package
      color.sty in LaTeX.}%
    \renewcommand\color[2][]{}%
  }%
  \providecommand\includegraphics[2][]{%
    \GenericError{(gnuplot) \space\space\space\@spaces}{%
      Package graphicx or graphics not loaded%
    }{See the gnuplot documentation for explanation.%
    }{The gnuplot epslatex terminal needs graphicx.sty or graphics.sty.}%
    \renewcommand\includegraphics[2][]{}%
  }%
  \providecommand\rotatebox[2]{#2}%
  \@ifundefined{ifGPcolor}{%
    \newif\ifGPcolor
    \GPcolortrue
  }{}%
  \@ifundefined{ifGPblacktext}{%
    \newif\ifGPblacktext
    \GPblacktextfalse
  }{}%
  \let\gplgaddtomacro\g@addto@macro
  \gdef\gplbacktext{}%
  \gdef\gplfronttext{}%
  \makeatother
  \ifGPblacktext
    \def\colorrgb#1{}%
    \def\colorgray#1{}%
  \else
    \ifGPcolor
      \def\colorrgb#1{\color[rgb]{#1}}%
      \def\colorgray#1{\color[gray]{#1}}%
      \expandafter\def\csname LTw\endcsname{\color{white}}%
      \expandafter\def\csname LTb\endcsname{\color{black}}%
      \expandafter\def\csname LTa\endcsname{\color{black}}%
      \expandafter\def\csname LT0\endcsname{\color[rgb]{1,0,0}}%
      \expandafter\def\csname LT1\endcsname{\color[rgb]{0,1,0}}%
      \expandafter\def\csname LT2\endcsname{\color[rgb]{0,0,1}}%
      \expandafter\def\csname LT3\endcsname{\color[rgb]{1,0,1}}%
      \expandafter\def\csname LT4\endcsname{\color[rgb]{0,1,1}}%
      \expandafter\def\csname LT5\endcsname{\color[rgb]{1,1,0}}%
      \expandafter\def\csname LT6\endcsname{\color[rgb]{0,0,0}}%
      \expandafter\def\csname LT7\endcsname{\color[rgb]{1,0.3,0}}%
      \expandafter\def\csname LT8\endcsname{\color[rgb]{0.5,0.5,0.5}}%
    \else
      \def\colorrgb#1{\color{black}}%
      \def\colorgray#1{\color[gray]{#1}}%
      \expandafter\def\csname LTw\endcsname{\color{white}}%
      \expandafter\def\csname LTb\endcsname{\color{black}}%
      \expandafter\def\csname LTa\endcsname{\color{black}}%
      \expandafter\def\csname LT0\endcsname{\color{black}}%
      \expandafter\def\csname LT1\endcsname{\color{black}}%
      \expandafter\def\csname LT2\endcsname{\color{black}}%
      \expandafter\def\csname LT3\endcsname{\color{black}}%
      \expandafter\def\csname LT4\endcsname{\color{black}}%
      \expandafter\def\csname LT5\endcsname{\color{black}}%
      \expandafter\def\csname LT6\endcsname{\color{black}}%
      \expandafter\def\csname LT7\endcsname{\color{black}}%
      \expandafter\def\csname LT8\endcsname{\color{black}}%
    \fi
  \fi
  \setlength{\unitlength}{0.0500bp}%
  \begin{picture}(8502.00,5668.00)%
    \gplgaddtomacro\gplbacktext{%
    }%
    \gplgaddtomacro\gplfronttext{%
      \csname LTb\endcsname%
      \put(1628,881){\makebox(0,0){\strut{} 20}}%
      \put(2211,881){\makebox(0,0){\strut{} 40}}%
      \put(2794,881){\makebox(0,0){\strut{} 60}}%
      \put(3377,881){\makebox(0,0){\strut{} 80}}%
      \put(3960,881){\makebox(0,0){\strut{} 100}}%
      \put(4542,881){\makebox(0,0){\strut{} 120}}%
      \put(5125,881){\makebox(0,0){\strut{} 140}}%
      \put(5708,881){\makebox(0,0){\strut{} 160}}%
      \put(6291,881){\makebox(0,0){\strut{} 180}}%
      \put(6874,881){\makebox(0,0){\strut{} 200}}%
      \put(4251,551){\makebox(0,0){\strut{}$m_{\widetilde{\chi}}$ [GeV]}}%
      \put(1165,1167){\makebox(0,0)[r]{\strut{} -18}}%
      \put(1165,1760){\makebox(0,0)[r]{\strut{} -16}}%
      \put(1165,2352){\makebox(0,0)[r]{\strut{} -14}}%
      \put(1165,2944){\makebox(0,0)[r]{\strut{} -12}}%
      \put(1165,3536){\makebox(0,0)[r]{\strut{} -10}}%
      \put(1165,4128){\makebox(0,0)[r]{\strut{} -8}}%
      \put(1165,4721){\makebox(0,0)[r]{\strut{} -6}}%
      \put(307,2944){\rotatebox{-270}{\makebox(0,0){\strut{}$\log (\sigma^{\rm SI})$ [pb]}}}%
      \put(7734,1166){\makebox(0,0)[l]{\strut{} 10}}%
      \put(7734,2943){\makebox(0,0)[l]{\strut{} 100}}%
      \put(7734,4721){\makebox(0,0)[l]{\strut{} 1000}}%
      \put(8460,2943){\rotatebox{-270}{\makebox(0,0){\strut{}$\Delta_{\rm tot}$}}}%
 \put(7065,4321){\makebox(0,0)[r]{\strut{} \scriptsize{ \rm \textbf{XENON100 (2011)}}}}%
 \put(7065,3921){\makebox(0,0)[r]{\strut{} \scriptsize{ \rm \textbf{XENON100 (2012)}}}}%
 \put(7005,3591){\makebox(0,0)[r]{\strut{} \scriptsize{ \rm \textbf{XENON 1t}}}}%
    }%
    \gplbacktext
    \put(0,0){\includegraphics{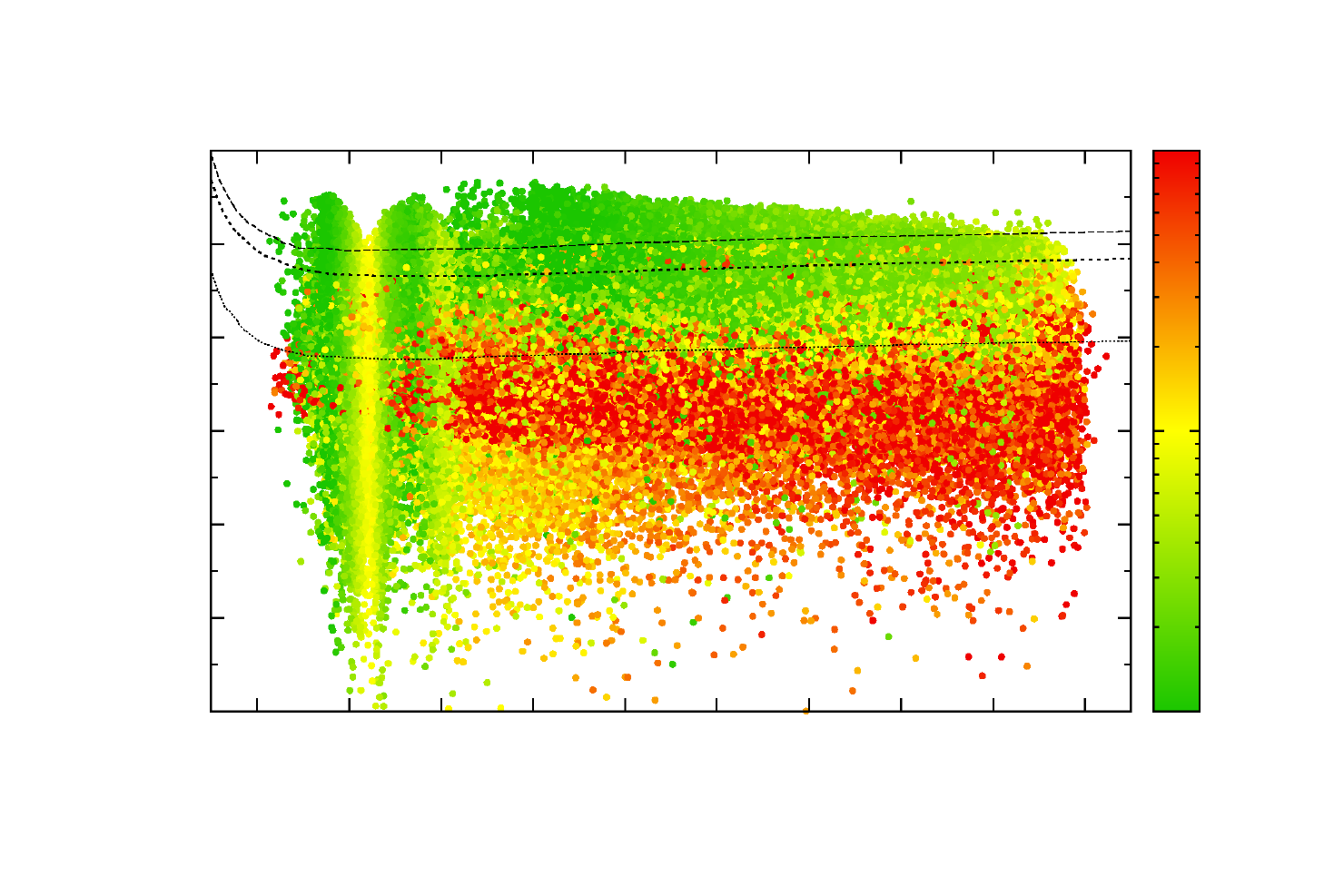}}%
    \gplfronttext
  \end{picture}%
\endgroup

%% file: pic/light_staus.tex
\begingroup
  \makeatletter
  \providecommand\color[2][]{%
    \GenericError{(gnuplot) \space\space\space\@spaces}{%
      Package color not loaded in conjunction with
      terminal option `colourtext'%
    }{See the gnuplot documentation for explanation.%
    }{Either use 'blacktext' in gnuplot or load the package
      color.sty in LaTeX.}%
    \renewcommand\color[2][]{}%
  }%
  \providecommand\includegraphics[2][]{%
    \GenericError{(gnuplot) \space\space\space\@spaces}{%
      Package graphicx or graphics not loaded%
    }{See the gnuplot documentation for explanation.%
    }{The gnuplot epslatex terminal needs graphicx.sty or graphics.sty.}%
    \renewcommand\includegraphics[2][]{}%
  }%
  \providecommand\rotatebox[2]{#2}%
  \@ifundefined{ifGPcolor}{%
    \newif\ifGPcolor
    \GPcolortrue
  }{}%
  \@ifundefined{ifGPblacktext}{%
    \newif\ifGPblacktext
    \GPblacktextfalse
  }{}%
  \let\gplgaddtomacro\g@addto@macro
  \gdef\gplbacktext{}%
  \gdef\gplfronttext{}%
  \makeatother
  \ifGPblacktext
    \def\colorrgb#1{}%
    \def\colorgray#1{}%
  \else
    \ifGPcolor
      \def\colorrgb#1{\color[rgb]{#1}}%
      \def\colorgray#1{\color[gray]{#1}}%
      \expandafter\def\csname LTw\endcsname{\color{white}}%
      \expandafter\def\csname LTb\endcsname{\color{black}}%
      \expandafter\def\csname LTa\endcsname{\color{black}}%
      \expandafter\def\csname LT0\endcsname{\color[rgb]{1,0,0}}%
      \expandafter\def\csname LT1\endcsname{\color[rgb]{0,1,0}}%
      \expandafter\def\csname LT2\endcsname{\color[rgb]{0,0,1}}%
      \expandafter\def\csname LT3\endcsname{\color[rgb]{1,0,1}}%
      \expandafter\def\csname LT4\endcsname{\color[rgb]{0,1,1}}%
      \expandafter\def\csname LT5\endcsname{\color[rgb]{1,1,0}}%
      \expandafter\def\csname LT6\endcsname{\color[rgb]{0,0,0}}%
      \expandafter\def\csname LT7\endcsname{\color[rgb]{1,0.3,0}}%
      \expandafter\def\csname LT8\endcsname{\color[rgb]{0.5,0.5,0.5}}%
    \else
      \def\colorrgb#1{\color{black}}%
      \def\colorgray#1{\color[gray]{#1}}%
      \expandafter\def\csname LTw\endcsname{\color{white}}%
      \expandafter\def\csname LTb\endcsname{\color{black}}%
      \expandafter\def\csname LTa\endcsname{\color{black}}%
      \expandafter\def\csname LT0\endcsname{\color{black}}%
      \expandafter\def\csname LT1\endcsname{\color{black}}%
      \expandafter\def\csname LT2\endcsname{\color{black}}%
      \expandafter\def\csname LT3\endcsname{\color{black}}%
      \expandafter\def\csname LT4\endcsname{\color{black}}%
      \expandafter\def\csname LT5\endcsname{\color{black}}%
      \expandafter\def\csname LT6\endcsname{\color{black}}%
      \expandafter\def\csname LT7\endcsname{\color{black}}%
      \expandafter\def\csname LT8\endcsname{\color{black}}%
    \fi
  \fi
  \setlength{\unitlength}{0.0500bp}%
  \begin{picture}(17006.00,6802.00)%
    \gplgaddtomacro\gplbacktext{%
    }%
    \gplgaddtomacro\gplfronttext{%
      \csname LTb\endcsname%
\put(2118,5850){\makebox(0,0){\strut{} \Large{$\mu > 0$}}}%
      \put(1954,395){\makebox(0,0){\strut{} \Large{20}}}%
      \put(2961,395){\makebox(0,0){\strut{} \Large{40}}}%
      \put(3969,395){\makebox(0,0){\strut{} \Large{60}}}%
      \put(4977,395){\makebox(0,0){\strut{} \Large{80}}}%
      \put(5983,395){\makebox(0,0){\strut{} \Large{100}}}%
      \put(6991,395){\makebox(0,0){\strut{} \Large{120}}}%
      \put(7998,395){\makebox(0,0){\strut{} \Large{140}}}%
      \put(5102,65){\makebox(0,0){\strut{}\Large{$m_{\widetilde{\chi}}$ [GeV]}}}%
      \put(1530,681){\makebox(0,0)[r]{\strut{} \Large{50}}}%
      \put(1530,1285){\makebox(0,0)[r]{\strut{} \Large{100}}}%
      \put(1530,1890){\makebox(0,0)[r]{\strut{} \Large{150}}}%
      \put(1530,2495){\makebox(0,0)[r]{\strut{} \Large{200}}}%
      \put(1530,3099){\makebox(0,0)[r]{\strut{} \Large{250}}}%
      \put(1530,3703){\makebox(0,0)[r]{\strut{} \Large{300}}}%
      \put(1530,4307){\makebox(0,0)[r]{\strut{} \Large{350}}}%
      \put(1530,4912){\makebox(0,0)[r]{\strut{} \Large{400}}}%
      \put(1530,5517){\makebox(0,0)[r]{\strut{} \Large{450}}}%
      \put(1530,6121){\makebox(0,0)[r]{\strut{} \Large{500}}}%
      \put(856,3401){\rotatebox{-270}{\makebox(0,0){\strut{}\Large{$m_{\widetilde{\tau}_1}$ [GeV]}}}}%
    }%
    \gplgaddtomacro\gplbacktext{%
    }%
    \gplgaddtomacro\gplfronttext{%
      \csname LTb\endcsname%
      \put(8756,395){\makebox(0,0){\strut{} \Large{20}}}%
\put(8920,5850){\makebox(0,0){\strut{} \Large{$\mu< 0$}}}%
      \put(9763,395){\makebox(0,0){\strut{} \Large{40}}}%
      \put(10771,395){\makebox(0,0){\strut{} \Large{60}}}%
      \put(11779,395){\makebox(0,0){\strut{} \Large{80}}}%
      \put(12785,395){\makebox(0,0){\strut{} \Large{100}}}%
      \put(13793,395){\makebox(0,0){\strut{} \Large{120}}}%
      \put(14800,395){\makebox(0,0){\strut{} \Large{140}}}%
      \put(11904,65){\makebox(0,0){\strut{}\Large{$m_{\widetilde{\chi}}$ [GeV]}}}%
      \put(15946,680){\makebox(0,0)[l]{\strut{} \Large{10}}}%
      \put(15946,3400){\makebox(0,0)[l]{\strut{} \Large{100}}}%
      \put(15946,6121){\makebox(0,0)[l]{\strut{} \Large{1000}}}%
 \put(16672,3417){\rotatebox{-270}{\makebox(0,0){\strut{}\Large{$\Delta_{\rm tot}$}}}}%
    }%
    \gplbacktext
    \put(0,0){\includegraphics{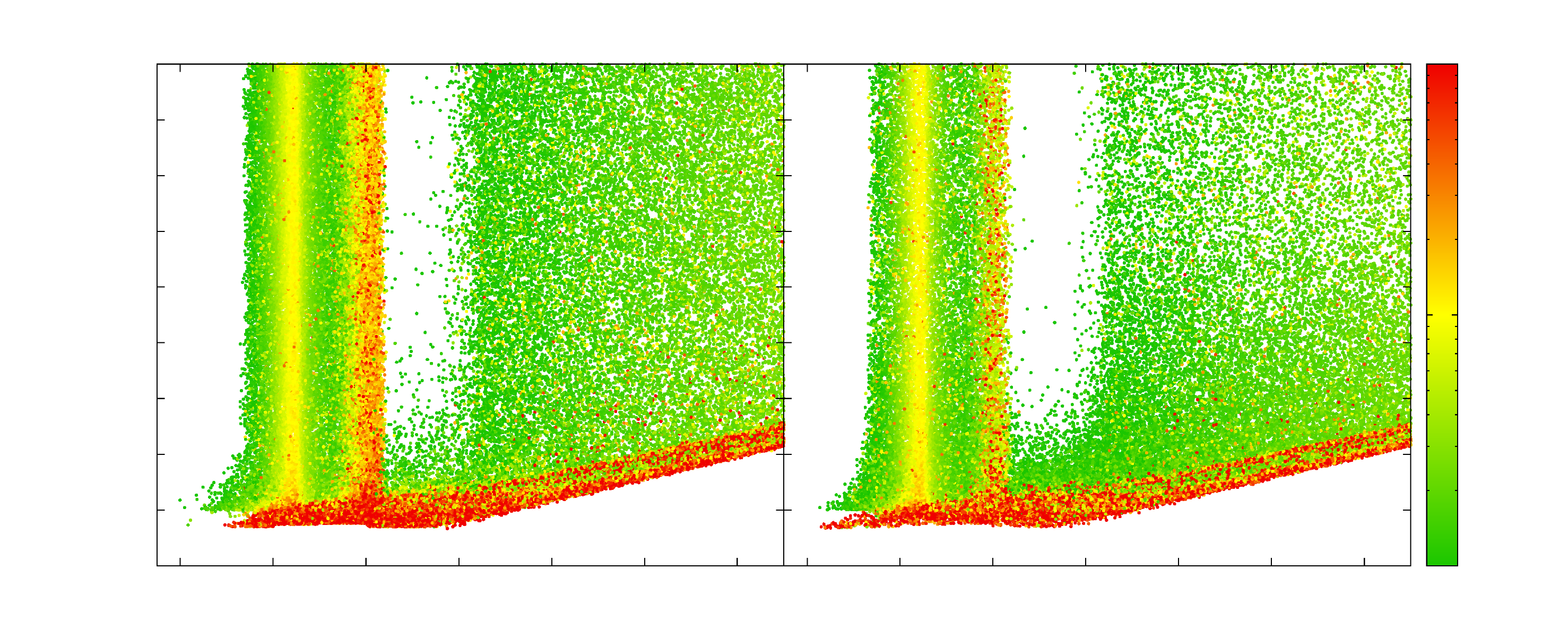}}%
    \gplfronttext
  \end{picture}%
\endgroup

%% file: pic/mu_dependence_sigma.tex
\begingroup
  \makeatletter
  \providecommand\color[2][]{%
    \GenericError{(gnuplot) \space\space\space\@spaces}{%
      Package color not loaded in conjunction with
      terminal option `colourtext'%
    }{See the gnuplot documentation for explanation.%
    }{Either use 'blacktext' in gnuplot or load the package
      color.sty in LaTeX.}%
    \renewcommand\color[2][]{}%
  }%
  \providecommand\includegraphics[2][]{%
    \GenericError{(gnuplot) \space\space\space\@spaces}{%
      Package graphicx or graphics not loaded%
    }{See the gnuplot documentation for explanation.%
    }{The gnuplot epslatex terminal needs graphicx.sty or graphics.sty.}%
    \renewcommand\includegraphics[2][]{}%
  }%
  \providecommand\rotatebox[2]{#2}%
  \@ifundefined{ifGPcolor}{%
    \newif\ifGPcolor
    \GPcolortrue
  }{}%
  \@ifundefined{ifGPblacktext}{%
    \newif\ifGPblacktext
    \GPblacktextfalse
  }{}%
  \let\gplgaddtomacro\g@addto@macro
  \gdef\gplbacktext{}%
  \gdef\gplfronttext{}%
  \makeatother
  \ifGPblacktext
    \def\colorrgb#1{}%
    \def\colorgray#1{}%
  \else
    \ifGPcolor
      \def\colorrgb#1{\color[rgb]{#1}}%
      \def\colorgray#1{\color[gray]{#1}}%
      \expandafter\def\csname LTw\endcsname{\color{white}}%
      \expandafter\def\csname LTb\endcsname{\color{black}}%
      \expandafter\def\csname LTa\endcsname{\color{black}}%
      \expandafter\def\csname LT0\endcsname{\color[rgb]{1,0,0}}%
      \expandafter\def\csname LT1\endcsname{\color[rgb]{0,1,0}}%
      \expandafter\def\csname LT2\endcsname{\color[rgb]{0,0,1}}%
      \expandafter\def\csname LT3\endcsname{\color[rgb]{1,0,1}}%
      \expandafter\def\csname LT4\endcsname{\color[rgb]{0,1,1}}%
      \expandafter\def\csname LT5\endcsname{\color[rgb]{1,1,0}}%
      \expandafter\def\csname LT6\endcsname{\color[rgb]{0,0,0}}%
      \expandafter\def\csname LT7\endcsname{\color[rgb]{1,0.3,0}}%
      \expandafter\def\csname LT8\endcsname{\color[rgb]{0.5,0.5,0.5}}%
    \else
      \def\colorrgb#1{\color{black}}%
      \def\colorgray#1{\color[gray]{#1}}%
      \expandafter\def\csname LTw\endcsname{\color{white}}%
      \expandafter\def\csname LTb\endcsname{\color{black}}%
      \expandafter\def\csname LTa\endcsname{\color{black}}%
      \expandafter\def\csname LT0\endcsname{\color{black}}%
      \expandafter\def\csname LT1\endcsname{\color{black}}%
      \expandafter\def\csname LT2\endcsname{\color{black}}%
      \expandafter\def\csname LT3\endcsname{\color{black}}%
      \expandafter\def\csname LT4\endcsname{\color{black}}%
      \expandafter\def\csname LT5\endcsname{\color{black}}%
      \expandafter\def\csname LT6\endcsname{\color{black}}%
      \expandafter\def\csname LT7\endcsname{\color{black}}%
      \expandafter\def\csname LT8\endcsname{\color{black}}%
    \fi
  \fi
  \setlength{\unitlength}{0.0500bp}%
  \begin{picture}(17006.00,6236.00)%
    \gplgaddtomacro\gplbacktext{%
    }%
    \gplgaddtomacro\gplfronttext{%
      \csname LTb\endcsname%
      \put(1702,339){\makebox(0,0){\strut{}\Large{-1000}}}%
      \put(2418,339){\makebox(0,0){\strut{}\Large{-900}}}%
      \put(3134,339){\makebox(0,0){\strut{}\Large{-800}}}%
      \put(3850,339){\makebox(0,0){\strut{}\Large{-700}}}%
      \put(4566,339){\makebox(0,0){\strut{}\Large{-600}}}%
      \put(5280,339){\makebox(0,0){\strut{}\Large{-500}}}%
      \put(5996,339){\makebox(0,0){\strut{}\Large{-400}}}%
      \put(6712,339){\makebox(0,0){\strut{}\Large{-300}}}%
      \put(7428,339){\makebox(0,0){\strut{}\Large{-200}}}%
      \put(8144,339){\makebox(0,0){\strut{}\Large{-100}}}%
      \put(5102,0){\makebox(0,0){\strut{} \Large{$\mu$ [GeV]}}}%
      \put(1530,625){\makebox(0,0)[r]{\strut{}\Large{-18}}}%
      \put(1530,1456){\makebox(0,0)[r]{\strut{}\Large{-16}}}%
      \put(1530,2287){\makebox(0,0)[r]{\strut{}\Large{-14}}}%
      \put(1530,3118){\makebox(0,0)[r]{\strut{}\Large{-12}}}%
      \put(1530,3949){\makebox(0,0)[r]{\strut{}\Large{-10}}}%
      \put(1530,4780){\makebox(0,0)[r]{\strut{}\Large{-8}}}%
      \put(1530,5611){\makebox(0,0)[r]{\strut{}\Large{-6}}}%
      \put(868,3118){\rotatebox{-270}{\makebox(0,0){\strut{} \Large{$\log (\sigma^{\rm SI})$ [pb]}}}}%
    }%
    \gplgaddtomacro\gplbacktext{%
    }%
    \gplgaddtomacro\gplfronttext{%
      \csname LTb\endcsname%
      \put(8862,339){\makebox(0,0){\strut{} \Large{100}}}%
      \put(9578,339){\makebox(0,0){\strut{} \Large{200}}}%
      \put(10294,339){\makebox(0,0){\strut{} \Large{300}}}%
      \put(11010,339){\makebox(0,0){\strut{} \Large{400}}}%
      \put(11726,339){\makebox(0,0){\strut{} \Large{500}}}%
      \put(12440,339){\makebox(0,0){\strut{} \Large{600}}}%
      \put(13156,339){\makebox(0,0){\strut{} \Large{700}}}%
      \put(13872,339){\makebox(0,0){\strut{} \Large{800}}}%
      \put(14588,339){\makebox(0,0){\strut{} \Large{900}}}%
      \put(15304,339){\makebox(0,0){\strut{} \Large{1000}}}%
      \put(11904,0){\makebox(0,0){\strut{} \Large{$\mu$ [GeV]}}}%
      \put(15946,624){\makebox(0,0)[l]{\strut{} \Large{10}}}%
      \put(15946,3117){\makebox(0,0)[l]{\strut{} \Large{100}}}%
      \put(15946,5611){\makebox(0,0)[l]{\strut{} \Large{1000}}}%
      \put(16672,3117){\rotatebox{-270}{\makebox(0,0){\strut{}\Large{$\Delta_{\rm tot}$}}}}%
    }%
    \gplbacktext
    \put(0,0){\includegraphics{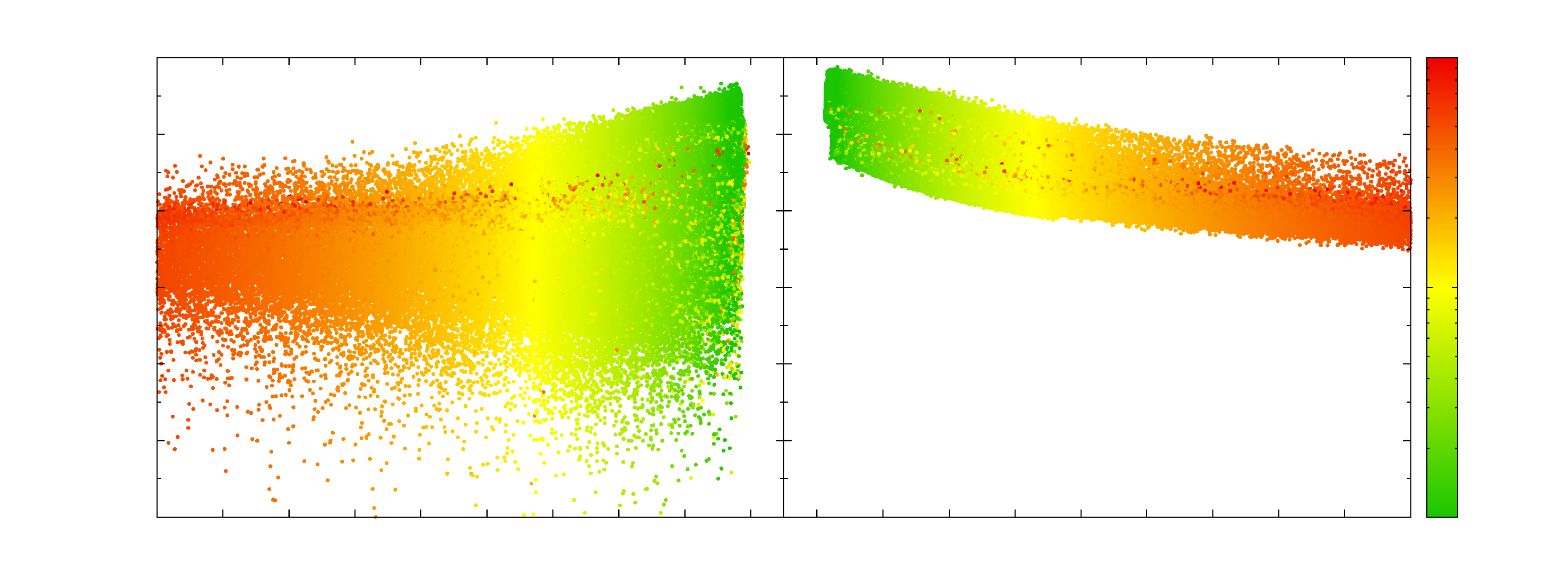}}%
    \gplfronttext
  \end{picture}%
\endgroup

%% file: pic/glt_wino.tex
\begingroup
  \makeatletter
  \providecommand\color[2][]{%
    \GenericError{(gnuplot) \space\space\space\@spaces}{%
      Package color not loaded in conjunction with
      terminal option `colourtext'%
    }{See the gnuplot documentation for explanation.%
    }{Either use 'blacktext' in gnuplot or load the package
      color.sty in LaTeX.}%
    \renewcommand\color[2][]{}%
  }%
  \providecommand\includegraphics[2][]{%
    \GenericError{(gnuplot) \space\space\space\@spaces}{%
      Package graphicx or graphics not loaded%
    }{See the gnuplot documentation for explanation.%
    }{The gnuplot epslatex terminal needs graphicx.sty or graphics.sty.}%
    \renewcommand\includegraphics[2][]{}%
  }%
  \providecommand\rotatebox[2]{#2}%
  \@ifundefined{ifGPcolor}{%
    \newif\ifGPcolor
    \GPcolortrue
  }{}%
  \@ifundefined{ifGPblacktext}{%
    \newif\ifGPblacktext
    \GPblacktextfalse
  }{}%
  \let\gplgaddtomacro\g@addto@macro
  \gdef\gplbacktext{}%
  \gdef\gplfronttext{}%
  \makeatother
  \ifGPblacktext
    \def\colorrgb#1{}%
    \def\colorgray#1{}%
  \else
    \ifGPcolor
      \def\colorrgb#1{\color[rgb]{#1}}%
      \def\colorgray#1{\color[gray]{#1}}%
      \expandafter\def\csname LTw\endcsname{\color{white}}%
      \expandafter\def\csname LTb\endcsname{\color{black}}%
      \expandafter\def\csname LTa\endcsname{\color{black}}%
      \expandafter\def\csname LT0\endcsname{\color[rgb]{1,0,0}}%
      \expandafter\def\csname LT1\endcsname{\color[rgb]{0,1,0}}%
      \expandafter\def\csname LT2\endcsname{\color[rgb]{0,0,1}}%
      \expandafter\def\csname LT3\endcsname{\color[rgb]{1,0,1}}%
      \expandafter\def\csname LT4\endcsname{\color[rgb]{0,1,1}}%
      \expandafter\def\csname LT5\endcsname{\color[rgb]{1,1,0}}%
      \expandafter\def\csname LT6\endcsname{\color[rgb]{0,0,0}}%
      \expandafter\def\csname LT7\endcsname{\color[rgb]{1,0.3,0}}%
      \expandafter\def\csname LT8\endcsname{\color[rgb]{0.5,0.5,0.5}}%
    \else
      \def\colorrgb#1{\color{black}}%
      \def\colorgray#1{\color[gray]{#1}}%
      \expandafter\def\csname LTw\endcsname{\color{white}}%
      \expandafter\def\csname LTb\endcsname{\color{black}}%
      \expandafter\def\csname LTa\endcsname{\color{black}}%
      \expandafter\def\csname LT0\endcsname{\color{black}}%
      \expandafter\def\csname LT1\endcsname{\color{black}}%
      \expandafter\def\csname LT2\endcsname{\color{black}}%
      \expandafter\def\csname LT3\endcsname{\color{black}}%
      \expandafter\def\csname LT4\endcsname{\color{black}}%
      \expandafter\def\csname LT5\endcsname{\color{black}}%
      \expandafter\def\csname LT6\endcsname{\color{black}}%
      \expandafter\def\csname LT7\endcsname{\color{black}}%
      \expandafter\def\csname LT8\endcsname{\color{black}}%
    \fi
  \fi
  \setlength{\unitlength}{0.0500bp}%
  \begin{picture}(8502.00,5668.00)%
    \gplgaddtomacro\gplbacktext{%
    }%
    \gplgaddtomacro\gplfronttext{%
      \csname LTb\endcsname%
      \put(1644,881){\makebox(0,0){\strut{}-1400}}%
      \put(2257,881){\makebox(0,0){\strut{}-1200}}%
      \put(2871,881){\makebox(0,0){\strut{}-1000}}%
      \put(3485,881){\makebox(0,0){\strut{}-800}}%
      \put(4098,881){\makebox(0,0){\strut{}-600}}%
      \put(4711,881){\makebox(0,0){\strut{}-400}}%
      \put(5324,881){\makebox(0,0){\strut{}-200}}%
      \put(5938,881){\makebox(0,0){\strut{} 0}}%
      \put(6551,881){\makebox(0,0){\strut{} 200}}%
      \put(7165,881){\makebox(0,0){\strut{} 400}}%
      \put(4251,551){\makebox(0,0){\strut{}$M_1 + \mu \sin 2 \beta$ [GeV]}}%
      \put(1165,1167){\makebox(0,0)[r]{\strut{} -22}}%
      \put(1165,1612){\makebox(0,0)[r]{\strut{} -20}}%
      \put(1165,2056){\makebox(0,0)[r]{\strut{} -18}}%
      \put(1165,2500){\makebox(0,0)[r]{\strut{} -16}}%
      \put(1165,2944){\makebox(0,0)[r]{\strut{} -14}}%
      \put(1165,3388){\makebox(0,0)[r]{\strut{} -12}}%
      \put(1165,3832){\makebox(0,0)[r]{\strut{} -10}}%
      \put(1165,4276){\makebox(0,0)[r]{\strut{} -8}}%
      \put(1165,4721){\makebox(0,0)[r]{\strut{} -6}}%
      \put(307,2944){\rotatebox{-270}{\makebox(0,0){\strut{}$\log (\sigma^{\rm SI})$ [pb]}}}%
      \put(7734,1165){\makebox(0,0)[l]{\strut{}-0.3}}%
      \put(7734,1610){\makebox(0,0)[l]{\strut{}-0.25}}%
      \put(7734,2054){\makebox(0,0)[l]{\strut{}-0.2}}%
      \put(7734,2499){\makebox(0,0)[l]{\strut{}-0.15}}%
      \put(7734,2943){\makebox(0,0)[l]{\strut{}-0.1}}%
      \put(7734,3387){\makebox(0,0)[l]{\strut{}-0.05}}%
      \put(7734,3832){\makebox(0,0)[l]{\strut{} 0}}%
      \put(7734,4276){\makebox(0,0)[l]{\strut{} 0.05}}%
      \put(7734,4720){\makebox(0,0)[l]{\strut{} 0.1}}%
      \put(8460,2943){\rotatebox{-270}{\makebox(0,0){\strut{}$N_{12}$}}}%
    }%
    \gplbacktext
    \put(0,0){\includegraphics{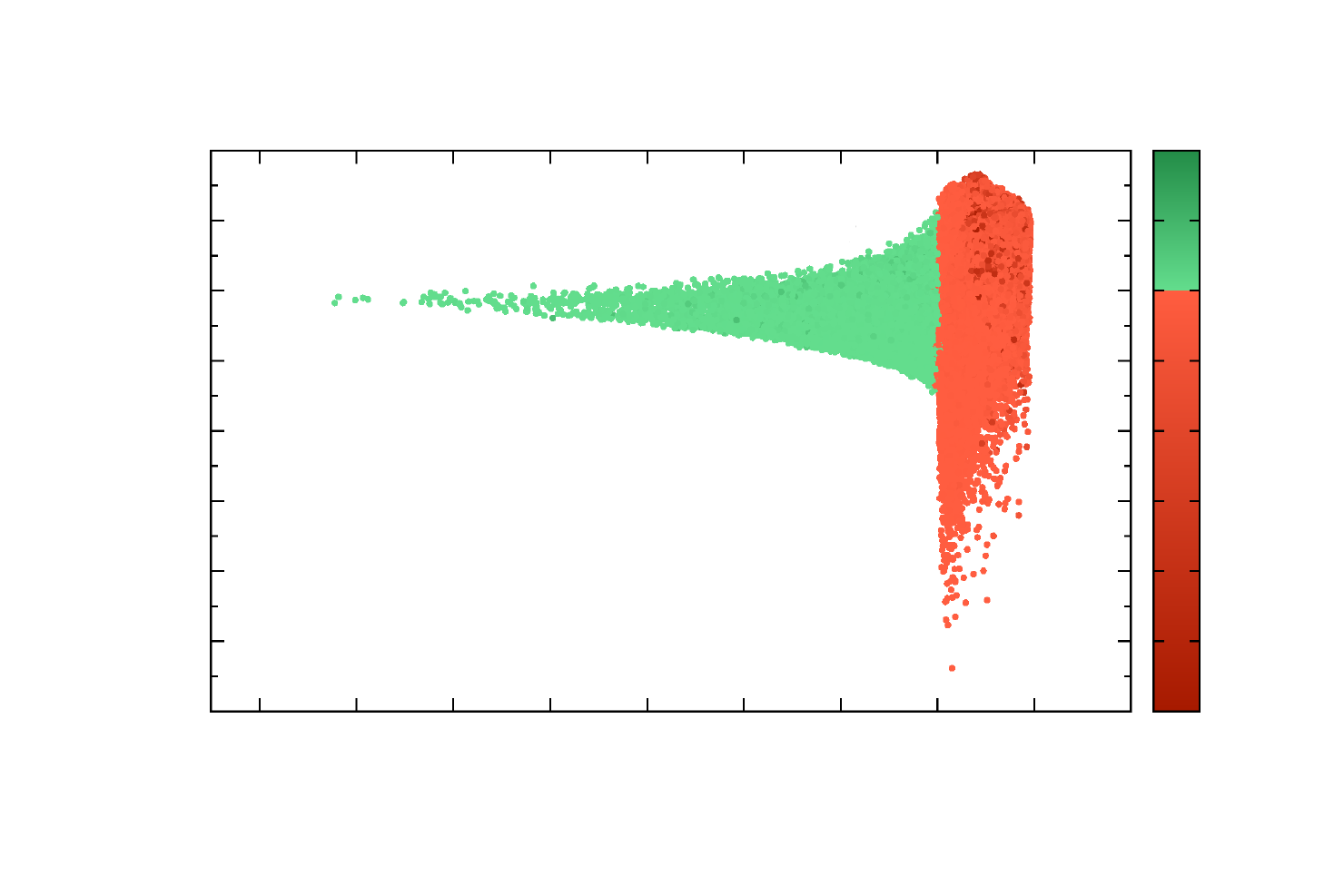}}%
    \gplfronttext
  \end{picture}%
\endgroup

%% file: pic/pos_g23_direct.tex
\begingroup
  \makeatletter
  \providecommand\color[2][]{%
    \GenericError{(gnuplot) \space\space\space\@spaces}{%
      Package color not loaded in conjunction with
      terminal option `colourtext'%
    }{See the gnuplot documentation for explanation.%
    }{Either use 'blacktext' in gnuplot or load the package
      color.sty in LaTeX.}%
    \renewcommand\color[2][]{}%
  }%
  \providecommand\includegraphics[2][]{%
    \GenericError{(gnuplot) \space\space\space\@spaces}{%
      Package graphicx or graphics not loaded%
    }{See the gnuplot documentation for explanation.%
    }{The gnuplot epslatex terminal needs graphicx.sty or graphics.sty.}%
    \renewcommand\includegraphics[2][]{}%
  }%
  \providecommand\rotatebox[2]{#2}%
  \@ifundefined{ifGPcolor}{%
    \newif\ifGPcolor
    \GPcolortrue
  }{}%
  \@ifundefined{ifGPblacktext}{%
    \newif\ifGPblacktext
    \GPblacktextfalse
  }{}%
  \let\gplgaddtomacro\g@addto@macro
  \gdef\gplbacktext{}%
  \gdef\gplfronttext{}%
  \makeatother
  \ifGPblacktext
    \def\colorrgb#1{}%
    \def\colorgray#1{}%
  \else
    \ifGPcolor
      \def\colorrgb#1{\color[rgb]{#1}}%
      \def\colorgray#1{\color[gray]{#1}}%
      \expandafter\def\csname LTw\endcsname{\color{white}}%
      \expandafter\def\csname LTb\endcsname{\color{black}}%
      \expandafter\def\csname LTa\endcsname{\color{black}}%
      \expandafter\def\csname LT0\endcsname{\color[rgb]{1,0,0}}%
      \expandafter\def\csname LT1\endcsname{\color[rgb]{0,1,0}}%
      \expandafter\def\csname LT2\endcsname{\color[rgb]{0,0,1}}%
      \expandafter\def\csname LT3\endcsname{\color[rgb]{1,0,1}}%
      \expandafter\def\csname LT4\endcsname{\color[rgb]{0,1,1}}%
      \expandafter\def\csname LT5\endcsname{\color[rgb]{1,1,0}}%
      \expandafter\def\csname LT6\endcsname{\color[rgb]{0,0,0}}%
      \expandafter\def\csname LT7\endcsname{\color[rgb]{1,0.3,0}}%
      \expandafter\def\csname LT8\endcsname{\color[rgb]{0.5,0.5,0.5}}%
    \else
      \def\colorrgb#1{\color{black}}%
      \def\colorgray#1{\color[gray]{#1}}%
      \expandafter\def\csname LTw\endcsname{\color{white}}%
      \expandafter\def\csname LTb\endcsname{\color{black}}%
      \expandafter\def\csname LTa\endcsname{\color{black}}%
      \expandafter\def\csname LT0\endcsname{\color{black}}%
      \expandafter\def\csname LT1\endcsname{\color{black}}%
      \expandafter\def\csname LT2\endcsname{\color{black}}%
      \expandafter\def\csname LT3\endcsname{\color{black}}%
      \expandafter\def\csname LT4\endcsname{\color{black}}%
      \expandafter\def\csname LT5\endcsname{\color{black}}%
      \expandafter\def\csname LT6\endcsname{\color{black}}%
      \expandafter\def\csname LT7\endcsname{\color{black}}%
      \expandafter\def\csname LT8\endcsname{\color{black}}%
    \fi
  \fi
  \setlength{\unitlength}{0.0500bp}%
  \begin{picture}(8502.00,5668.00)%
    \gplgaddtomacro\gplbacktext{%
    }%
    \gplgaddtomacro\gplfronttext{%
      \csname LTb\endcsname%
      \put(1628,881){\makebox(0,0){\strut{} 20}}%
      \put(2211,881){\makebox(0,0){\strut{} 40}}%
      \put(2794,881){\makebox(0,0){\strut{} 60}}%
      \put(3377,881){\makebox(0,0){\strut{} 80}}%
      \put(3960,881){\makebox(0,0){\strut{} 100}}%
      \put(4542,881){\makebox(0,0){\strut{} 120}}%
      \put(5125,881){\makebox(0,0){\strut{} 140}}%
      \put(5708,881){\makebox(0,0){\strut{} 160}}%
      \put(6291,881){\makebox(0,0){\strut{} 180}}%
      \put(6874,881){\makebox(0,0){\strut{} 200}}%
      \put(4251,551){\makebox(0,0){\strut{}$m_{\widetilde{\chi}}$ [GeV]}}%
      \put(1165,1167){\makebox(0,0)[r]{\strut{} -12}}%
      \put(1165,1760){\makebox(0,0)[r]{\strut{} -11}}%
      \put(1165,2352){\makebox(0,0)[r]{\strut{} -10}}%
      \put(1165,2944){\makebox(0,0)[r]{\strut{} -9}}%
      \put(1165,3536){\makebox(0,0)[r]{\strut{} -8}}%
      \put(1165,4128){\makebox(0,0)[r]{\strut{} -7}}%
      \put(1165,4721){\makebox(0,0)[r]{\strut{} -6}}%
      \put(307,2944){\rotatebox{-270}{\makebox(0,0){\strut{}$\log (\sigma^{\rm SI})$ [pb]}}}%
      \put(7734,1166){\makebox(0,0)[l]{\strut{} 10}}%
      \put(7734,2943){\makebox(0,0)[l]{\strut{} 100}}%
      \put(7734,4721){\makebox(0,0)[l]{\strut{} 1000}}%
      \put(8460,2943){\rotatebox{-270}{\makebox(0,0){\strut{}$\Delta_{\rm tot}$}}}%
\put(7065,3781){\makebox(0,0)[r]{\strut{} \scriptsize{ \rm \textbf{XENON100 (2011)}}}}%
\put(7065,3200){\makebox(0,0)[r]{\strut{} \scriptsize{ \rm \textbf{XENON100 (2012)}}}}%
 \put(6865,2391){\makebox(0,0)[r]{\strut{} \scriptsize{ \rm \textbf{XENON 1t}}}}%

    }%
    \gplbacktext
    \put(0,0){\includegraphics{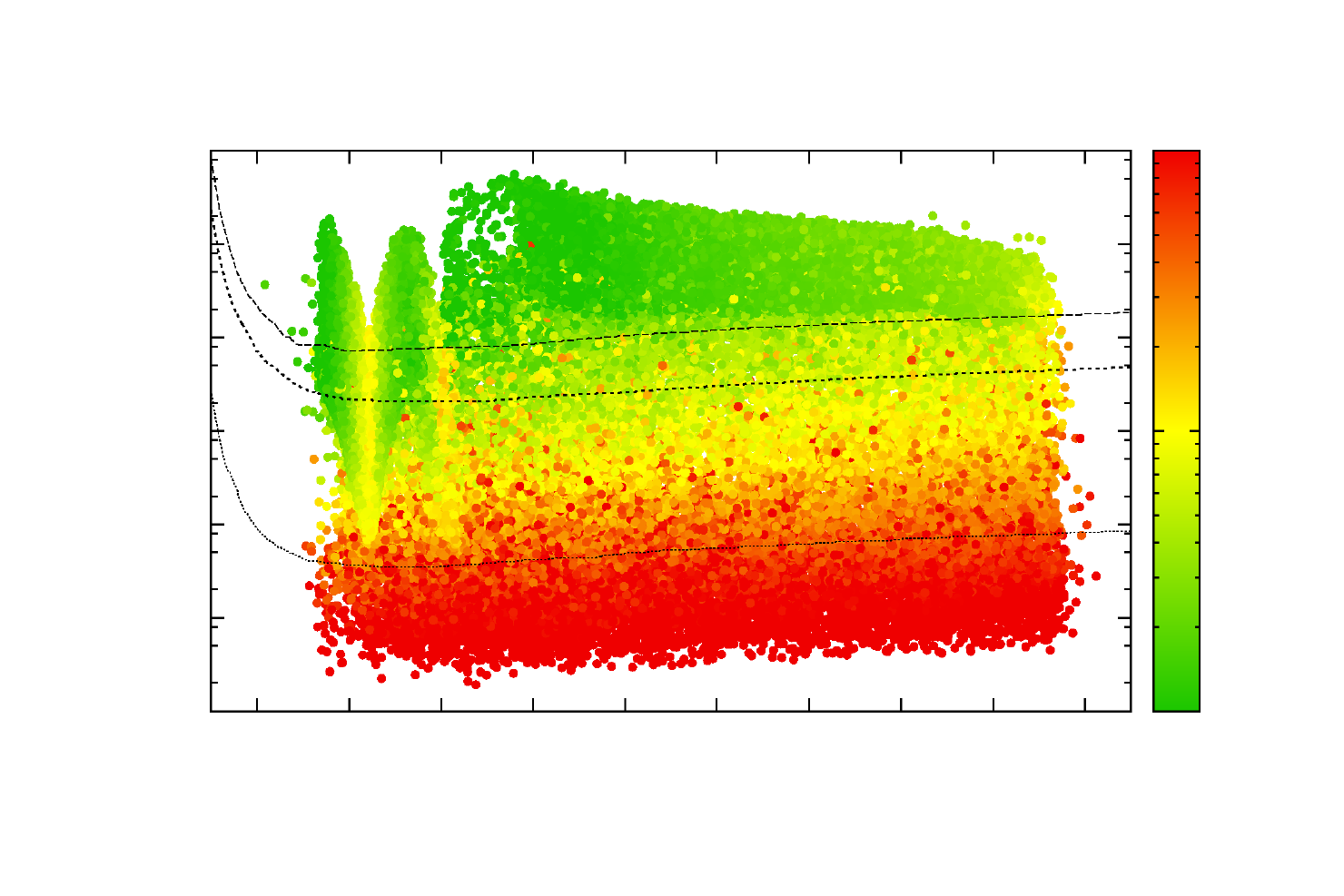}}%
    \gplfronttext
  \end{picture}%
\endgroup

%% file: pic/assumption_g2.tex
\begingroup
  \makeatletter
  \providecommand\color[2][]{%
    \GenericError{(gnuplot) \space\space\space\@spaces}{%
      Package color not loaded in conjunction with
      terminal option `colourtext'%
    }{See the gnuplot documentation for explanation.%
    }{Either use 'blacktext' in gnuplot or load the package
      color.sty in LaTeX.}%
    \renewcommand\color[2][]{}%
  }%
  \providecommand\includegraphics[2][]{%
    \GenericError{(gnuplot) \space\space\space\@spaces}{%
      Package graphicx or graphics not loaded%
    }{See the gnuplot documentation for explanation.%
    }{The gnuplot epslatex terminal needs graphicx.sty or graphics.sty.}%
    \renewcommand\includegraphics[2][]{}%
  }%
  \providecommand\rotatebox[2]{#2}%
  \@ifundefined{ifGPcolor}{%
    \newif\ifGPcolor
    \GPcolortrue
  }{}%
  \@ifundefined{ifGPblacktext}{%
    \newif\ifGPblacktext
    \GPblacktextfalse
  }{}%
  \let\gplgaddtomacro\g@addto@macro
  \gdef\gplbacktext{}%
  \gdef\gplfronttext{}%
  \makeatother
  \ifGPblacktext
    \def\colorrgb#1{}%
    \def\colorgray#1{}%
  \else
    \ifGPcolor
      \def\colorrgb#1{\color[rgb]{#1}}%
      \def\colorgray#1{\color[gray]{#1}}%
      \expandafter\def\csname LTw\endcsname{\color{white}}%
      \expandafter\def\csname LTb\endcsname{\color{black}}%
      \expandafter\def\csname LTa\endcsname{\color{black}}%
      \expandafter\def\csname LT0\endcsname{\color[rgb]{1,0,0}}%
      \expandafter\def\csname LT1\endcsname{\color[rgb]{0,1,0}}%
      \expandafter\def\csname LT2\endcsname{\color[rgb]{0,0,1}}%
      \expandafter\def\csname LT3\endcsname{\color[rgb]{1,0,1}}%
      \expandafter\def\csname LT4\endcsname{\color[rgb]{0,1,1}}%
      \expandafter\def\csname LT5\endcsname{\color[rgb]{1,1,0}}%
      \expandafter\def\csname LT6\endcsname{\color[rgb]{0,0,0}}%
      \expandafter\def\csname LT7\endcsname{\color[rgb]{1,0.3,0}}%
      \expandafter\def\csname LT8\endcsname{\color[rgb]{0.5,0.5,0.5}}%
    \else
      \def\colorrgb#1{\color{black}}%
      \def\colorgray#1{\color[gray]{#1}}%
      \expandafter\def\csname LTw\endcsname{\color{white}}%
      \expandafter\def\csname LTb\endcsname{\color{black}}%
      \expandafter\def\csname LTa\endcsname{\color{black}}%
      \expandafter\def\csname LT0\endcsname{\color{black}}%
      \expandafter\def\csname LT1\endcsname{\color{black}}%
      \expandafter\def\csname LT2\endcsname{\color{black}}%
      \expandafter\def\csname LT3\endcsname{\color{black}}%
      \expandafter\def\csname LT4\endcsname{\color{black}}%
      \expandafter\def\csname LT5\endcsname{\color{black}}%
      \expandafter\def\csname LT6\endcsname{\color{black}}%
      \expandafter\def\csname LT7\endcsname{\color{black}}%
      \expandafter\def\csname LT8\endcsname{\color{black}}%
    \fi
  \fi
  \setlength{\unitlength}{0.0500bp}%
  \begin{picture}(5102.00,3968.00)%
    \gplgaddtomacro\gplbacktext{%
    }%
    \gplgaddtomacro\gplfronttext{%
      \csname LTb\endcsname%
      \put(766,508){\makebox(0,0){\strut{} 0.01}}%
      \put(1722,508){\makebox(0,0){\strut{} 0.1}}%
      \put(2678,508){\makebox(0,0){\strut{} 1}}%
      \put(3634,508){\makebox(0,0){\strut{} 10}}%
      \put(4590,508){\makebox(0,0){\strut{} 100}}%
      \put(2678,178){\makebox(0,0){\strut{}$m_{\widetilde{l}_L}/m_{\widetilde{l}_R}$}}%
      \put(594,994){\makebox(0,0)[r]{\strut{} 1}}%
      \put(594,2281){\makebox(0,0)[r]{\strut{} 10}}%
      \put(594,3570){\makebox(0,0)[r]{\strut{} 100}}%
      \put(0,2182){\rotatebox{-270}{\makebox(0,0){\strut{}$M_2/M_1$}}}%
    }%
    \gplbacktext
    \put(0,0){\includegraphics{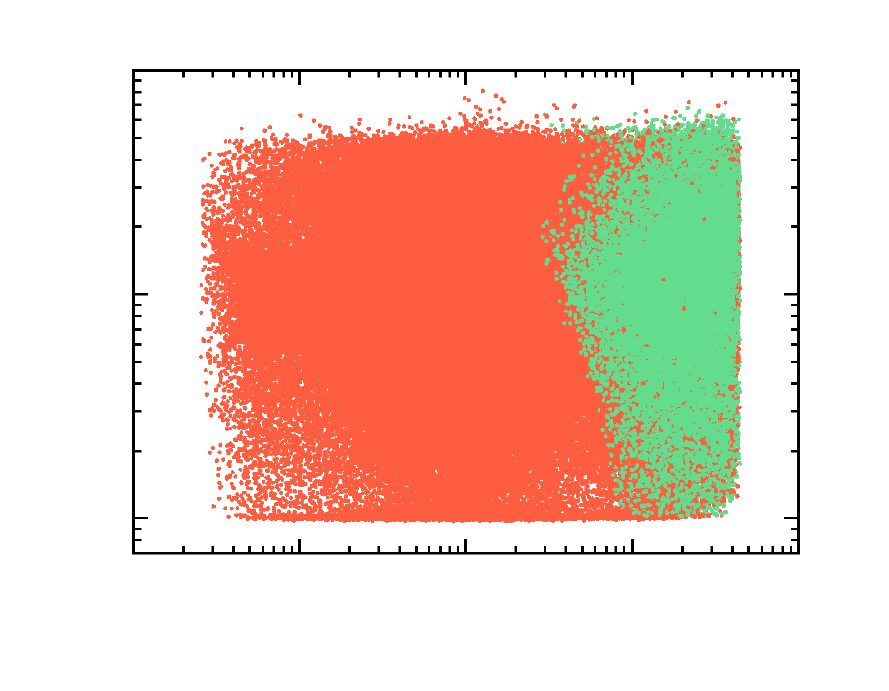}}%
    \gplfronttext
  \end{picture}%
\endgroup

%% file: pic/g22_mu.tex
\begingroup
  \makeatletter
  \providecommand\color[2][]{%
    \GenericError{(gnuplot) \space\space\space\@spaces}{%
      Package color not loaded in conjunction with
      terminal option `colourtext'%
    }{See the gnuplot documentation for explanation.%
    }{Either use 'blacktext' in gnuplot or load the package
      color.sty in LaTeX.}%
    \renewcommand\color[2][]{}%
  }%
  \providecommand\includegraphics[2][]{%
    \GenericError{(gnuplot) \space\space\space\@spaces}{%
      Package graphicx or graphics not loaded%
    }{See the gnuplot documentation for explanation.%
    }{The gnuplot epslatex terminal needs graphicx.sty or graphics.sty.}%
    \renewcommand\includegraphics[2][]{}%
  }%
  \providecommand\rotatebox[2]{#2}%
  \@ifundefined{ifGPcolor}{%
    \newif\ifGPcolor
    \GPcolortrue
  }{}%
  \@ifundefined{ifGPblacktext}{%
    \newif\ifGPblacktext
    \GPblacktextfalse
  }{}%
  \let\gplgaddtomacro\g@addto@macro
  \gdef\gplbacktext{}%
  \gdef\gplfronttext{}%
  \makeatother
  \ifGPblacktext
    \def\colorrgb#1{}%
    \def\colorgray#1{}%
  \else
    \ifGPcolor
      \def\colorrgb#1{\color[rgb]{#1}}%
      \def\colorgray#1{\color[gray]{#1}}%
      \expandafter\def\csname LTw\endcsname{\color{white}}%
      \expandafter\def\csname LTb\endcsname{\color{black}}%
      \expandafter\def\csname LTa\endcsname{\color{black}}%
      \expandafter\def\csname LT0\endcsname{\color[rgb]{1,0,0}}%
      \expandafter\def\csname LT1\endcsname{\color[rgb]{0,1,0}}%
      \expandafter\def\csname LT2\endcsname{\color[rgb]{0,0,1}}%
      \expandafter\def\csname LT3\endcsname{\color[rgb]{1,0,1}}%
      \expandafter\def\csname LT4\endcsname{\color[rgb]{0,1,1}}%
      \expandafter\def\csname LT5\endcsname{\color[rgb]{1,1,0}}%
      \expandafter\def\csname LT6\endcsname{\color[rgb]{0,0,0}}%
      \expandafter\def\csname LT7\endcsname{\color[rgb]{1,0.3,0}}%
      \expandafter\def\csname LT8\endcsname{\color[rgb]{0.5,0.5,0.5}}%
    \else
      \def\colorrgb#1{\color{black}}%
      \def\colorgray#1{\color[gray]{#1}}%
      \expandafter\def\csname LTw\endcsname{\color{white}}%
      \expandafter\def\csname LTb\endcsname{\color{black}}%
      \expandafter\def\csname LTa\endcsname{\color{black}}%
      \expandafter\def\csname LT0\endcsname{\color{black}}%
      \expandafter\def\csname LT1\endcsname{\color{black}}%
      \expandafter\def\csname LT2\endcsname{\color{black}}%
      \expandafter\def\csname LT3\endcsname{\color{black}}%
      \expandafter\def\csname LT4\endcsname{\color{black}}%
      \expandafter\def\csname LT5\endcsname{\color{black}}%
      \expandafter\def\csname LT6\endcsname{\color{black}}%
      \expandafter\def\csname LT7\endcsname{\color{black}}%
      \expandafter\def\csname LT8\endcsname{\color{black}}%
    \fi
  \fi
  \setlength{\unitlength}{0.0500bp}%
  \begin{picture}(5102.00,3968.00)%
    \gplgaddtomacro\gplbacktext{%
    }%
    \gplgaddtomacro\gplfronttext{%
      \csname LTb\endcsname%
      \put(766,508){\makebox(0,0){\strut{}-2000}}%
      \put(1747,508){\makebox(0,0){\strut{}-1500}}%
      \put(2727,508){\makebox(0,0){\strut{}-1000}}%
      \put(3708,508){\makebox(0,0){\strut{}-500}}%
      \put(2678,178){\makebox(0,0){\strut{}$\mu$ [GeV]}}%
      \put(594,794){\makebox(0,0)[r]{\strut{}-8}}%
      \put(594,1191){\makebox(0,0)[r]{\strut{}-6}}%
      \put(594,1587){\makebox(0,0)[r]{\strut{}-4}}%
      \put(594,1984){\makebox(0,0)[r]{\strut{}-2}}%
      \put(594,2380){\makebox(0,0)[r]{\strut{} 0}}%
      \put(594,2777){\makebox(0,0)[r]{\strut{} 2}}%
      \put(594,3173){\makebox(0,0)[r]{\strut{} 4}}%
      \put(594,3570){\makebox(0,0)[r]{\strut{} 6}}%
      \put(214,2182){\rotatebox{-270}{\makebox(0,0){\strut{}$a_{\mu}\times 10^{9}$}}}%
    }%
    \gplbacktext
    \put(0,0){\includegraphics{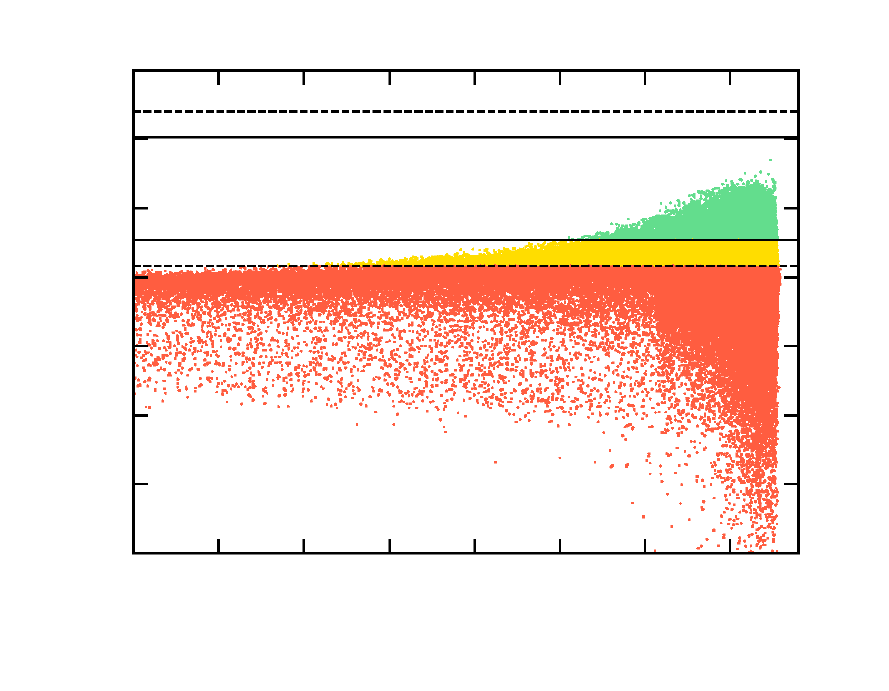}}%
    \gplfronttext
  \end{picture}%
\endgroup

%% file: pic/neg_g23_direct.tex
\begingroup
  \makeatletter
  \providecommand\color[2][]{%
    \GenericError{(gnuplot) \space\space\space\@spaces}{%
      Package color not loaded in conjunction with
      terminal option `colourtext'%
    }{See the gnuplot documentation for explanation.%
    }{Either use 'blacktext' in gnuplot or load the package
      color.sty in LaTeX.}%
    \renewcommand\color[2][]{}%
  }%
  \providecommand\includegraphics[2][]{%
    \GenericError{(gnuplot) \space\space\space\@spaces}{%
      Package graphicx or graphics not loaded%
    }{See the gnuplot documentation for explanation.%
    }{The gnuplot epslatex terminal needs graphicx.sty or graphics.sty.}%
    \renewcommand\includegraphics[2][]{}%
  }%
  \providecommand\rotatebox[2]{#2}%
  \@ifundefined{ifGPcolor}{%
    \newif\ifGPcolor
    \GPcolortrue
  }{}%
  \@ifundefined{ifGPblacktext}{%
    \newif\ifGPblacktext
    \GPblacktextfalse
  }{}%
  \let\gplgaddtomacro\g@addto@macro
  \gdef\gplbacktext{}%
  \gdef\gplfronttext{}%
  \makeatother
  \ifGPblacktext
    \def\colorrgb#1{}%
    \def\colorgray#1{}%
  \else
    \ifGPcolor
      \def\colorrgb#1{\color[rgb]{#1}}%
      \def\colorgray#1{\color[gray]{#1}}%
      \expandafter\def\csname LTw\endcsname{\color{white}}%
      \expandafter\def\csname LTb\endcsname{\color{black}}%
      \expandafter\def\csname LTa\endcsname{\color{black}}%
      \expandafter\def\csname LT0\endcsname{\color[rgb]{1,0,0}}%
      \expandafter\def\csname LT1\endcsname{\color[rgb]{0,1,0}}%
      \expandafter\def\csname LT2\endcsname{\color[rgb]{0,0,1}}%
      \expandafter\def\csname LT3\endcsname{\color[rgb]{1,0,1}}%
      \expandafter\def\csname LT4\endcsname{\color[rgb]{0,1,1}}%
      \expandafter\def\csname LT5\endcsname{\color[rgb]{1,1,0}}%
      \expandafter\def\csname LT6\endcsname{\color[rgb]{0,0,0}}%
      \expandafter\def\csname LT7\endcsname{\color[rgb]{1,0.3,0}}%
      \expandafter\def\csname LT8\endcsname{\color[rgb]{0.5,0.5,0.5}}%
    \else
      \def\colorrgb#1{\color{black}}%
      \def\colorgray#1{\color[gray]{#1}}%
      \expandafter\def\csname LTw\endcsname{\color{white}}%
      \expandafter\def\csname LTb\endcsname{\color{black}}%
      \expandafter\def\csname LTa\endcsname{\color{black}}%
      \expandafter\def\csname LT0\endcsname{\color{black}}%
      \expandafter\def\csname LT1\endcsname{\color{black}}%
      \expandafter\def\csname LT2\endcsname{\color{black}}%
      \expandafter\def\csname LT3\endcsname{\color{black}}%
      \expandafter\def\csname LT4\endcsname{\color{black}}%
      \expandafter\def\csname LT5\endcsname{\color{black}}%
      \expandafter\def\csname LT6\endcsname{\color{black}}%
      \expandafter\def\csname LT7\endcsname{\color{black}}%
      \expandafter\def\csname LT8\endcsname{\color{black}}%
    \fi
  \fi
  \setlength{\unitlength}{0.0500bp}%
  \begin{picture}(8502.00,5668.00)%
    \gplgaddtomacro\gplbacktext{%
    }%
    \gplgaddtomacro\gplfronttext{%
      \csname LTb\endcsname%
      \put(1628,881){\makebox(0,0){\strut{} 20}}%
      \put(2211,881){\makebox(0,0){\strut{} 40}}%
      \put(2794,881){\makebox(0,0){\strut{} 60}}%
      \put(3377,881){\makebox(0,0){\strut{} 80}}%
      \put(3960,881){\makebox(0,0){\strut{} 100}}%
      \put(4542,881){\makebox(0,0){\strut{} 120}}%
      \put(5125,881){\makebox(0,0){\strut{} 140}}%
      \put(5708,881){\makebox(0,0){\strut{} 160}}%
      \put(6291,881){\makebox(0,0){\strut{} 180}}%
      \put(6874,881){\makebox(0,0){\strut{} 200}}%
      \put(4251,551){\makebox(0,0){\strut{}$m_{\widetilde{\chi}}$ [GeV]}}%
      \put(1165,1167){\makebox(0,0)[r]{\strut{} -18}}%
      \put(1165,1760){\makebox(0,0)[r]{\strut{} -16}}%
      \put(1165,2352){\makebox(0,0)[r]{\strut{} -14}}%
      \put(1165,2944){\makebox(0,0)[r]{\strut{} -12}}%
      \put(1165,3536){\makebox(0,0)[r]{\strut{} -10}}%
      \put(1165,4128){\makebox(0,0)[r]{\strut{} -8}}%
      \put(1165,4721){\makebox(0,0)[r]{\strut{} -6}}%
      \put(307,2944){\rotatebox{-270}{\makebox(0,0){\strut{}$\log (\sigma^{\rm SI})$ [pb]}}}%
      \put(7734,1166){\makebox(0,0)[l]{\strut{} 10}}%
      \put(7734,2943){\makebox(0,0)[l]{\strut{} 100}}%
      \put(7734,4721){\makebox(0,0)[l]{\strut{} 1000}}%
      \put(8460,2943){\rotatebox{-270}{\makebox(0,0){\strut{}$\Delta_{\rm tot}$}}}%
 \put(7065,4321){\makebox(0,0)[r]{\strut{} \scriptsize{ \rm \textbf{XENON100 (2011)}}}}%
 \put(7065,3921){\makebox(0,0)[r]{\strut{} \scriptsize{ \rm \textbf{XENON100 (2012)}}}}%
 \put(7005,3591){\makebox(0,0)[r]{\strut{} \scriptsize{ \rm \textbf{XENON 1t}}}}%
    }%
    \gplbacktext
    \put(0,0){\includegraphics{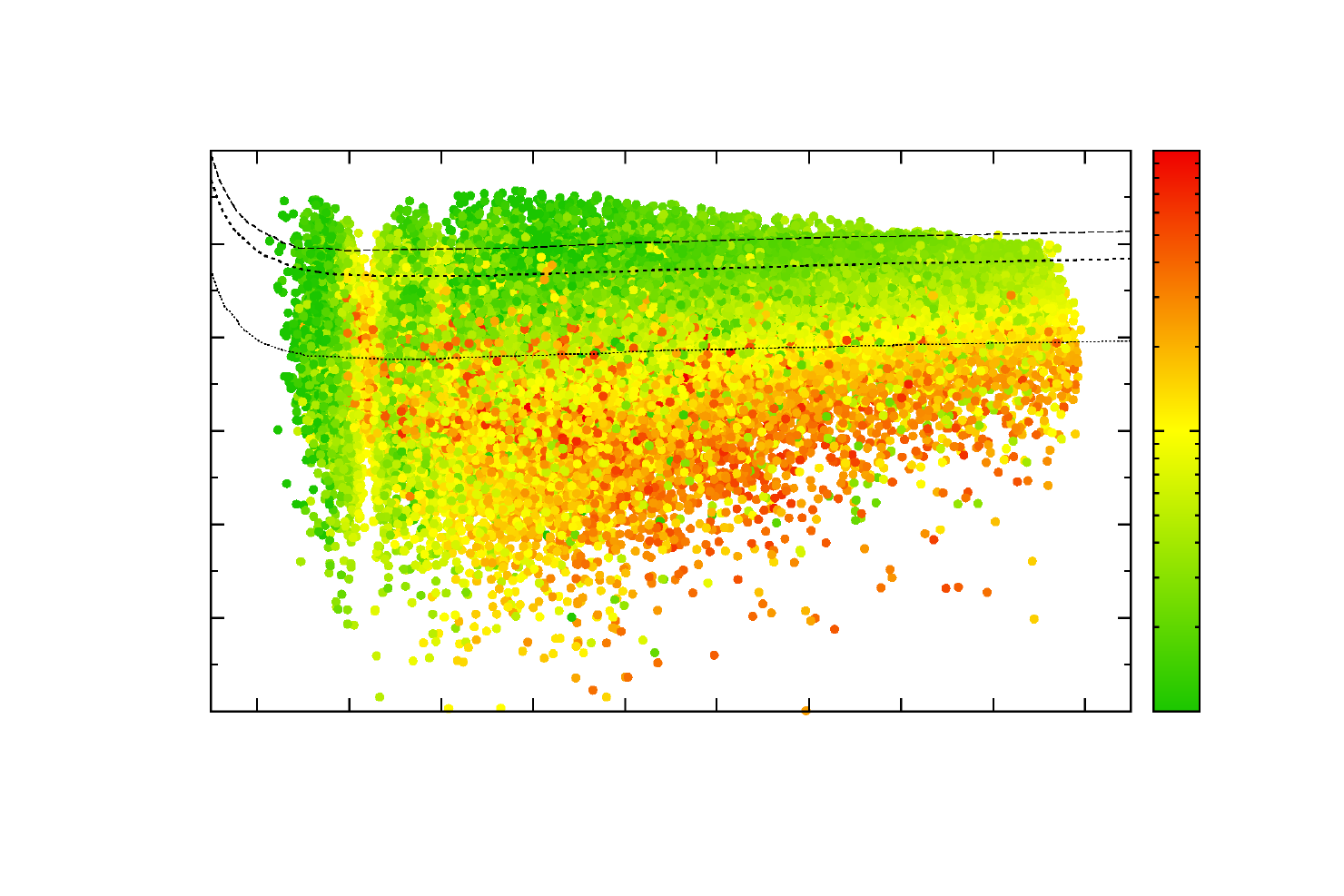}}%
    \gplfronttext
  \end{picture}%
\endgroup

%% file: pic/neg_g22.tex
\begingroup
  \makeatletter
  \providecommand\color[2][]{%
    \GenericError{(gnuplot) \space\space\space\@spaces}{%
      Package color not loaded in conjunction with
      terminal option `colourtext'%
    }{See the gnuplot documentation for explanation.%
    }{Either use 'blacktext' in gnuplot or load the package
      color.sty in LaTeX.}%
    \renewcommand\color[2][]{}%
  }%
  \providecommand\includegraphics[2][]{%
    \GenericError{(gnuplot) \space\space\space\@spaces}{%
      Package graphicx or graphics not loaded%
    }{See the gnuplot documentation for explanation.%
    }{The gnuplot epslatex terminal needs graphicx.sty or graphics.sty.}%
    \renewcommand\includegraphics[2][]{}%
  }%
  \providecommand\rotatebox[2]{#2}%
  \@ifundefined{ifGPcolor}{%
    \newif\ifGPcolor
    \GPcolortrue
  }{}%
  \@ifundefined{ifGPblacktext}{%
    \newif\ifGPblacktext
    \GPblacktextfalse
  }{}%
  \let\gplgaddtomacro\g@addto@macro
  \gdef\gplbacktext{}%
  \gdef\gplfronttext{}%
  \makeatother
  \ifGPblacktext
    \def\colorrgb#1{}%
    \def\colorgray#1{}%
  \else
    \ifGPcolor
      \def\colorrgb#1{\color[rgb]{#1}}%
      \def\colorgray#1{\color[gray]{#1}}%
      \expandafter\def\csname LTw\endcsname{\color{white}}%
      \expandafter\def\csname LTb\endcsname{\color{black}}%
      \expandafter\def\csname LTa\endcsname{\color{black}}%
      \expandafter\def\csname LT0\endcsname{\color[rgb]{1,0,0}}%
      \expandafter\def\csname LT1\endcsname{\color[rgb]{0,1,0}}%
      \expandafter\def\csname LT2\endcsname{\color[rgb]{0,0,1}}%
      \expandafter\def\csname LT3\endcsname{\color[rgb]{1,0,1}}%
      \expandafter\def\csname LT4\endcsname{\color[rgb]{0,1,1}}%
      \expandafter\def\csname LT5\endcsname{\color[rgb]{1,1,0}}%
      \expandafter\def\csname LT6\endcsname{\color[rgb]{0,0,0}}%
      \expandafter\def\csname LT7\endcsname{\color[rgb]{1,0.3,0}}%
      \expandafter\def\csname LT8\endcsname{\color[rgb]{0.5,0.5,0.5}}%
    \else
      \def\colorrgb#1{\color{black}}%
      \def\colorgray#1{\color[gray]{#1}}%
      \expandafter\def\csname LTw\endcsname{\color{white}}%
      \expandafter\def\csname LTb\endcsname{\color{black}}%
      \expandafter\def\csname LTa\endcsname{\color{black}}%
      \expandafter\def\csname LT0\endcsname{\color{black}}%
      \expandafter\def\csname LT1\endcsname{\color{black}}%
      \expandafter\def\csname LT2\endcsname{\color{black}}%
      \expandafter\def\csname LT3\endcsname{\color{black}}%
      \expandafter\def\csname LT4\endcsname{\color{black}}%
      \expandafter\def\csname LT5\endcsname{\color{black}}%
      \expandafter\def\csname LT6\endcsname{\color{black}}%
      \expandafter\def\csname LT7\endcsname{\color{black}}%
      \expandafter\def\csname LT8\endcsname{\color{black}}%
    \fi
  \fi
  \setlength{\unitlength}{0.0500bp}%
  \begin{picture}(8502.00,5668.00)%
    \gplgaddtomacro\gplbacktext{%
    }%
    \gplgaddtomacro\gplfronttext{%
      \csname LTb\endcsname%
       \put(1628,881){\makebox(0,0){\strut{} 20}}%
      \put(2211,881){\makebox(0,0){\strut{} 40}}%
      \put(2794,881){\makebox(0,0){\strut{} 60}}%
      \put(3377,881){\makebox(0,0){\strut{} 80}}%
      \put(3960,881){\makebox(0,0){\strut{} 100}}%
      \put(4542,881){\makebox(0,0){\strut{} 120}}%
      \put(5125,881){\makebox(0,0){\strut{} 140}}%
      \put(5708,881){\makebox(0,0){\strut{} 160}}%
      \put(6291,881){\makebox(0,0){\strut{} 180}}%
      \put(6874,881){\makebox(0,0){\strut{} 200}}%
      \put(4251,551){\makebox(0,0){\strut{}$m_{\widetilde{\chi}}$ [GeV]}}%
      \put(1165,1167){\makebox(0,0)[r]{\strut{} -18}}%
      \put(1165,1760){\makebox(0,0)[r]{\strut{} -16}}%
      \put(1165,2352){\makebox(0,0)[r]{\strut{} -14}}%
      \put(1165,2944){\makebox(0,0)[r]{\strut{} -12}}%
      \put(1165,3536){\makebox(0,0)[r]{\strut{} -10}}%
      \put(1165,4128){\makebox(0,0)[r]{\strut{} -8}}%
      \put(1165,4721){\makebox(0,0)[r]{\strut{} -6}}%
      \put(307,2944){\rotatebox{-270}{\makebox(0,0){\strut{}$\log (\sigma^{\rm SI})$ [pb]}}}%
      \put(7734,1166){\makebox(0,0)[l]{\strut{} 10}}%
      \put(7734,2943){\makebox(0,0)[l]{\strut{} 100}}%
      \put(7734,4721){\makebox(0,0)[l]{\strut{} 1000}}%
      \put(8460,2943){\rotatebox{-270}{\makebox(0,0){\strut{}$\Delta_{\rm tot}$}}}%
 \put(7065,4321){\makebox(0,0)[r]{\strut{} \scriptsize{ \rm \textbf{XENON100 (2011)}}}}%
 \put(7065,3921){\makebox(0,0)[r]{\strut{} \scriptsize{ \rm \textbf{XENON100 (2012)}}}}%
 \put(7005,3591){\makebox(0,0)[r]{\strut{} \scriptsize{ \rm \textbf{XENON 1t}}}}%
    }%
    \gplbacktext
    \put(0,0){\includegraphics{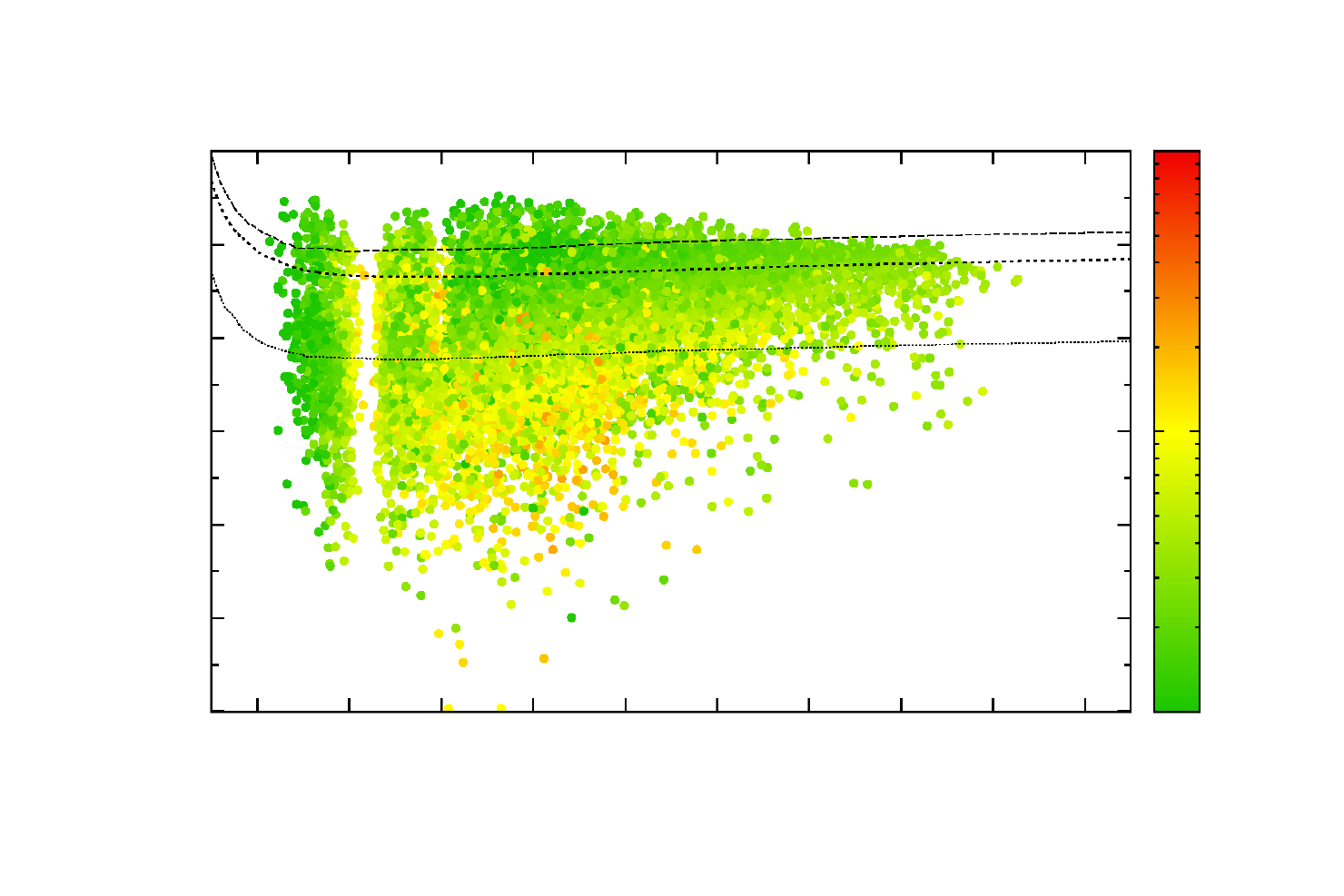}}%
    \gplfronttext
  \end{picture}%
\endgroup

%% file: pic/FT_sigma_neg_nog2_direct.tex
\begingroup
  \makeatletter
  \providecommand\color[2][]{%
    \GenericError{(gnuplot) \space\space\space\@spaces}{%
      Package color not loaded in conjunction with
      terminal option `colourtext'%
    }{See the gnuplot documentation for explanation.%
    }{Either use 'blacktext' in gnuplot or load the package
      color.sty in LaTeX.}%
    \renewcommand\color[2][]{}%
  }%
  \providecommand\includegraphics[2][]{%
    \GenericError{(gnuplot) \space\space\space\@spaces}{%
      Package graphicx or graphics not loaded%
    }{See the gnuplot documentation for explanation.%
    }{The gnuplot epslatex terminal needs graphicx.sty or graphics.sty.}%
    \renewcommand\includegraphics[2][]{}%
  }%
  \providecommand\rotatebox[2]{#2}%
  \@ifundefined{ifGPcolor}{%
    \newif\ifGPcolor
    \GPcolortrue
  }{}%
  \@ifundefined{ifGPblacktext}{%
    \newif\ifGPblacktext
    \GPblacktextfalse
  }{}%
  \let\gplgaddtomacro\g@addto@macro
  \gdef\gplbacktext{}%
  \gdef\gplfronttext{}%
  \makeatother
  \ifGPblacktext
    \def\colorrgb#1{}%
    \def\colorgray#1{}%
  \else
    \ifGPcolor
      \def\colorrgb#1{\color[rgb]{#1}}%
      \def\colorgray#1{\color[gray]{#1}}%
      \expandafter\def\csname LTw\endcsname{\color{white}}%
      \expandafter\def\csname LTb\endcsname{\color{black}}%
      \expandafter\def\csname LTa\endcsname{\color{black}}%
      \expandafter\def\csname LT0\endcsname{\color[rgb]{1,0,0}}%
      \expandafter\def\csname LT1\endcsname{\color[rgb]{0,1,0}}%
      \expandafter\def\csname LT2\endcsname{\color[rgb]{0,0,1}}%
      \expandafter\def\csname LT3\endcsname{\color[rgb]{1,0,1}}%
      \expandafter\def\csname LT4\endcsname{\color[rgb]{0,1,1}}%
      \expandafter\def\csname LT5\endcsname{\color[rgb]{1,1,0}}%
      \expandafter\def\csname LT6\endcsname{\color[rgb]{0,0,0}}%
      \expandafter\def\csname LT7\endcsname{\color[rgb]{1,0.3,0}}%
      \expandafter\def\csname LT8\endcsname{\color[rgb]{0.5,0.5,0.5}}%
    \else
      \def\colorrgb#1{\color{black}}%
      \def\colorgray#1{\color[gray]{#1}}%
      \expandafter\def\csname LTw\endcsname{\color{white}}%
      \expandafter\def\csname LTb\endcsname{\color{black}}%
      \expandafter\def\csname LTa\endcsname{\color{black}}%
      \expandafter\def\csname LT0\endcsname{\color{black}}%
      \expandafter\def\csname LT1\endcsname{\color{black}}%
      \expandafter\def\csname LT2\endcsname{\color{black}}%
      \expandafter\def\csname LT3\endcsname{\color{black}}%
      \expandafter\def\csname LT4\endcsname{\color{black}}%
      \expandafter\def\csname LT5\endcsname{\color{black}}%
      \expandafter\def\csname LT6\endcsname{\color{black}}%
      \expandafter\def\csname LT7\endcsname{\color{black}}%
      \expandafter\def\csname LT8\endcsname{\color{black}}%
    \fi
  \fi
  \setlength{\unitlength}{0.0500bp}%
  \begin{picture}(8502.00,5668.00)%
    \gplgaddtomacro\gplbacktext{%
    }%
    \gplgaddtomacro\gplfronttext{%
      \csname LTb\endcsname%
      \put(1331,564){\makebox(0,0){\strut{} \Large{20}}}%
      \put(1951,564){\makebox(0,0){\strut{} \Large{40}}}%
      \put(2572,564){\makebox(0,0){\strut{} \Large{60}}}%
      \put(3193,564){\makebox(0,0){\strut{} \Large{80}}}%
      \put(3813,564){\makebox(0,0){\strut{} \Large{100}}}%
      \put(4433,564){\makebox(0,0){\strut{} \Large{120}}}%
      \put(5053,564){\makebox(0,0){\strut{} \Large{140}}}%
      \put(5674,564){\makebox(0,0){\strut{} \Large{160}}}%
      \put(6295,564){\makebox(0,0){\strut{} \Large{180}}}%
      \put(6915,564){\makebox(0,0){\strut{} \Large{200}}}%
      \put(4123,150){\makebox(0,0){\strut{}\Large{$m_{\widetilde{\chi}}$ [GeV]}}}%
      \put(849,850){\makebox(0,0)[r]{\strut{} \Large{-18}}}%
      \put(849,1606){\makebox(0,0)[r]{\strut{} \Large{-16}}}%
      \put(849,2362){\makebox(0,0)[r]{\strut{} \Large{-14}}}%
      \put(849,3117){\makebox(0,0)[r]{\strut{} \Large{-12}}}%
      \put(849,3872){\makebox(0,0)[r]{\strut{} \Large{-10}}}%
      \put(849,4628){\makebox(0,0)[r]{\strut{} \Large{-8}}}%
      \put(849,5384){\makebox(0,0)[r]{\strut{} \Large{-6}}}%
      \put(200,3117){\rotatebox{-270}{\makebox(0,0){\strut{}\Large{$\log (\sigma^{\rm SI})$ [pb]}}}}%
      \put(7823,849){\makebox(0,0)[l]{\strut{} \Large{10}}}%
      \put(7823,3116){\makebox(0,0)[l]{\strut{} \Large{100}}}%
      \put(7823,5384){\makebox(0,0)[l]{\strut{} \Large{1000}}}%
      \put(8549,3116){\rotatebox{-270}{\makebox(0,0){\strut{}\Large{ $\mbox{ \Large $ \varSigma $}_{\rm FT} $} }}}%
 \put(7065,4851){\makebox(0,0)[r]{\strut{} \footnotesize{ \rm \textbf{XENON100 (2011)}}}}%
\put(7065,4351){\makebox(0,0)[r]{\strut{} \footnotesize{ \rm \textbf{XENON100 (2012)}}}}%
 \put(6605,3951){\makebox(0,0)[r]{\strut{} \footnotesize{ \rm \textbf{XENON 1t}}}}%
    }%
    \gplbacktext
    \put(0,0){\includegraphics{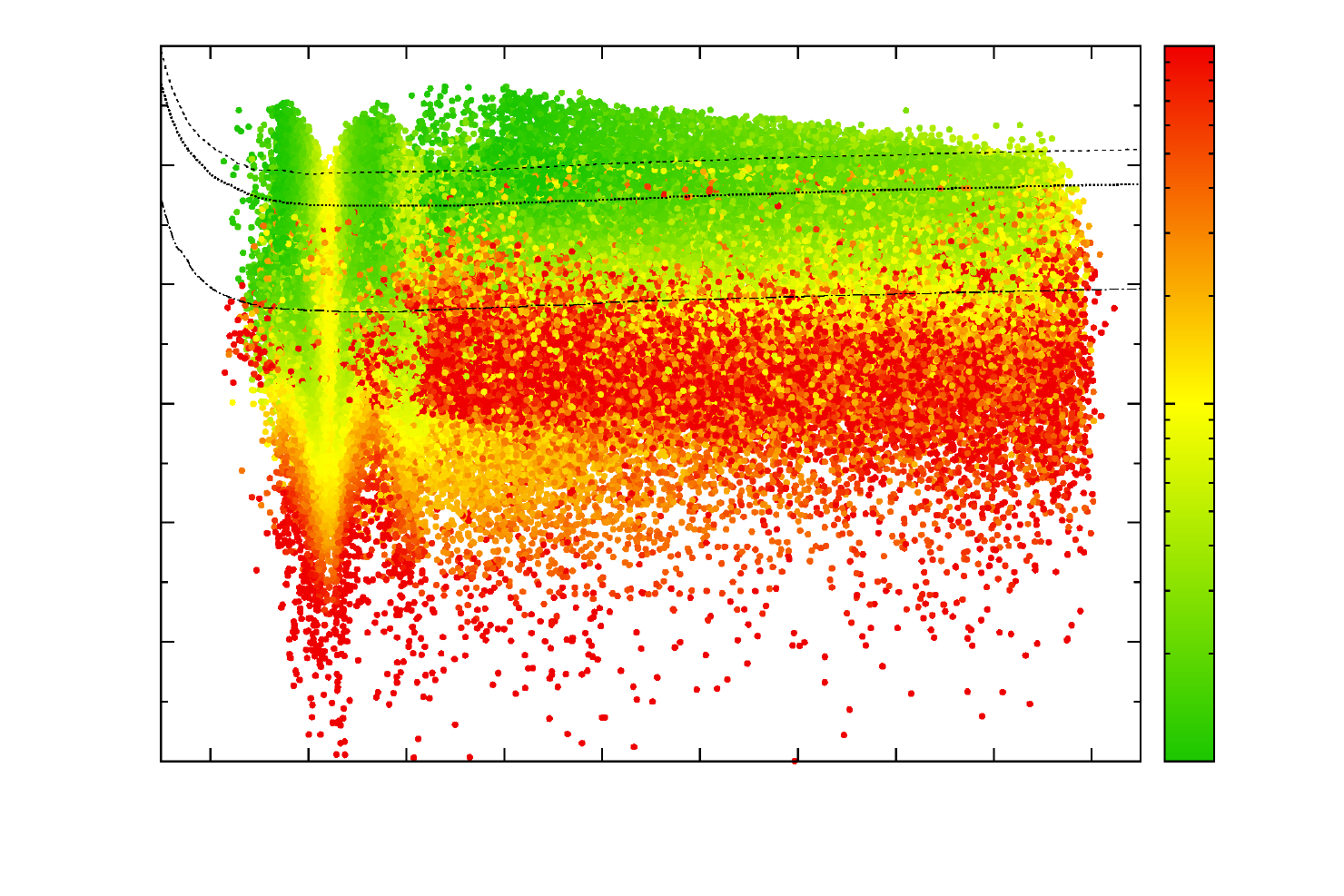}}%
    \gplfronttext
  \end{picture}%
\endgroup

%% file: pic/FT_sigma_neg_g23_direct.tex
\begingroup
  \makeatletter
  \providecommand\color[2][]{%
    \GenericError{(gnuplot) \space\space\space\@spaces}{%
      Package color not loaded in conjunction with
      terminal option `colourtext'%
    }{See the gnuplot documentation for explanation.%
    }{Either use 'blacktext' in gnuplot or load the package
      color.sty in LaTeX.}%
    \renewcommand\color[2][]{}%
  }%
  \providecommand\includegraphics[2][]{%
    \GenericError{(gnuplot) \space\space\space\@spaces}{%
      Package graphicx or graphics not loaded%
    }{See the gnuplot documentation for explanation.%
    }{The gnuplot epslatex terminal needs graphicx.sty or graphics.sty.}%
    \renewcommand\includegraphics[2][]{}%
  }%
  \providecommand\rotatebox[2]{#2}%
  \@ifundefined{ifGPcolor}{%
    \newif\ifGPcolor
    \GPcolortrue
  }{}%
  \@ifundefined{ifGPblacktext}{%
    \newif\ifGPblacktext
    \GPblacktextfalse
  }{}%
  \let\gplgaddtomacro\g@addto@macro
  \gdef\gplbacktext{}%
  \gdef\gplfronttext{}%
  \makeatother
  \ifGPblacktext
    \def\colorrgb#1{}%
    \def\colorgray#1{}%
  \else
    \ifGPcolor
      \def\colorrgb#1{\color[rgb]{#1}}%
      \def\colorgray#1{\color[gray]{#1}}%
      \expandafter\def\csname LTw\endcsname{\color{white}}%
      \expandafter\def\csname LTb\endcsname{\color{black}}%
      \expandafter\def\csname LTa\endcsname{\color{black}}%
      \expandafter\def\csname LT0\endcsname{\color[rgb]{1,0,0}}%
      \expandafter\def\csname LT1\endcsname{\color[rgb]{0,1,0}}%
      \expandafter\def\csname LT2\endcsname{\color[rgb]{0,0,1}}%
      \expandafter\def\csname LT3\endcsname{\color[rgb]{1,0,1}}%
      \expandafter\def\csname LT4\endcsname{\color[rgb]{0,1,1}}%
      \expandafter\def\csname LT5\endcsname{\color[rgb]{1,1,0}}%
      \expandafter\def\csname LT6\endcsname{\color[rgb]{0,0,0}}%
      \expandafter\def\csname LT7\endcsname{\color[rgb]{1,0.3,0}}%
      \expandafter\def\csname LT8\endcsname{\color[rgb]{0.5,0.5,0.5}}%
    \else
      \def\colorrgb#1{\color{black}}%
      \def\colorgray#1{\color[gray]{#1}}%
      \expandafter\def\csname LTw\endcsname{\color{white}}%
      \expandafter\def\csname LTb\endcsname{\color{black}}%
      \expandafter\def\csname LTa\endcsname{\color{black}}%
      \expandafter\def\csname LT0\endcsname{\color{black}}%
      \expandafter\def\csname LT1\endcsname{\color{black}}%
      \expandafter\def\csname LT2\endcsname{\color{black}}%
      \expandafter\def\csname LT3\endcsname{\color{black}}%
      \expandafter\def\csname LT4\endcsname{\color{black}}%
      \expandafter\def\csname LT5\endcsname{\color{black}}%
      \expandafter\def\csname LT6\endcsname{\color{black}}%
      \expandafter\def\csname LT7\endcsname{\color{black}}%
      \expandafter\def\csname LT8\endcsname{\color{black}}%
    \fi
  \fi
  \setlength{\unitlength}{0.0500bp}%
  \begin{picture}(8502.00,5668.00)%
    \gplgaddtomacro\gplbacktext{%
    }%
    \gplgaddtomacro\gplfronttext{%
      \csname LTb\endcsname%
      \put(1331,564){\makebox(0,0){\strut{} \Large{20}}}%
      \put(1951,564){\makebox(0,0){\strut{} \Large{40}}}%
      \put(2572,564){\makebox(0,0){\strut{} \Large{60}}}%
      \put(3193,564){\makebox(0,0){\strut{} \Large{80}}}%
      \put(3813,564){\makebox(0,0){\strut{} \Large{100}}}%
      \put(4433,564){\makebox(0,0){\strut{} \Large{120}}}%
      \put(5053,564){\makebox(0,0){\strut{} \Large{140}}}%
      \put(5674,564){\makebox(0,0){\strut{} \Large{160}}}%
      \put(6295,564){\makebox(0,0){\strut{} \Large{180}}}%
      \put(6915,564){\makebox(0,0){\strut{} \Large{200}}}%
      \put(4123,150){\makebox(0,0){\strut{}\Large{$m_{\widetilde{\chi}}$ [GeV]}}}%
      \put(849,850){\makebox(0,0)[r]{\strut{} \Large{-18}}}%
      \put(849,1606){\makebox(0,0)[r]{\strut{} \Large{-16}}}%
      \put(849,2362){\makebox(0,0)[r]{\strut{} \Large{-14}}}%
      \put(849,3117){\makebox(0,0)[r]{\strut{} \Large{-12}}}%
      \put(849,3872){\makebox(0,0)[r]{\strut{} \Large{-10}}}%
      \put(849,4628){\makebox(0,0)[r]{\strut{} \Large{-8}}}%
      \put(849,5384){\makebox(0,0)[r]{\strut{} \Large{-6}}}%
      \put(200,3117){\rotatebox{-270}{\makebox(0,0){\strut{}\Large{$\log (\sigma^{\rm SI})$ [pb]}}}}%
      \put(7823,849){\makebox(0,0)[l]{\strut{} \Large{10}}}%
      \put(7823,3116){\makebox(0,0)[l]{\strut{} \Large{100}}}%
      \put(7823,5384){\makebox(0,0)[l]{\strut{} \Large{1000}}}%
      \put(8549,3116){\rotatebox{-270}{\makebox(0,0){\strut{}\Large{ $\mbox{ \Large $ \varSigma $}_{\rm FT} $} }}}%
 \put(7065,4851){\makebox(0,0)[r]{\strut{} \footnotesize{ \rm \textbf{XENON100 (2011)}}}}%
\put(7065,4351){\makebox(0,0)[r]{\strut{} \footnotesize{ \rm \textbf{XENON100 (2012)}}}}%
 \put(6605,3951){\makebox(0,0)[r]{\strut{} \footnotesize{ \rm \textbf{XENON 1t}}}}%
    }%
    \gplbacktext
    \put(0,0){\includegraphics{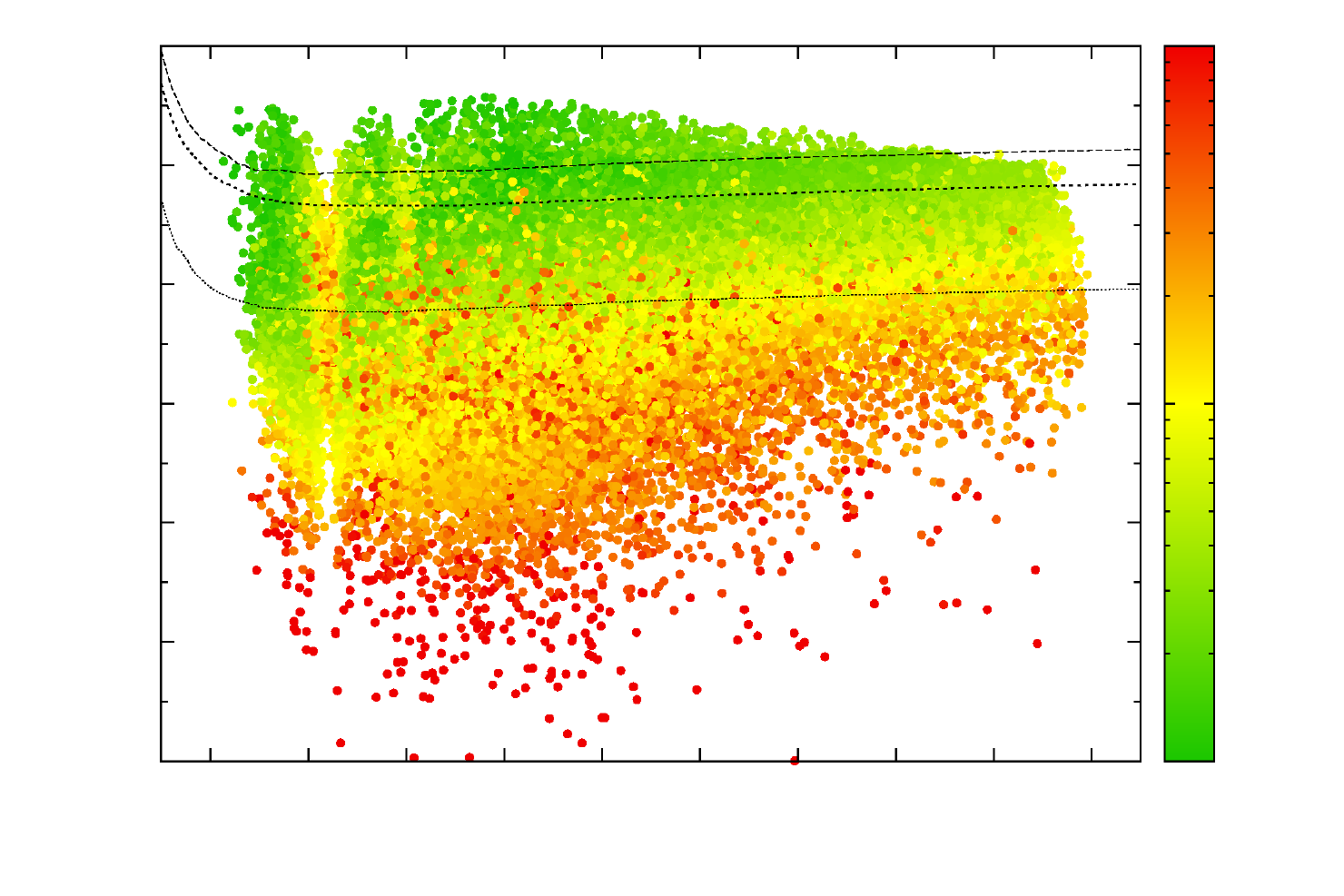}}%
    \gplfronttext
  \end{picture}%
\endgroup

%% file: pic/pos_omega_tuning.tex
\begingroup
  \makeatletter
  \providecommand\color[2][]{%
    \GenericError{(gnuplot) \space\space\space\@spaces}{%
      Package color not loaded in conjunction with
      terminal option `colourtext'%
    }{See the gnuplot documentation for explanation.%
    }{Either use 'blacktext' in gnuplot or load the package
      color.sty in LaTeX.}%
    \renewcommand\color[2][]{}%
  }%
  \providecommand\includegraphics[2][]{%
    \GenericError{(gnuplot) \space\space\space\@spaces}{%
      Package graphicx or graphics not loaded%
    }{See the gnuplot documentation for explanation.%
    }{The gnuplot epslatex terminal needs graphicx.sty or graphics.sty.}%
    \renewcommand\includegraphics[2][]{}%
  }%
  \providecommand\rotatebox[2]{#2}%
  \@ifundefined{ifGPcolor}{%
    \newif\ifGPcolor
    \GPcolortrue
  }{}%
  \@ifundefined{ifGPblacktext}{%
    \newif\ifGPblacktext
    \GPblacktextfalse
  }{}%
  \let\gplgaddtomacro\g@addto@macro
  \gdef\gplbacktext{}%
  \gdef\gplfronttext{}%
  \makeatother
  \ifGPblacktext
    \def\colorrgb#1{}%
    \def\colorgray#1{}%
  \else
    \ifGPcolor
      \def\colorrgb#1{\color[rgb]{#1}}%
      \def\colorgray#1{\color[gray]{#1}}%
      \expandafter\def\csname LTw\endcsname{\color{white}}%
      \expandafter\def\csname LTb\endcsname{\color{black}}%
      \expandafter\def\csname LTa\endcsname{\color{black}}%
      \expandafter\def\csname LT0\endcsname{\color[rgb]{1,0,0}}%
      \expandafter\def\csname LT1\endcsname{\color[rgb]{0,1,0}}%
      \expandafter\def\csname LT2\endcsname{\color[rgb]{0,0,1}}%
      \expandafter\def\csname LT3\endcsname{\color[rgb]{1,0,1}}%
      \expandafter\def\csname LT4\endcsname{\color[rgb]{0,1,1}}%
      \expandafter\def\csname LT5\endcsname{\color[rgb]{1,1,0}}%
      \expandafter\def\csname LT6\endcsname{\color[rgb]{0,0,0}}%
      \expandafter\def\csname LT7\endcsname{\color[rgb]{1,0.3,0}}%
      \expandafter\def\csname LT8\endcsname{\color[rgb]{0.5,0.5,0.5}}%
    \else
      \def\colorrgb#1{\color{black}}%
      \def\colorgray#1{\color[gray]{#1}}%
      \expandafter\def\csname LTw\endcsname{\color{white}}%
      \expandafter\def\csname LTb\endcsname{\color{black}}%
      \expandafter\def\csname LTa\endcsname{\color{black}}%
      \expandafter\def\csname LT0\endcsname{\color{black}}%
      \expandafter\def\csname LT1\endcsname{\color{black}}%
      \expandafter\def\csname LT2\endcsname{\color{black}}%
      \expandafter\def\csname LT3\endcsname{\color{black}}%
      \expandafter\def\csname LT4\endcsname{\color{black}}%
      \expandafter\def\csname LT5\endcsname{\color{black}}%
      \expandafter\def\csname LT6\endcsname{\color{black}}%
      \expandafter\def\csname LT7\endcsname{\color{black}}%
      \expandafter\def\csname LT8\endcsname{\color{black}}%
    \fi
  \fi
  \setlength{\unitlength}{0.0500bp}%
  \begin{picture}(8502.00,5668.00)%
    \gplgaddtomacro\gplbacktext{%
    }%
    \gplgaddtomacro\gplfronttext{%
      \csname LTb\endcsname%
      \put(1628,881){\makebox(0,0){\strut{} 20}}%
      \put(2211,881){\makebox(0,0){\strut{} 40}}%
      \put(2794,881){\makebox(0,0){\strut{} 60}}%
      \put(3377,881){\makebox(0,0){\strut{} 80}}%
      \put(3960,881){\makebox(0,0){\strut{} 100}}%
      \put(4542,881){\makebox(0,0){\strut{} 120}}%
      \put(5125,881){\makebox(0,0){\strut{} 140}}%
      \put(5708,881){\makebox(0,0){\strut{} 160}}%
      \put(6291,881){\makebox(0,0){\strut{} 180}}%
      \put(6874,881){\makebox(0,0){\strut{} 200}}%
      \put(4251,551){\makebox(0,0){\strut{}$m_{\widetilde{\chi}}$ [GeV]}}%
      \put(1165,1167){\makebox(0,0)[r]{\strut{}-12}}%
      \put(1165,1760){\makebox(0,0)[r]{\strut{}-11}}%
      \put(1165,2352){\makebox(0,0)[r]{\strut{}-10}}%
      \put(1165,2944){\makebox(0,0)[r]{\strut{}-9}}%
      \put(1165,3536){\makebox(0,0)[r]{\strut{}-8}}%
      \put(1165,4128){\makebox(0,0)[r]{\strut{}-7}}%
      \put(1165,4721){\makebox(0,0)[r]{\strut{}-6}}%
      \put(703,2944){\rotatebox{90}{\makebox(0,0){\strut{}$\log (\sigma^{\rm SI})$ [pb]}}}%
      \put(7734,1166){\makebox(0,0)[l]{\strut{} 10}}%
      \put(7734,2943){\makebox(0,0)[l]{\strut{} 100}}%
      \put(7734,4721){\makebox(0,0)[l]{\strut{} 1000}}%
      \put(8592,2943){\rotatebox{90}{\makebox(0,0){\strut{} \Large{$\widetilde{\varSigma}$}$_{\rm FT}$}}}%
   \put(7065,3850){\makebox(0,0)[r]{\strut{} \footnotesize{ \rm \textbf{XENON100 (2011)}}}}%
\put(7065,3095){\makebox(0,0)[r]{\strut{} \footnotesize{ \rm \textbf{XENON100 (2012)}}}}%
 \put(6605,2365){\makebox(0,0)[r]{\strut{} \footnotesize{ \rm \textbf{XENON 1t}}}}%
  }%
    \gplbacktext
    \put(0,0){\includegraphics{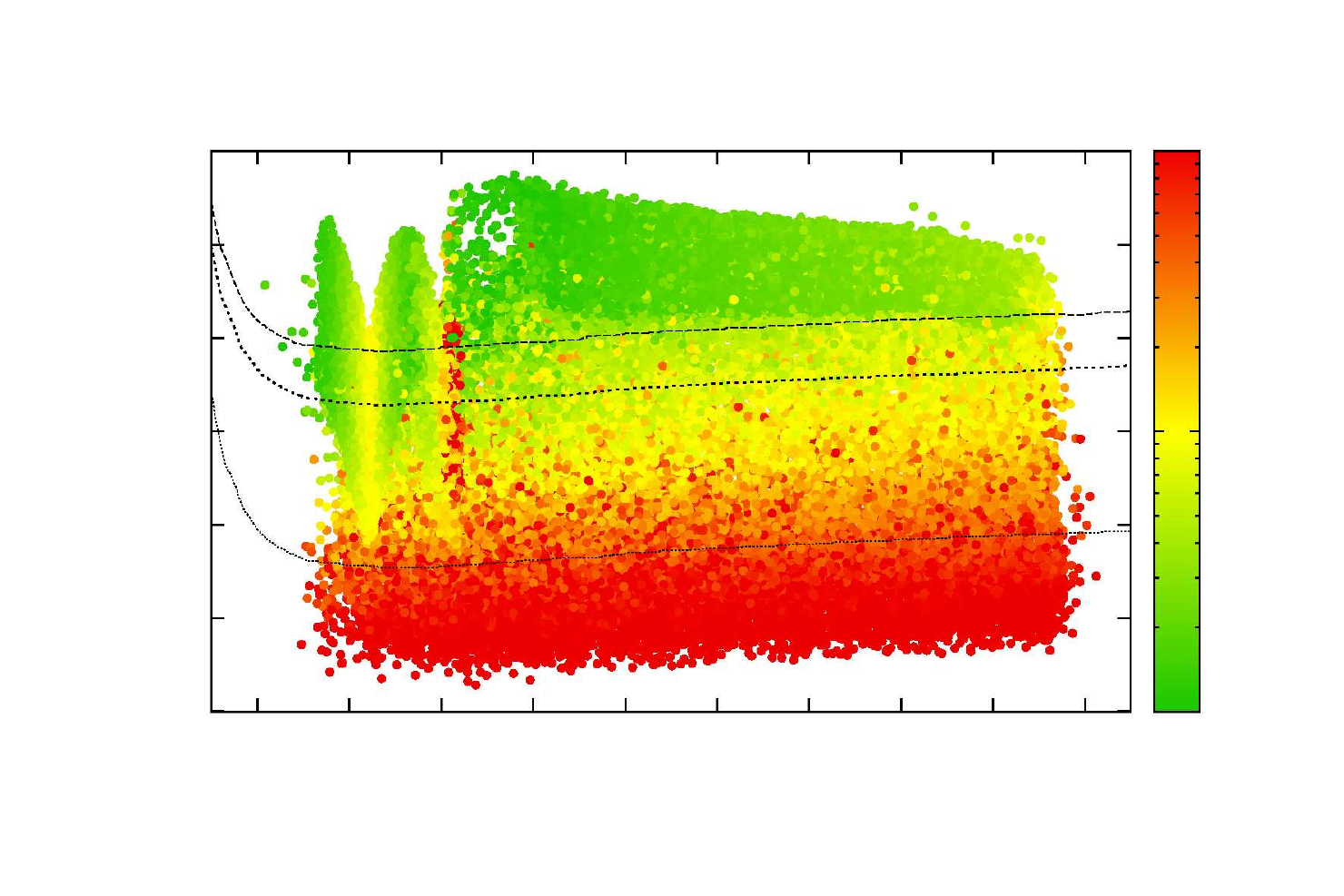}}%
    \gplfronttext
  \end{picture}%
\endgroup

%% file: pic/neg_omega_tuning.tex
\begingroup
  \makeatletter
  \providecommand\color[2][]{%
    \GenericError{(gnuplot) \space\space\space\@spaces}{%
      Package color not loaded in conjunction with
      terminal option `colourtext'%
    }{See the gnuplot documentation for explanation.%
    }{Either use 'blacktext' in gnuplot or load the package
      color.sty in LaTeX.}%
    \renewcommand\color[2][]{}%
  }%
  \providecommand\includegraphics[2][]{%
    \GenericError{(gnuplot) \space\space\space\@spaces}{%
      Package graphicx or graphics not loaded%
    }{See the gnuplot documentation for explanation.%
    }{The gnuplot epslatex terminal needs graphicx.sty or graphics.sty.}%
    \renewcommand\includegraphics[2][]{}%
  }%
  \providecommand\rotatebox[2]{#2}%
  \@ifundefined{ifGPcolor}{%
    \newif\ifGPcolor
    \GPcolortrue
  }{}%
  \@ifundefined{ifGPblacktext}{%
    \newif\ifGPblacktext
    \GPblacktextfalse
  }{}%
  \let\gplgaddtomacro\g@addto@macro
  \gdef\gplbacktext{}%
  \gdef\gplfronttext{}%
  \makeatother
  \ifGPblacktext
    \def\colorrgb#1{}%
    \def\colorgray#1{}%
  \else
    \ifGPcolor
      \def\colorrgb#1{\color[rgb]{#1}}%
      \def\colorgray#1{\color[gray]{#1}}%
      \expandafter\def\csname LTw\endcsname{\color{white}}%
      \expandafter\def\csname LTb\endcsname{\color{black}}%
      \expandafter\def\csname LTa\endcsname{\color{black}}%
      \expandafter\def\csname LT0\endcsname{\color[rgb]{1,0,0}}%
      \expandafter\def\csname LT1\endcsname{\color[rgb]{0,1,0}}%
      \expandafter\def\csname LT2\endcsname{\color[rgb]{0,0,1}}%
      \expandafter\def\csname LT3\endcsname{\color[rgb]{1,0,1}}%
      \expandafter\def\csname LT4\endcsname{\color[rgb]{0,1,1}}%
      \expandafter\def\csname LT5\endcsname{\color[rgb]{1,1,0}}%
      \expandafter\def\csname LT6\endcsname{\color[rgb]{0,0,0}}%
      \expandafter\def\csname LT7\endcsname{\color[rgb]{1,0.3,0}}%
      \expandafter\def\csname LT8\endcsname{\color[rgb]{0.5,0.5,0.5}}%
    \else
      \def\colorrgb#1{\color{black}}%
      \def\colorgray#1{\color[gray]{#1}}%
      \expandafter\def\csname LTw\endcsname{\color{white}}%
      \expandafter\def\csname LTb\endcsname{\color{black}}%
      \expandafter\def\csname LTa\endcsname{\color{black}}%
      \expandafter\def\csname LT0\endcsname{\color{black}}%
      \expandafter\def\csname LT1\endcsname{\color{black}}%
      \expandafter\def\csname LT2\endcsname{\color{black}}%
      \expandafter\def\csname LT3\endcsname{\color{black}}%
      \expandafter\def\csname LT4\endcsname{\color{black}}%
      \expandafter\def\csname LT5\endcsname{\color{black}}%
      \expandafter\def\csname LT6\endcsname{\color{black}}%
      \expandafter\def\csname LT7\endcsname{\color{black}}%
      \expandafter\def\csname LT8\endcsname{\color{black}}%
    \fi
  \fi
  \setlength{\unitlength}{0.0500bp}%
  \begin{picture}(8502.00,5668.00)%
    \gplgaddtomacro\gplbacktext{%
    }%
    \gplgaddtomacro\gplfronttext{%
      \csname LTb\endcsname%
      \put(1628,881){\makebox(0,0){\strut{} 20}}%
      \put(2211,881){\makebox(0,0){\strut{} 40}}%
      \put(2794,881){\makebox(0,0){\strut{} 60}}%
      \put(3377,881){\makebox(0,0){\strut{} 80}}%
      \put(3960,881){\makebox(0,0){\strut{} 100}}%
      \put(4542,881){\makebox(0,0){\strut{} 120}}%
      \put(5125,881){\makebox(0,0){\strut{} 140}}%
      \put(5708,881){\makebox(0,0){\strut{} 160}}%
      \put(6291,881){\makebox(0,0){\strut{} 180}}%
      \put(6874,881){\makebox(0,0){\strut{} 200}}%
      \put(4251,551){\makebox(0,0){\strut{}$m_{\widetilde{\chi}}$ [GeV]}}%
      \put(1165,1167){\makebox(0,0)[r]{\strut{}-18}}%
      \put(1165,1760){\makebox(0,0)[r]{\strut{}-16}}%
      \put(1165,2352){\makebox(0,0)[r]{\strut{}-14}}%
      \put(1165,2944){\makebox(0,0)[r]{\strut{}-12}}%
      \put(1165,3536){\makebox(0,0)[r]{\strut{}-10}}%
      \put(1165,4128){\makebox(0,0)[r]{\strut{}-8}}%
      \put(1165,4721){\makebox(0,0)[r]{\strut{}-6}}%
      \put(703,2944){\rotatebox{90}{\makebox(0,0){\strut{}$\log (\sigma^{\rm SI})$ [pb]}}}%
      \put(7734,1166){\makebox(0,0)[l]{\strut{} 10}}%
      \put(7734,2943){\makebox(0,0)[l]{\strut{} 100}}%
      \put(7734,4721){\makebox(0,0)[l]{\strut{} 1000}}%
      \put(8592,2943){\rotatebox{90}{\makebox(0,0){\strut{} \Large{$\widetilde{\varSigma}$}$_{\rm FT}$}}}%
 \put(7065,4351){\makebox(0,0)[r]{\strut{} \footnotesize{ \rm \textbf{XENON100 (2011)}}}}%
\put(7065,3875){\makebox(0,0)[r]{\strut{} \footnotesize{ \rm \textbf{XENON100 (2012)}}}}%
 \put(6605,3350){\makebox(0,0)[r]{\strut{} \footnotesize{ \rm
       \textbf{XENON 1t}}}}%
    }%
    \gplbacktext
    \put(0,0){\includegraphics{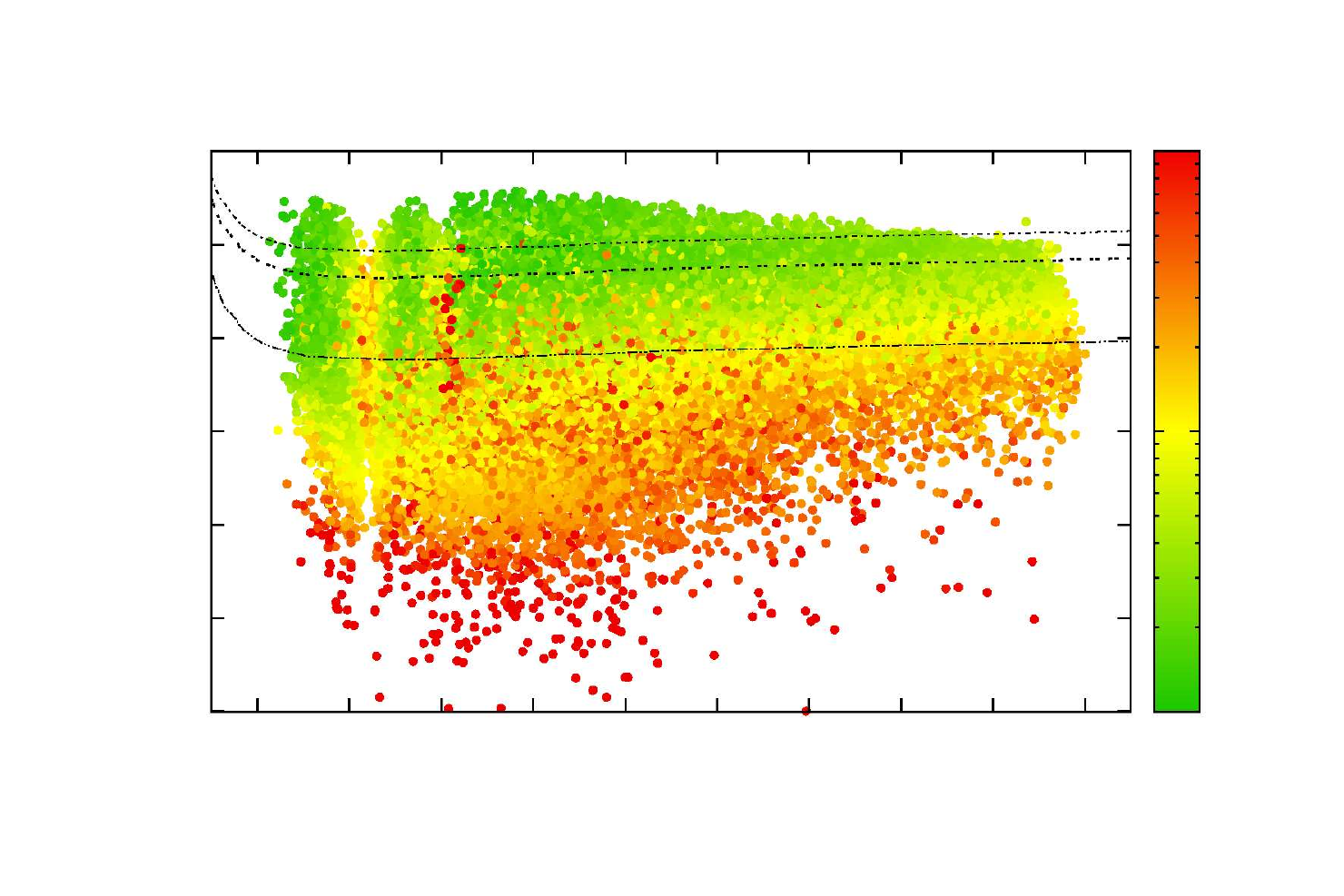}}%
    \gplfronttext
  \end{picture}%
\endgroup

%% file: pic/probability_distribution_pos_g23.tex
\begingroup
  \makeatletter
  \providecommand\color[2][]{%
    \GenericError{(gnuplot) \space\space\space\@spaces}{%
      Package color not loaded in conjunction with
      terminal option `colourtext'%
    }{See the gnuplot documentation for explanation.%
    }{Either use 'blacktext' in gnuplot or load the package
      color.sty in LaTeX.}%
    \renewcommand\color[2][]{}%
  }%
  \providecommand\includegraphics[2][]{%
    \GenericError{(gnuplot) \space\space\space\@spaces}{%
      Package graphicx or graphics not loaded%
    }{See the gnuplot documentation for explanation.%
    }{The gnuplot epslatex terminal needs graphicx.sty or graphics.sty.}%
    \renewcommand\includegraphics[2][]{}%
  }%
  \providecommand\rotatebox[2]{#2}%
  \@ifundefined{ifGPcolor}{%
    \newif\ifGPcolor
    \GPcolortrue
  }{}%
  \@ifundefined{ifGPblacktext}{%
    \newif\ifGPblacktext
    \GPblacktextfalse
  }{}%
  \let\gplgaddtomacro\g@addto@macro
  \gdef\gplbacktext{}%
  \gdef\gplfronttext{}%
  \makeatother
  \ifGPblacktext
    \def\colorrgb#1{}%
    \def\colorgray#1{}%
  \else
    \ifGPcolor
      \def\colorrgb#1{\color[rgb]{#1}}%
      \def\colorgray#1{\color[gray]{#1}}%
      \expandafter\def\csname LTw\endcsname{\color{white}}%
      \expandafter\def\csname LTb\endcsname{\color{black}}%
      \expandafter\def\csname LTa\endcsname{\color{black}}%
      \expandafter\def\csname LT0\endcsname{\color[rgb]{1,0,0}}%
      \expandafter\def\csname LT1\endcsname{\color[rgb]{0,1,0}}%
      \expandafter\def\csname LT2\endcsname{\color[rgb]{0,0,1}}%
      \expandafter\def\csname LT3\endcsname{\color[rgb]{1,0,1}}%
      \expandafter\def\csname LT4\endcsname{\color[rgb]{0,1,1}}%
      \expandafter\def\csname LT5\endcsname{\color[rgb]{1,1,0}}%
      \expandafter\def\csname LT6\endcsname{\color[rgb]{0,0,0}}%
      \expandafter\def\csname LT7\endcsname{\color[rgb]{1,0.3,0}}%
      \expandafter\def\csname LT8\endcsname{\color[rgb]{0.5,0.5,0.5}}%
    \else
      \def\colorrgb#1{\color{black}}%
      \def\colorgray#1{\color[gray]{#1}}%
      \expandafter\def\csname LTw\endcsname{\color{white}}%
      \expandafter\def\csname LTb\endcsname{\color{black}}%
      \expandafter\def\csname LTa\endcsname{\color{black}}%
      \expandafter\def\csname LT0\endcsname{\color{black}}%
      \expandafter\def\csname LT1\endcsname{\color{black}}%
      \expandafter\def\csname LT2\endcsname{\color{black}}%
      \expandafter\def\csname LT3\endcsname{\color{black}}%
      \expandafter\def\csname LT4\endcsname{\color{black}}%
      \expandafter\def\csname LT5\endcsname{\color{black}}%
      \expandafter\def\csname LT6\endcsname{\color{black}}%
      \expandafter\def\csname LT7\endcsname{\color{black}}%
      \expandafter\def\csname LT8\endcsname{\color{black}}%
    \fi
  \fi
  \setlength{\unitlength}{0.0500bp}%
  \begin{picture}(8502.00,5668.00)%
    \gplgaddtomacro\gplbacktext{%
    }%
    \gplgaddtomacro\gplfronttext{%
      \csname LTb\endcsname%
      \put(1331,564){\makebox(0,0){\strut{} \Large{20}}}%
      \put(1951,564){\makebox(0,0){\strut{} \Large{40}}}%
      \put(2572,564){\makebox(0,0){\strut{} \Large{60}}}%
      \put(3193,564){\makebox(0,0){\strut{} \Large{80}}}%
      \put(3813,564){\makebox(0,0){\strut{} \Large{100}}}%
      \put(4433,564){\makebox(0,0){\strut{} \Large{120}}}%
      \put(5053,564){\makebox(0,0){\strut{} \Large{140}}}%
      \put(5674,564){\makebox(0,0){\strut{} \Large{160}}}%
      \put(6295,564){\makebox(0,0){\strut{} \Large{180}}}%
      \put(6915,564){\makebox(0,0){\strut{} \Large{200}}}%
      \put(4123,234){\makebox(0,0){\strut{}\Large{$m_{\widetilde{\chi}}$ [GeV]}}}%
      \put(849,850){\makebox(0,0)[r]{\strut{} \Large{-12}}}%
      \put(849,1606){\makebox(0,0)[r]{\strut{} \Large{-11}}}%
      \put(849,2362){\makebox(0,0)[r]{\strut{} \Large{-10}}}%
      \put(849,3117){\makebox(0,0)[r]{\strut{} \Large{-9}}}%
      \put(849,3872){\makebox(0,0)[r]{\strut{} \Large{-8}}}%
      \put(849,4628){\makebox(0,0)[r]{\strut{} \Large{-7}}}%
      \put(849,5384){\makebox(0,0)[r]{\strut{} \Large{-6}}}%
      \put(-9,3117){\rotatebox{-270}{\makebox(0,0){\strut{} \Large{$\log (\sigma^{\rm SI})$ [pb]}}}}%
      \put(7823,849){\makebox(0,0)[l]{\strut{} \Large{10}}}%
      \put(7823,3116){\makebox(0,0)[l]{\strut{} \Large{100}}}%
      \put(7823,5384){\makebox(0,0)[l]{\strut{} \Large{1000}}}%
      \put(8549,3116){\rotatebox{-270}{\makebox(0,0){\strut{} \Large{$\widetilde{\varSigma}$}$_{\rm FT}$}}}%
\put(7065,4181){\makebox(0,0)[r]{\strut{} \footnotesize{ \rm \textbf{XENON100 (2011)}}}}%
\put(7065,3401){\makebox(0,0)[r]{\strut{} \footnotesize{ \rm \textbf{XENON100 (2012)}}}}%
 \put(6865,2391){\makebox(0,0)[r]{\strut{} \footnotesize{ \rm \textbf{XENON 1t}}}}%
    }%
    \gplbacktext
    \put(0,0){\includegraphics{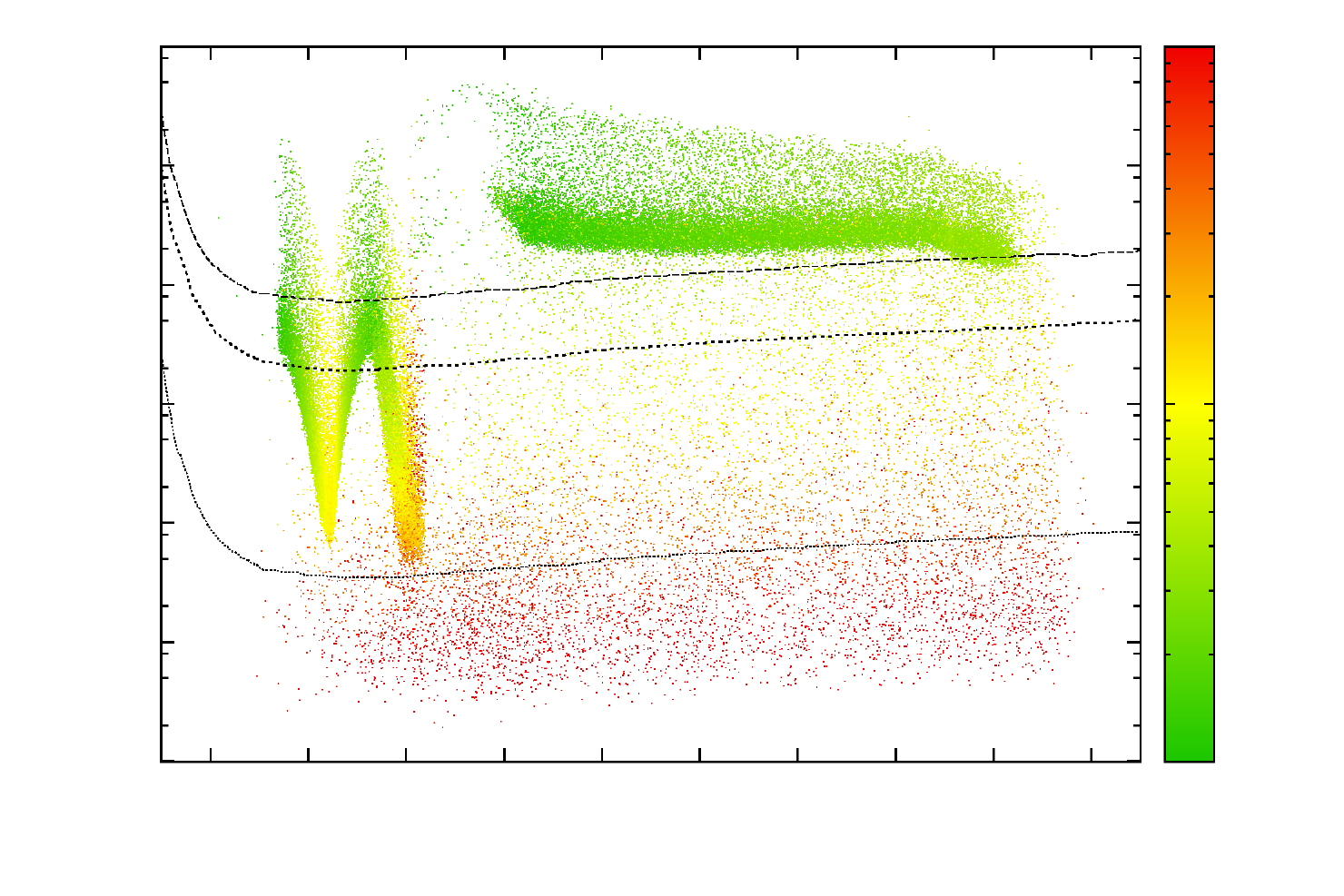}}%
    \gplfronttext
  \end{picture}%
\endgroup

%% file: pic/probability_distribution_neg_g23.tex
\begingroup
  \makeatletter
  \providecommand\color[2][]{%
    \GenericError{(gnuplot) \space\space\space\@spaces}{%
      Package color not loaded in conjunction with
      terminal option `colourtext'%
    }{See the gnuplot documentation for explanation.%
    }{Either use 'blacktext' in gnuplot or load the package
      color.sty in LaTeX.}%
    \renewcommand\color[2][]{}%
  }%
  \providecommand\includegraphics[2][]{%
    \GenericError{(gnuplot) \space\space\space\@spaces}{%
      Package graphicx or graphics not loaded%
    }{See the gnuplot documentation for explanation.%
    }{The gnuplot epslatex terminal needs graphicx.sty or graphics.sty.}%
    \renewcommand\includegraphics[2][]{}%
  }%
  \providecommand\rotatebox[2]{#2}%
  \@ifundefined{ifGPcolor}{%
    \newif\ifGPcolor
    \GPcolortrue
  }{}%
  \@ifundefined{ifGPblacktext}{%
    \newif\ifGPblacktext
    \GPblacktextfalse
  }{}%
  \let\gplgaddtomacro\g@addto@macro
  \gdef\gplbacktext{}%
  \gdef\gplfronttext{}%
  \makeatother
  \ifGPblacktext
    \def\colorrgb#1{}%
    \def\colorgray#1{}%
  \else
    \ifGPcolor
      \def\colorrgb#1{\color[rgb]{#1}}%
      \def\colorgray#1{\color[gray]{#1}}%
      \expandafter\def\csname LTw\endcsname{\color{white}}%
      \expandafter\def\csname LTb\endcsname{\color{black}}%
      \expandafter\def\csname LTa\endcsname{\color{black}}%
      \expandafter\def\csname LT0\endcsname{\color[rgb]{1,0,0}}%
      \expandafter\def\csname LT1\endcsname{\color[rgb]{0,1,0}}%
      \expandafter\def\csname LT2\endcsname{\color[rgb]{0,0,1}}%
      \expandafter\def\csname LT3\endcsname{\color[rgb]{1,0,1}}%
      \expandafter\def\csname LT4\endcsname{\color[rgb]{0,1,1}}%
      \expandafter\def\csname LT5\endcsname{\color[rgb]{1,1,0}}%
      \expandafter\def\csname LT6\endcsname{\color[rgb]{0,0,0}}%
      \expandafter\def\csname LT7\endcsname{\color[rgb]{1,0.3,0}}%
      \expandafter\def\csname LT8\endcsname{\color[rgb]{0.5,0.5,0.5}}%
    \else
      \def\colorrgb#1{\color{black}}%
      \def\colorgray#1{\color[gray]{#1}}%
      \expandafter\def\csname LTw\endcsname{\color{white}}%
      \expandafter\def\csname LTb\endcsname{\color{black}}%
      \expandafter\def\csname LTa\endcsname{\color{black}}%
      \expandafter\def\csname LT0\endcsname{\color{black}}%
      \expandafter\def\csname LT1\endcsname{\color{black}}%
      \expandafter\def\csname LT2\endcsname{\color{black}}%
      \expandafter\def\csname LT3\endcsname{\color{black}}%
      \expandafter\def\csname LT4\endcsname{\color{black}}%
      \expandafter\def\csname LT5\endcsname{\color{black}}%
      \expandafter\def\csname LT6\endcsname{\color{black}}%
      \expandafter\def\csname LT7\endcsname{\color{black}}%
      \expandafter\def\csname LT8\endcsname{\color{black}}%
    \fi
  \fi
  \setlength{\unitlength}{0.0500bp}%
  \begin{picture}(8502.00,5668.00)%
    \gplgaddtomacro\gplbacktext{%
    }%
    \gplgaddtomacro\gplfronttext{%
      \csname LTb\endcsname%
      \put(1331,564){\makebox(0,0){\strut{} \Large{20}}}%
      \put(1951,564){\makebox(0,0){\strut{} \Large{40}}}%
      \put(2572,564){\makebox(0,0){\strut{} \Large{60}}}%
      \put(3193,564){\makebox(0,0){\strut{} \Large{80}}}%
      \put(3813,564){\makebox(0,0){\strut{} \Large{100}}}%
      \put(4433,564){\makebox(0,0){\strut{} \Large{120}}}%
      \put(5053,564){\makebox(0,0){\strut{} \Large{140}}}%
      \put(5674,564){\makebox(0,0){\strut{} \Large{160}}}%
      \put(6295,564){\makebox(0,0){\strut{} \Large{180}}}%
      \put(6915,564){\makebox(0,0){\strut{} \Large{200}}}%
      \put(4123,150){\makebox(0,0){\strut{}\Large{$m_{\widetilde{\chi}}$ [GeV]}}}%
      \put(849,850){\makebox(0,0)[r]{\strut{} \Large{-18}}}%
      \put(849,1606){\makebox(0,0)[r]{\strut{} \Large{-16}}}%
      \put(849,2362){\makebox(0,0)[r]{\strut{} \Large{-14}}}%
      \put(849,3117){\makebox(0,0)[r]{\strut{} \Large{-12}}}%
      \put(849,3872){\makebox(0,0)[r]{\strut{} \Large{-10}}}%
      \put(849,4628){\makebox(0,0)[r]{\strut{} \Large{-8}}}%
      \put(849,5384){\makebox(0,0)[r]{\strut{} \Large{-6}}}%
      \put(200,3117){\rotatebox{-270}{\makebox(0,0){\strut{}\Large{$\log (\sigma^{\rm SI})$ [pb]}}}}%
      \put(7823,849){\makebox(0,0)[l]{\strut{} \Large{10}}}%
      \put(7823,3116){\makebox(0,0)[l]{\strut{} \Large{100}}}%
      \put(7823,5384){\makebox(0,0)[l]{\strut{} \Large{1000}}}%
      \put(8549,3116){\rotatebox{-270}{\makebox(0,0){\strut{}\Large{$\widetilde{\varSigma}$}$_{\rm FT}$}}}%
 \put(7065,4851){\makebox(0,0)[r]{\strut{} \footnotesize{ \rm \textbf{XENON100 (2011)}}}}%
\put(7065,4351){\makebox(0,0)[r]{\strut{} \footnotesize{ \rm \textbf{XENON100 (2012)}}}}%
 \put(6605,3951){\makebox(0,0)[r]{\strut{} \footnotesize{ \rm \textbf{XENON 1t}}}}%
    }%
    \gplbacktext
    \put(0,0){\includegraphics{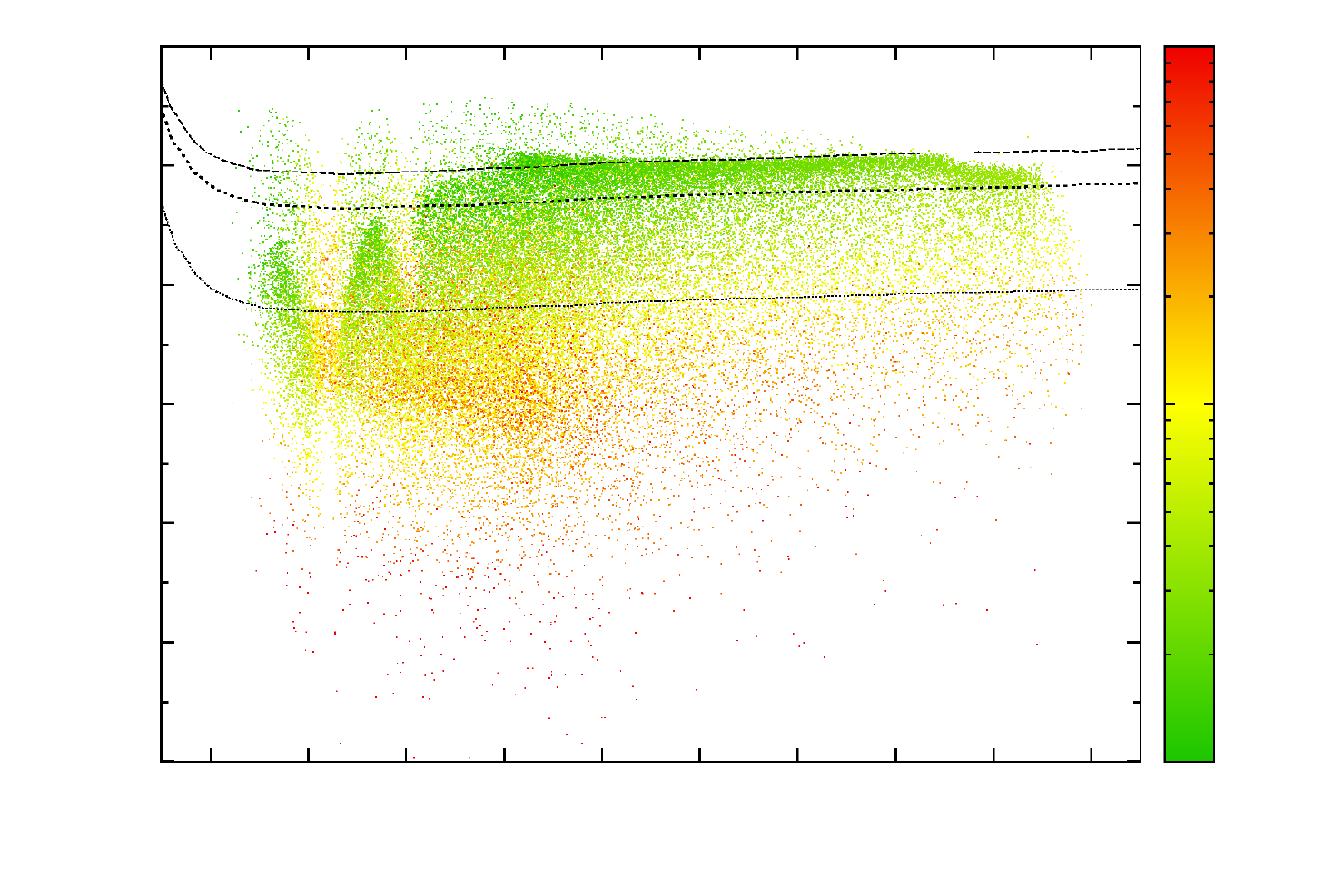}}%
    \gplfronttext
  \end{picture}%
\endgroup

%% file: gltresub11jhep.bbl
\begin{thebibliography}{99}

\bibitem{xenon2012}
E.~Aprile, ``Latest XENON100 Results", Talk given at Dark Attack 2012,
18 July 2012, Ascona, Switzerland

\bibitem{ATLASseminarHiggs}
F.~Gianotti, ``Status of Standard Model
  Higgs searches in ATLAS'', at Latest update in the search for the
  Higgs boson, 4 July 2012, CERN, Geneva, Schwitzerland.
%
\bibitem{CMSseminarHiggs}
J.~Incandela, ``Status of Standard Model
  Higgs searches in CMS'', at Latest update in the search for the
  Higgs boson, 4 July 2012, CERN, Geneva, Schwitzerland.

\bibitem{Ellis:1990nz}
  J.~R.~Ellis, G.~Ridolfi and F.~Zwirner,
  Phys.\ Lett.\ B {\bf 257} (1991) 83.

\bibitem{Ellis:1991zd}
  J.~R.~Ellis, G.~Ridolfi and F.~Zwirner,
  Phys.\ Lett.\ B {\bf 262} (1991) 477.


\bibitem{Okada:1990vk}
  Y.~Okada, M.~Yamaguchi and T.~Yanagida,
  Prog.\ Theor.\ Phys.\  {\bf 85} (1991) 1.

\bibitem{Haber:1990aw}
  H.~E.~Haber and R.~Hempfling,
  Phys.\ Rev.\ Lett.\  {\bf 66} (1991) 1815.

\bibitem{Drees:1991mx}
  M.~Drees and M.~M.~Nojiri,
  Phys.\ Rev.\ D {\bf 45} (1992) 2482.

\bibitem{Degrassi:2002fi}
  G.~Degrassi, S.~Heinemeyer, W.~Hollik, P.~Slavich and G.~Weiglein,
  Eur.\ Phys.\ J.\ C {\bf 28} (2003) 133
  [hep-ph/0212020].

\bibitem{Buchmueller:2009fn}
  O.~Buchmueller, R.~Cavanaugh, A.~De Roeck, J.~R.~Ellis, H.~Flacher, S.~Heinemeyer, G.~Isidori 
and K.~A.~Olive {\it et al.},
  Eur.\ Phys.\ J.\ C {\bf 64} (2009) 391
  [arXiv:0907.5568 [hep-ph]].

\bibitem{Komatsu:2010fb}
  E.~Komatsu {\it et al.}  [WMAP Collaboration],
  Astrophys.\ J.\ Suppl.\  {\bf 192} (2011) 18
  [arXiv:1001.4538 [astro-ph.CO]].

\bibitem{Goldberg:1983nd}
  H.~Goldberg,
  Phys.\ Rev.\ Lett.\  {\bf 50} (1983) 1419
   [Erratum-ibid.\  {\bf 103} (2009) 099905].

\bibitem{Ellis:1983ew}
  J.~R.~Ellis, J.~S.~Hagelin, D.~V.~Nanopoulos, K.~A.~Olive and M.~Srednicki,
  Nucl.\ Phys.\ B {\bf 238} (1984) 453.

\bibitem{Jungman:1995df}
  G.~Jungman, M.~Kamionkowski and K.~Griest,
  Phys.\ Rept.\  {\bf 267} (1996) 195
  [hep-ph/9506380].

\bibitem{Bergstrom:2000pn}
  L.~Bergstr{\"o}m,
  Rept.\ Prog.\ Phys.\  {\bf 63} (2000) 793
  [hep-ph/0002126].

\bibitem{Nilles:1983ge}
  H.~P.~Nilles,
  Phys.\ Rept.\  {\bf 110} (1984) 1.

\bibitem{Haber:1984rc}
  H.~E.~Haber and G.~L.~Kane,
  Phys.\ Rept.\  {\bf 117} (1985) 75.

\bibitem{Martin:1997ns}
  S.~P.~Martin,
  ``A Supersymmetry primer'', hep-ph/9709356.

\bibitem{Ellis:1986yg}
  J.~R.~Ellis, K.~Enqvist, D.~V.~Nanopoulos and F.~Zwirner,
  Mod.\ Phys.\ Lett.\ A {\bf 1} (1986) 57.

\bibitem{Barbieri:1987fn}
  R.~Barbieri and G.~F.~Giudice,
  Nucl.\ Phys.\ B {\bf 306} (1988) 63.

\bibitem{Farina:2011bh}
  M.~Farina, M.~Kadastik, D.~Pappadopulo, J.~Pata, M.~Raidal and A.~Strumia,
  Nucl.\ Phys.\ B {\bf 853} (2011) 607
  [arXiv:1104.3572 [hep-ph]].

\bibitem{Buchmueller:2011sw}
  O.~Buchmueller, R.~Cavanaugh, A.~De Roeck, M.~J.~Dolan, J.~R.~Ellis, H.~Flacher, S.~Heinemeyer and G.~Isidori {\it et al.},
  Eur.\ Phys.\ J.\ C {\bf 72} (2012) 1878
  [arXiv:1110.3568 [hep-ph]].

\bibitem{Buchmueller:2011ab}
  O.~Buchmueller, R.~Cavanaugh, A.~De Roeck, M.~J.~Dolan, J.~R.~Ellis, H.~Flacher, S.~Heinemeyer and G.~Isidori {\it et al.},
  arXiv:1112.3564 [hep-ph].


\bibitem{AbdusSalam:2009qd}
  S.~S.~AbdusSalam, B.~C.~Allanach, F.~Quevedo, F.~Feroz and M.~Hobson,
  Phys.\ Rev.\ D {\bf 81} (2010) 095012
  [arXiv:0904.2548 [hep-ph]].


\bibitem{Sekmen:2011cz}
  S.~Sekmen, S.~Kraml, J.~Lykken, F.~Moortgat, S.~Padhi, L.~Pape, M.~Pierini and H.~B.~Prosper {\it et al.},
  JHEP {\bf 1202} (2012) 075
  [arXiv:1109.5119 [hep-ph]].

\bibitem{Arbey:2011un}
  A.~Arbey, M.~Battaglia and F.~Mahmoudi,
  Eur.\ Phys.\ J.\ C {\bf 72} (2012) 1847
  [arXiv:1110.3726 [hep-ph]].

\bibitem{AlbornozVasquez:2012px}
  D.~Albornoz V{\'a}squez, G.~B{\'e}langer, J.~Billard and F.~Mayet,
  Phys.\ Rev.\ D {\bf 85} (2012) 055023
  [arXiv:1201.6150 [hep-ph]].

\bibitem{CahillRowley:2012rv}
  M.~W.~Cahill-Rowley, J.~L.~Hewett, A.~Ismail and T.~G.~Rizzo,
  arXiv:1206.5800 [hep-ph].

\bibitem{Bernabei:2010mq}
  R.~Bernabei {\it et al.}  [DAMA and LIBRA Collaborations],
  Eur.\ Phys.\ J.\ C {\bf 67} (2010) 39
  [arXiv:1002.1028 [astro-ph.GA]].

\bibitem{Aalseth:2010vx}
  C.~E.~Aalseth {\it et al.}  [CoGeNT Collaboration],
  Phys.\ Rev.\ Lett.\  {\bf 106} (2011) 131301
  [arXiv:1002.4703 [astro-ph.CO]].

\bibitem{Hooper:2002nq}
  D.~Hooper and T.~Plehn,
  Phys.\ Lett.\ B {\bf 562} (2003) 18
  [hep-ph/0212226].

\bibitem{Bottino:2002ry}
  A.~Bottino, N.~Fornengo and S.~Scopel,
  Phys.\ Rev.\ D {\bf 67} (2003) 063519
  [hep-ph/0212379].


\bibitem{Dreiner:2009ic}
  H.~K.~Dreiner, S.~Heinemeyer, O.~Kittel, U.~Langenfeld, A.~M.~Weber and G.~Weiglein,
  Eur.\ Phys.\ J.\ C {\bf 62} (2009) 547
  [arXiv:0901.3485 [hep-ph]].

\bibitem{Kuflik:2010ah}
  E.~Kuflik, A.~Pierce and K.~M.~Zurek,
  Phys.\ Rev.\ D {\bf 81} (2010) 111701
  [arXiv:1003.0682 [hep-ph]].

\bibitem{Feldman:2010ke}
  D.~Feldman, Z.~Liu and P.~Nath,
  Phys.\ Rev.\ D {\bf 81} (2010) 117701
  [arXiv:1003.0437 [hep-ph]].

\bibitem{Vasquez:2010ru}
  D.~A.~V{\'a}squez, G.~B{\'e}langer, C.~B{\oe}hm, A.~Pukhov and J.~Silk,
  Phys.\ Rev.\ D {\bf 82} (2010) 115027
  [arXiv:1009.4380 [hep-ph]].


\bibitem{Fornengo:2010mk}
  N.~Fornengo, S.~Scopel and A.~Bottino,
  Phys.\ Rev.\ D {\bf 83} (2011) 015001
  [arXiv:1011.4743 [hep-ph]].


\bibitem{Calibbi:2011ug}
  L.~Calibbi, T.~Ota and Y.~Takanishi,
  JHEP {\bf 1107} (2011) 013
  [arXiv:1104.1134 [hep-ph]].

\bibitem{Arbey:2012na}
  A.~Arbey, M.~Battaglia and F.~Mahmoudi,
  arXiv:1205.2557 [hep-ph].

\bibitem{Angloher:2011uu}
  G.~Angloher, M.~Bauer, I.~Bavykina, A.~Bento, C.~Bucci, C.~Ciemniak, G.~Deuter and F.~von Feilitzsch {\it et al.},
  Eur.\ Phys.\ J.\ C {\bf 72} (2012) 1971
  [arXiv:1109.0702 [astro-ph.CO]].

\bibitem{Calibbi:2011un}
  L.~Calibbi, T.~Ota and Y.~Takanishi,
  arXiv:1112.0219 [hep-ph].

\bibitem{Kopp:2011yr}
  J.~Kopp, T.~Schwetz and J.~Zupan,
  JCAP {\bf 1203} (2012) 001
  [arXiv:1110.2721 [hep-ph]].

\bibitem{Aprile:2011hi}
  E.~Aprile {\it et al.}  [XENON100 Collaboration],
  Phys.\ Rev.\ Lett.\  {\bf 107} (2011) 131302
  [arXiv:1104.2549 [astro-ph.CO]].

\bibitem{Aprile:2010um}
  E.~Aprile {\it et al.}  [XENON100 Collaboration],
  Phys.\ Rev.\ Lett.\  {\bf 105} (2010) 131302
  [arXiv:1005.0380 [astro-ph.CO]].

\bibitem{Angle:2007uj}
  J.~Angle {\it et al.}  [XENON Collaboration],
  Phys.\ Rev.\ Lett.\  {\bf 100} (2008) 021303
  [arXiv:0706.0039 [astro-ph]].

\bibitem{Ahmed:2010wy}
  Z.~Ahmed {\it et al.}  [CDMS-II Collaboration],
  Phys.\ Rev.\ Lett.\  {\bf 106} (2011) 131302
  [arXiv:1011.2482 [astro-ph.CO]].



\bibitem{Perelstein:2007nx}
  M.~Perelstein and C.~Spethmann,
  JHEP {\bf 0704} (2007) 070
  [hep-ph/0702038].

\bibitem{Hall:2011a}
  L.~J.~Hall, D.~Pinner and J.~T.~Ruderman,
  JHEP {\bf 1204}, 131 (2012)
  [arXiv:1112.2703 [hep-ph]].

\bibitem{Ellis:2001zk}
  J.~R.~Ellis and K.~A.~Olive,
  Phys.\ Lett.\ B {\bf 514} (2001) 114
  [hep-ph/0105004].

\bibitem{Kitano:2006gv}
  R.~Kitano and Y.~Nomura,
  Phys.\ Rev.\ D {\bf 73} (2006) 095004
  [hep-ph/0602096].

\bibitem{Cassel:2010px}
  S.~Cassel, D.~M.~Ghilencea and G.~G.~Ross,
  Nucl.\ Phys.\ B {\bf 835} (2010) 110
  [arXiv:1001.3884 [hep-ph]].

\bibitem{Ghilencea:2012gz}
  D.~M.~Ghilencea, H.~M.~Lee and M.~Park,
  arXiv:1203.0569 [hep-ph].

\bibitem{Perelstein:2011tg}
  M.~Perelstein and B.~Shakya,
  JHEP {\bf 1110} (2011) 142
  [arXiv:1107.5048 [hep-ph]].
%


\bibitem{AlbornozVasquez:2011yq}
  D.~Albornoz V{\'a}squez, G.~B{\'e}langer and C.~B{\oe}hm,
  Phys.\ Rev.\ D {\bf 84} (2011) 095015
  [arXiv:1108.1338 [hep-ph]].

\bibitem{MT1} 
  M.~Matsumoto and T.~Nishimura, ``Mersenne Twister: a
  623-dimensionally equidistributed uniform pseudorandom number
  generator'', ACM Transactions on Modeling and Computer Simulation
  {\bf 8} (1998) 3.

\bibitem{MT2} 
     T.~Nishimura, ``Tables of 64-bit Mersenne Twisters'', ACM
     Transactions on Modeling and Computer Simulation {\bf 10} (2000)
     348.


\bibitem{Dumont:2012ee}
  B.~Dumont, G.~B{\'e}langer, S.~Fichet, S.~Kraml and T.~Schwetz,
  arXiv:1206.1521 [hep-ph].

\bibitem{ATLAS:2012ae}
  G.~Aad {\it et al.}  [ATLAS Collaboration],
  Phys.\ Lett.\ B {\bf 710} (2012) 49
  [arXiv:1202.1408 [hep-ex]].



\bibitem{Chatrchyan:2012tx}
  S.~Chatrchyan {\it et al.}  [CMS Collaboration],
  Phys.\ Lett.\ B {\bf 710} (2012) 26
  [arXiv:1202.1488 [hep-ex]].


\bibitem{ATLAS:2012ad} 

G.~Aad {\it et al.}  [ATLAS Collaboration],
  Phys.\ Rev.\ Lett.\  {\bf 108} (2012) 111803
  [arXiv:1202.1414 [hep-ex]].


\bibitem{Chatrchyan:2012tw}
  S.~Chatrchyan {\it et al.}  [CMS Collaboration],
  Phys.\ Lett.\ B {\bf 710} (2012) 403
  [arXiv:1202.1487 [hep-ex]].



\bibitem{ATLAS-CONF-2012-019} 
  ATLAS Collaboration, ``An update to the combined search for the
  Standard Model Higgs boson with the ATLAS detector at the LHC using
  up to 4.9~fb$^{-1}$ of $pp$ collision data at $\sqrt{s}=7$~TeV'',
  ATLAS-CONF-2012-019.

\bibitem{CMS-PAS-HIG-12-008} 
  CMS Collaboration, ``Combined results of searches for a Higgs boson
  in the context of the standard model and beyond-standard models'',
  CMS-PAS-HIG-12-008.

\bibitem{Barberio:2008fa}
  E.~Barberio {\it et al.} [Heavy Flavor Averaging Group Collaboration],
  arXiv:0808.1297 [hep-ex].


\bibitem{Aaij:2012ac}
  R.~Aaij {\it et al.}  [LHCb Collaboration],
  arXiv:1203.4493 [hep-ex].
\bibitem{Asner:2010qj}
  D.~Asner {\it et al.}  [Heavy Flavor Averaging Group Collaboration],
  arXiv:1010.1589 [hep-ex].

\bibitem{Antonelli:2008jg}
  M.~Antonelli {\it et al.} [FlaviaNet Working Group on Kaon Decays Collaboration],
arXiv:0801.1817 [hep-ph].
%
\bibitem{Bennett:2006fi}
  G.~W.~Bennett {\it et al.}  [Muon G-2 Collaboration],
  Phys.\ Rev.\ D {\bf 73} (2006) 072003
  [hep-ex/0602035].

\bibitem{ALEPH:2005ema}
  [ALEPH and DELPHI and L3 and OPAL and SLD and LEP Electroweak Working Group and SLD Electroweak 
Group and SLD Heavy Flavour Group Collaborations],
  Phys.\ Rept.\  {\bf 427 } (2006)  257 [hep-ex/0509008].



\bibitem{Aad:2011ib}
  G.~Aad {\it et al.}  [ATLAS Collaboration],
  Phys.\ Lett.\ B {\bf 710} (2012) 67
  [arXiv:1109.6572 [hep-ex]].


\bibitem{Chatrchyan:2011zy}
  S.~Chatrchyan {\it et al.}  [CMS Collaboration],
  Phys.\ Rev.\ Lett.\  {\bf 107} (2011) 221804
  [arXiv:1109.2352 [hep-ex]].

\bibitem{atlas_conf} 
  G.~Aad {\it et al.}  [ATLAS Collaboration], ``Search for squarks and
  gluinos with the ATLAS detector using final states with jets and
  missing transverse momentum and 4.7 fb$^{-1}$ of $\sqrt{s}$ = 7 TeV
  proton-proton collision data'', ATLAS-CONF-2012-033.

\bibitem{parker:2012}
A.~Parker, ``SUSY Searches (ATLAS/CMS): the Lady Vanishes'', ICHEP 2012, 4-11 July 2012, Melbourne, Australia.

\bibitem{Aad:2011rv}
  G.~Aad {\it et al.}  [ATLAS Collaboration],
  Phys.\ Lett.\ B {\bf 705} (2011) 174
  [arXiv:1107.5003 [hep-ex]].


\bibitem{Chatrchyan:2012vp}
  S.~Chatrchyan {\it et al.}  [CMS Collaboration],
  Phys.\ Lett.\ B {\bf 713} (2012) 68
  [arXiv:1202.4083 [hep-ex]].


\bibitem{Abbiendi:2003sc}
  G.~Abbiendi {\it et al.} [OPAL Collaboration],
  Eur.\ Phys.\ J.\  {\bf C35 } (2004)  1 [hep-ex/0401026].

\bibitem{Nakamura:2010zzi}
  K.~Nakamura {\it et al.}  [Particle Data Group Collaboration],
  J.\ Phys.\ G G {\bf 37} (2010) 075021.

\bibitem{Djouadi:2002ze}
  A.~Djouadi, J.-L.~Kneur and G.~Moultaka,
  Comput.\ Phys.\ Commun.\  {\bf 176} (2007) 426
  [hep-ph/0211331].

\bibitem{Belanger:2006is}
  G.~B{\'e}langer, F.~Boudjema, A.~Pukhov and A.~Semenov,
  Comput.\ Phys.\ Commun.\  {\bf 176} (2007) 367
  [hep-ph/0607059].

\bibitem{Belanger:2010gh}
  G.~B{\'e}langer, F.~Boudjema, P.~Brun, A.~Pukhov, S.~Rosier-Lees, P.~Salati and A.~Semenov,
  Comput.\ Phys.\ Commun.\  {\bf 182} (2011) 842
  [arXiv:1004.1092 [hep-ph]].

\bibitem{Mahmoudi:2008tp}
  F.~Mahmoudi,
  Comput.\ Phys.\ Commun.\  {\bf 180} (2009) 1579
  [arXiv:0808.3144 [hep-ph]].


\bibitem{Belanger:2008sj}
  G.~B{\'e}langer, F.~Boudjema, A.~Pukhov and A.~Semenov,
  Comput.\ Phys.\ Commun.\  {\bf 180} (2009) 747
  [arXiv:0803.2360 [hep-ph]].



\bibitem{Weniger:2012tx}
  C.~Weniger,
  arXiv:1204.2797 [hep-ph].



\bibitem{Nihei:2002ij}
  T.~Nihei, L.~Roszkowski and R.~Ruiz de Austri,
  JHEP {\bf 0203} (2002) 031
  [hep-ph/0202009].


\bibitem{Dreiner:2012}
  H.~K.~Dreiner, J.~S.~Kim and O.~Lebedev,
  arXiv:1206.3096 [hep-ph].

\bibitem{Falk:1998xj}
  T.~Falk, A.~Ferstl and K.~A.~Olive,
  Phys.\ Rev.\ D {\bf 59} (1999) 055009
   [Erratum-ibid.\ D {\bf 60} (1999) 119904]
  [hep-ph/9806413].

\bibitem{Bottino:1998jx}
  A.~Bottino, F.~Donato, N.~Fornengo and S.~Scopel,
  Phys.\ Rev.\ D {\bf 59} (1999) 095003
  [hep-ph/9808456].

\bibitem{Choi:2001ww}
  S.~Y.~Choi, J.~Kalinowski, G.~A.~Moortgat-Pick and P.~M.~Zerwas,
  Eur.\ Phys.\ J.\ C {\bf 22} (2001) 563
   [Addendum-ibid.\ C {\bf 23} (2002) 769]
  [hep-ph/0108117].

\bibitem{Kitano:2006ws}
  R.~Kitano and Y.~Nomura,
  hep-ph/0606134.
%

\bibitem{Cho:2011rk}
  G.-C.~Cho, K.~Hagiwara, Y.~Matsumoto and D.~Nomura,
  JHEP {\bf 1111} (2011) 068
  [arXiv:1104.1769 [hep-ph]].
%

\bibitem{Kim:2002cy} 
  Y.~G.~Kim, T.~Nihei, L.~Roszkowski and R.~Ruiz de Austri,
  JHEP {\bf 0212} (2002) 034
  [hep-ph/0208069].
%
\bibitem{LHCb:2012ct}
  R.~Aaij {\it et al.}  [LHCb Collaboration],
   arXiv:1211.2674 [hep-ex].


\bibitem{Aad:2012fqa}
  G.~Aad {\it et al.}  [ATLAS Collaboration],
  Phys.\ Rev.\ D {\bf 87} (2013) 012008
  [arXiv:1208.0949 [hep-ex]]; ATLAS-CONF-2013-007.

\bibitem{Cheung:2012qy}
  C.~Cheung, L.~J.~Hall, D.~Pinner and J.~T.~Ruderman,
  arXiv:1211.4873 [hep-ph].


\bibitem{Perelstein:2012qg}
  M.~Perelstein and B.~Shakya,
  arXiv:1208.0833 [hep-ph].





\end{thebibliography}
